\documentclass[
  aps,
  prx,
  reprint,
  superscriptaddress,
  longbibliography,
  nofootinbib
]{revtex4-1}

\usepackage[T1]{fontenc}
\usepackage[utf8]{inputenc}
\usepackage{amsmath,amssymb,bm,mathtools}
 
\IfFileExists{newtxtext.sty}{
  \usepackage{newtxtext}
  \usepackage{newtxmath}
}{
  \usepackage{mathptmx}
}
\usepackage{microtype}

\usepackage{graphicx}
\usepackage{booktabs}
\usepackage{dcolumn}
\usepackage{xcolor}
\usepackage{hyperref}
\usepackage[nameinlink,noabbrev]{cleveref}

\hypersetup{
  colorlinks=true,
  linkcolor=blue!45!black,
  citecolor=blue!55!black,
  urlcolor=blue!55!black,
  pdfauthor={Jamir Marino},
  pdftitle={Minimal quantum scaling of critical active matter}
}

\newcommand{\dd}{\mathop{}\!\mathrm{d}}
\newcommand{\ii}{\mathrm{i}}
\newcommand{\ee}{\mathrm{e}}
 \newcommand{\order}{\mathcal{O}}

 \DeclareRobustCommand{\revision}[1]{#1}
 \DeclareRobustCommand{\newtext}[1]{#1}
\DeclareRobustCommand{\newedit}[1]{#1}
\DeclareRobustCommand{\newchange}[1]{#1}
\DeclareRobustCommand{\figedit}[1]{#1}
\DeclareRobustCommand{\minorchange}[1]{#1}
\DeclareRobustCommand{\auditedit}[1]{#1}
 \DeclareRobustCommand{\secviedit}[1]{#1}

\DeclareRobustCommand{\introfix}[1]{#1}
\newcommand{\longdir}{\parallel}
\DeclareRobustCommand{\vthirtytwoedit}[1]{#1}
\newcommand{\onedir}{x}
\DeclareRobustCommand{\vthirtythreeedit}[1]{#1}

 \expandafter\def\csname v31bib@hofferberth2008\endcsname{}
 \newcommand{\markvthirtybib}[1]{\expandafter\def\csname v30bib@#1\endcsname{}}
\markvthirtybib{metelmannClerk2015}
\markvthirtybib{belyansky2025}
\markvthirtybib{nadolny2025}
 \newcommand{\markintrobib}[1]{}
 \newcommand{\markrevisionbib}[1]{%
  \expandafter\def\csname revisionbib@#1\endcsname{}}

\newcommand{\markqcnewbib}[1]{}

 \let\manuscriptbibliography\bibliography
\renewcommand{\bibliography}[1]{\clearpage\manuscriptbibliography{#1}}

\begin{document}

\title{Quantum critical scaling in convective active matter}

\author{Jamir Marino}
\affiliation{Department of Physics, The State University of New York at Buffalo, Buffalo, New York 14260, USA}

\begin{abstract}
{We construct a minimal field theory for critical quantum active matter, describing a nonconserved scalar order parameter with Landau-Ginzburg ordering and self-advection. When diffusive noise dominates, higher-order quantum vertices and non-Gaussian noise fluctuations remain relevant alongside the deterministic nonlinearities, with an upper critical dimension $d_c=2$. A one-loop renormalization-group analysis in $d=2-\epsilon$ reveals a purely dissipative fixed point, where nonlinear noise suppresses convection, and a quantum convective saddle, where   interactions, convection, and non-Gaussian noise coexist. We determine the critical exponents, show that the convective regime is distinct from conventional Burgers hydrodynamics, and discuss how it could govern transient universal scaling in microscopic models of quantum active matter before a   crossover to semi-classical  dynamics.}
\end{abstract}

\maketitle

\section{Introduction}\label{sec:introduction}

Active matter refers to systems composed of constituents that continuously consume or extract energy from their environment and convert it into autonomous motion, forces, or stresses, thereby maintaining the system out of thermodynamic equilibrium.
{For example, bacteria use chemical energy to swim, while molecular motors consume ATP to slide cytoskeletal filaments and generate stresses. Energy is converted locally into motion or force, rather than being continuously pumped by an external driving agent~\cite{marchetti2013}.}

In the  {past few years} there have been attempts to merge the field of active matter with quantum mechanics. The basic approach is to understand how quantum coherence and entanglement can  {intertwine} with the   mechanisms for non-equilibrium dynamics in active systems,  {with the long-term goal of engineering systems with no counterpart in either AMO quantum simulation or quantum materials.}  If successful, these efforts would bring together quantum many-body dynamics and the physics of living and complex systems -- a synthesis that could open new avenues in the search for exotic emergent quantum phenomena.

 Microscopic  models that may mimic 'activity' in a quantum setting are blossoming. On a lattice, spin-dependent dissipative hopping and local alignment produce quantum flocks with spatial coherence~\cite{khasseh2025,radhakrishnan2026}. In non-Hermitian models, spin-dependent asymmetric hopping can induce collective ordering, including ferromagnetism without an underlying microscopic ferromagnetic interaction~\cite{adachi2022,takasan2024}. Other proposals mimic self-propulsion using a moving trap~\cite{zheng2025}, convert heat absorbed by internal states into motion through spin--orbit coupling~\cite{penner2025}, or use oscillating magnetic fields in spin systems~\cite{hardt2025propelling}. These extensions encompass also monitored spin systems as demonstrated in Ref.~\cite{steiner2026}.
 Non-reciprocal   interactions  are     among the most promising directions for grounding quantum active matter into microscopic models~\cite{metelmannClerk2015,belyansky2025, nadolny2025,hanai2025photoinduced}. 

As is apparent from this list, these attempts span diverse research areas, reflecting the field's
early stage of development, and making it   natural     to search for a  systematic classification that relies on a minimal set of ingredients, with the flavor of universality. 
{The theory of critical phenomena traditionally provides a standard route to achieve this goal. Near a phase transition, microscopic models can share critical exponents and scaling functions because only a few couplings remain relevant under coarse graining. For dynamics, the Hohenberg--Halperin classification illustrates which additional information is needed~\cite{hohenbergHalperin1977}. Beyond the symmetry and number of components of the order parameter, one must specify whether the order parameter is conserved and whether it couples to other slow densities or hydrodynamic modes. Together with spatial dimension, these properties determine the nonlinear mechanisms that can affect critical relaxation and universality beyond simple Gaussian or mean-field models. Detailed balance and its associated fluctuation-dissipation symmetry provide a further distinction: microscopic violations may remain RG relevant or disappear at large scales~\cite{tauberAkkineniSantos2002,sieberer2015equilibrium,tauber2017}, opening the door to new universality classes out of equilibrium~\cite{sieberer2025}.}

The construction of field  theories for classical active matter is a developed field~\cite{marchetti2013,chate2020}.  The most iconic example is Toner--Tu hydrodynamics, where convection can stabilize polar order in two dimensions~\cite{tonerTu1995,tonerTu1998,tonerTuRamaswamy2005}, as a result of  {the} breakdown of  {the} Mermin-Wagner theorem out of equilibrium. The Toner-Tu model captures the non-equilibrium physics of self-propelled particles through the interplay of a Landau-Ginzburg ordering potential and nonlinear convection, through which the particles advect their own polar order. This coupling between spontaneous ordering and self-propelled transport is a defining feature of  active matter.
Because of this groundbreaking result and the simplicity of the model, Toner-Tu is considered the
'Ising model' of classical active matter~\cite{RevModPhys.85.1143,ramaswamy2017active,bowick2022symmetry,chen2025inconvenient}.
More broadly, classical active field theories are organized by the symmetries and conservation laws obeyed by their coarse-grained fields~\cite{teVrugtLiebchenCates2025,teVrugtWittkowski2025}. In dry active matter, momentum is not conserved, while particle number usually is: flocking hydrodynamics thus couples a conserved density to nonconserved polar order. In the scalar sector, 'Active Models B' and 'B+' extend  Model B of the Halperin-Hohenberg classification, capturing phenomena such as motility-induced phase separation~\cite{wittkowski2014,tjhung2018}. Conservation laws thus constrain the slow hydrodynamic modes and allowed nonlinearities, which, together with broken detailed balance, shape the possible non-equilibrium universality classes of classical active matter.

{For quantum active matter, the corresponding task  {of identifying} which quantum fluctuations remain relevant in a continuum description and how they change the renormalization-group (RG) flow  {remains} currently unaddressed. This question belongs to the broader effort to understand universality far from  equilibrium in quantum systems~\cite{polkovnikov2011,sieberer2025}. Isolated quantum gases display turbulent transport and self-similar evolution near nonthermal fixed points~\cite{bergesRothkopf2008,nowak2011,prufer2018,erne2018, rodriguez2022far,marcuzzi2016prethermalization}. Driven condensates exhibit universal critical dynamics and nonequilibrium phase fluctuations~\cite{sieberer2013,altman2015,fontaine2022}, while periodic driving supports phases of matter without equilibrium  {counterparts}~\cite{else2016}, including   anomalous topological edge states~\cite{rudner2013}.  }
{However, the mere presence of  microscopic quantum dynamics does not generically ensure that quantum fluctuations will control the asymptotic scaling regime. Thermalization and dephasing can make local observables in isolated systems insensitive to their initial coherence, even though the full state evolves unitarily~\cite{dalessio2016}. In open systems,  noise can induce emergent thermalization in the infrared, which trivializes  scaling behavior at large distances and long timescales at criticality~\introfix{\cite{sieberer2016,sieberer2025,dallaTorre2013Dicke,maghrebiGorshkov2016,kelly2020exploring}}. } Yet, on pre-thermal timescales, quantum fluctuations could reshape critical exponents and give
rise to a theory of universality genuinely different from its nonequilibrium classical counterpart.

{In this work we construct a minimal quantum field theory model for critical quantum active matter, inspired by a one-dimensional scalar version of Toner--Tu hydrodynamics  (along similar lines, dynamical quantum field theories inspired by the van der Pol oscillator have been recently put forward in Refs.~\cite{zelle2024,daviet2024,zelle2026goldstone}). In our model, a single real, non-conserved order parameter combines    Landau--Ginzburg ordering with self-advection. In one dimension (as \introfix{often also happens} at equilibrium \introfix{\cite{sachdev2011,gring2012}}) ordering, convection, and quantum fluctuations compete on equal footing.   }
{The associated RG flow contains two interacting fixed points. At the purely dissipative (PD) fixed point, non-Gaussian noise survives while convection and quartic nonlinearities of the Landau-Ginzburg potential landscape  are suppressed. At the quantum convective (QC) fixed point, the quartic nonlinearity, convection, and nonlinear noise instead coexist. After   mass perturbations are tuned, the QC fixed point has one relevant interaction direction: its stable manifold separates trajectories attracted toward PD from a runaway flow beyond perturbative control. The critical exponents and the scaling of the noise-to-diffusion ratio distinguish this QC fixed point from the conventional driven-diffusive regime. Despite being a RG saddle, we argue that the QC fixed point can govern an extended scaling regime before  emergent classical dynamics take over, offering a first instance of universal quantum   scaling of   active matter. 

 }

\section{Synopsis: model and key results}\label{sec:synopsis}

\subsection{Model and quantum scaling}\label{sec:model}

We consider a real scalar order parameter $\phi(x,t)$ with one-dimensional noisy dynamics
\begin{equation}
  \left(\gamma\partial_t-K\partial_{\onedir}^2+r\right)\phi
  -\lambda_c\phi\,\partial_{\onedir}\phi
  +\frac{u_c}{6}\phi^3
  = \xi .
  \label{eq:minimal-active-model}
\end{equation}
The coefficient $K>0$ controls spatial diffusion, while $\gamma$ is responsible for friction. The   mass $r$ and the nonlinearity $u_c$ establish the onset of ordering in the system, as in conventional statistical mechanics. Self-advection of the Burgers type~\cite{kardar1986,canet2011}, with coefficient $\lambda_c$,
couples the field to its own gradient and transports it with a velocity proportional to its local value. This nonlinear transport makes different parts of the profile of $\phi(x,t)$ move at different velocities, deforming it  and  steepening its gradients.  On the other hand, the   mass and quartic potential terms break  Galilean invariance in the Burgers equation allowing  $\lambda_c$ to receive fluctuation corrections~\cite{forster1977,canet2011}. This absence of Galilean protection is also familiar in  flocking theories and Toner-Tu hydrodynamics~\cite{tonerTu1995,tonerTu1998,tonerTuRamaswamy2005,marchetti2013}, where the interplay between ordering (promoted by the Landau-Ginzburg non-linearities) and self-propelled transport is at the root of 'active' behavior.  Together with the noise introduced in the following, the convective term $\lambda_c$ is the main non-equilibrium agent in our model.

The quadratic noise spectrum is short-ranged in time (Markovian) and contains both
a momentum-independent and a diffusive contribution:
\begin{align}
  \langle \xi(q,\omega)\xi(q',\omega')\rangle
  &=
  2D(q)(2\pi)^2\delta(q+q')\delta(\omega+\omega'),
  \nonumber\\
  D(q)&=\Gamma_0+\Gamma_1q^2 .
  \label{eq:noise-spectrum}
\end{align}
Here $\Gamma_0$ is the flat-noise strength and $\Gamma_1$ the
diffusive-noise coefficient (as it occurs in model B of the Hohenberg--Halperin classification~\cite{hohenbergHalperin1977}, in Langevin driven-diffusive systems~\cite{janssen1986field}, or in   quantum critical  driven Bose-Einstein condensates~\cite{marinoDiehl2016,marinoDiehlPRL2016}).   The choice $\Gamma_1=0$ gives the usual flat noise present in scalar Toner--Tu
models~\cite{tonerTu1995,tonerTu1998,chate2020}.  Suppressing $\Gamma_0$ instead changes the power counting
at criticality~\cite{marinoDiehl2016,marinoDiehlPRL2016}: this is the knob we will use  to enter quantum critical scaling in the dynamics of~\eqref{eq:minimal-active-model} when the mass is   tuned to zero ($r\to0$).  
At $r=0$, the relaxation rate is $Kq^2/\gamma$, giving the Gaussian
dynamical exponent $z=2$.  For $\Gamma_0=0$, the noise vanishes with the
same power of momentum.  Response and fluctuations then retain equal
canonical weight, and   higher quantum vertices in Keldysh field theory, that are suppressed
at  finite-noise   critical points (where a semi-classical description is sufficient), must be retained~\cite{marinoDiehl2016,sieberer2016, sieberer2013,maghrebiGorshkov2016,dallaTorre2013Dicke}.  These extra terms are typically higher order cumulants of the noise that encode a mixture of classical and quantum correlations: the former are introduced by   noise  in Eq.~\eqref{eq:minimal-active-model}, while   quantum correlations will depend on the specific microscopic model considered (see Sec.~\ref{sec:microscopic}). These    non-Gaussian terms  break detailed balance and drive the evolution of~\eqref{eq:minimal-active-model}  away from its semi-classical limit. In solving the noisy dynamics of many-body systems one should usually keep the full hierarchy of correlations (BBGKY hierarchy~\cite{huang2008statistical}), but close to non-equilibrium quantum critical points one can truncate such hierarchy to a few RG relevant terms. These are the only ones  responsible for determining the universality class of dynamics.

This is
the sense in which diffusive noise supports quantum scaling in a driven
open system~\cite{marinoDiehl2016,sieberer2025}.   Its role is to permit quantum RG operators beyond semi-classics   to survive coarse graining.
Residual flat noise limits this regime.  Equating the two terms in
$D(q)$ gives the \auditedit{bare} estimate
\begin{equation}
  q_M\simeq\sqrt{\Gamma_0/\Gamma_1}.
  \label{eq:markov-crossover}
\end{equation}
 
Diffusive-noise scaling requires $q_M\ll q\ll q_G$, where
$q_G$ is the scale below which nonlinear fluctuations become appreciable (known as the Ginzburg scale~\cite{marinoDiehl2016,altlandSimons2023}).
At $q\ll q_M$, flat noise dominates and the dynamics crosses into a
semiclassical regime~\cite{marinoDiehl2016,marinoDiehlPRL2016} (cf. Sec.~\ref{sec:semiclassics}).  Thus $\Gamma_0$ plays a role analogous to finite
temperature near an equilibrium quantum critical point~\cite{marinoDiehl2016}, although here its value is
itself renormalized since the system is out of equilibrium and there is no temperature protected by fluctuation-dissipation~\minorchange{\cite{sieberer2014,sieberer2016}}.   The tadpole corrections in
Sec.~\ref{sec:technical-kinetic} and the mass flow in
Sec.~\ref{sec:rg-masses}  explicitly compute the renormalization of $q_M$, which we address in Sec.~\minorchange{\ref{sec:infrared-crossover}}.

\begin{figure}[t]
  \centering
  \includegraphics[width=\columnwidth]{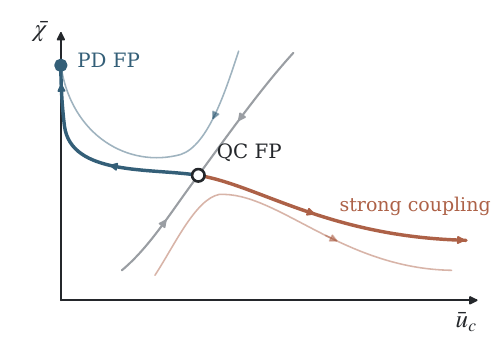}
  \caption{Schematic RG flow projected onto the dimensionless quartic
  coordinate \revision{$\bar u_c$} and non-Gaussian-noise strength $\bar\chi$.
  Arrows point toward longer length scales following the RG process.  The QC FP is a saddle:
  trajectories passing near it exhibit a common critical crossover before
  flowing toward the PD FP or leaving the perturbative regime.  At PD,
  the quartic Landau-Ginzburg term and both advective couplings become irrelevant, while
  nonlinear noise remains.   }
  \label{fig:rg-cartoon}
\end{figure}

Let us now discuss the extra  nonlinear terms that become RG relevant in the new quantum scaling regime of our scalar analog of the Toner--Tu equation.  The majority of them ($\Gamma_2$, $\lambda_q$, $\chi$) can be described by looking at higher order noise cumulants:  
\begin{align}
  \langle\xi_1\xi_2\rangle_{c}
  &=\left[\left(\Gamma_0-\Gamma_1\partial_{\onedir_1}^2\right)
    +\frac{\Gamma_2}{4}\phi_1^2\right.\nonumber\\[-2pt]
  &\hspace{3.2cm}\left.-\frac{\lambda_q}{2}\partial_{\onedir_1}\phi_1\right]\delta_{12},
  \nonumber\\
  \langle\xi_1\xi_2\xi_3\xi_4\rangle_{c}
  &=-\chi\delta_{12}\delta_{13}\delta_{14},
  \label{eq:nonlinear-noise-cumulants}
\end{align}
where $\xi_1\equiv\xi(x_1,t_1)$,
$\phi_1\equiv\phi(x_1,t_1)$, 
$\delta_{ij}\equiv\delta(x_i-x_j)\delta(t_i-t_j)$ and 'c' denotes connected correlations. 
The multiplicative-noise coupling $\Gamma_2$ makes the variance depend
on the squared field, while the non-Gaussian-noise
strength $\chi$ specifies the fourth cumulant (kurtosis).    The quantum convective noise coupling
$\lambda_q$ makes the variance depend on the field gradient through
$-\lambda_q\partial_{\onedir}\phi$.
This exotic fluctuation-induced counterpart of advection is absent from conventional hydrodynamics and,
to the best of our knowledge, has not previously been addressed in the literature. Finally, there is also the quantum counterpart of the quartic Landau--Ginzburg potential (later parametrized by $u_q$). This vertex is already present in unitary scalar quantum field theories, in or out of equilibrium~\cite{kamenev2023,mesterhazy2015quantum}, and captures fluctuations on top of the double well potential.  
 
 Diffusive noise   promotes higher response-field vertices as RG relevant effects. In a microscopic quantum model, these vertices can receive contributions both from coherent and dissipative processes, enabling dynamics beyond the usual semiclassical truncation to influence critical scaling (see also Sec.~\ref{sec:semiclassics}). Higher response-field vertices can however also occur in   non-Gaussian stochastic field theories with no residual quantum coherence~\cite{lin2023non}.  The purpose of our work is to   determine the   RG structure of this enlarged field theory, which would   describe both quantum active matter or an equally rich non-Gaussian stochastic theory of emergently classical convective  matter. Only the specific microscopic  realizations and  their non-equilibrium dynamics   can discriminate  whether quantum coherence and entanglement will survive at the critical point. If this occurs, then the proper RG framework to capture this scenario   is the one provided by this work. At the QC fixed point  we find a non-vanishing    coupling $u_q$, which is a RG operator that  intrinsically carries information about coherent quantum  fluctuations on top of the Landau-Ginzburg potential. In this sense, although not conclusive about each specific microscopic   model related to the field theory presented here, our findings promote the idea that the RG flow can support quantum fluctuations in the vicinity of its fixed points.

The  scaling limit   considered places $\lambda_q$ on the same
footing as $\lambda_c$, and $\Gamma_2,u_q,\chi$ on the same footing as
$u_c$, at the level of canonical power counting.  These two cubic and four quartic vertices form the augmented 
nonlinear field theory that we will use to compute RG corrections close to the upper critical dimension $d=2$. To see why the upper critical dimension is reduced compared to standard Toner-Tu~\cite{tonerTu1995}, recall that diffusive noise and spatial relaxation both vanish as $q^2$. With $z=2$, the physical and response fields then have the same canonical dimension, $d/2$, so the quartic couplings have dimension $2-d$ and the convective couplings have dimension $(2-d)/2$. Both become marginal at $d=2$. At this level, the quartic power counting can be written as $2-d=4-(d+z)$: one spatial dimension with $z=2$ has effective dimension three, which already hints at the possibility of spontaneous symmetry breaking and ordering of the field, although the system lives in one spatial dimension. Diffusive noise also removes the Gaussian infrared enhancement of equal-time fluctuations, since $D(q)/(\gamma Kq^2)$ tends to a constant.  

Following the analysis and remarks of this section, we  will  regard Eq.~\eqref{eq:minimal-active-model} as a minimal quantum field theory of  convective active matter in the following.

\subsubsection{Limiting cases}\label{sec:limiting-cases}

Before moving further with a synopsis of the results, we briefly outline some limiting cases of our model in order to make contact with established literature.  First, $r=u_c=0$ removes the local   Landau-Ginzburg potential responsible for the onset of a mean-field expectation value of $\phi$.  When the extra nonlinearities present in the quantum scaling limit are omitted by setting  
$\Gamma_2=u_q=\chi=\lambda_q=0$, the resulting noisy Burgers equation
has Galilean symmetry.  If also $\Gamma_0=0$,
the noise is a fluctuating current, $\xi=\partial_{\onedir}\eta_{\mathrm{KPZ}}$, with
$\langle\eta_{\mathrm{KPZ}}(x,t)\eta_{\mathrm{KPZ}}(x',t')\rangle=2\Gamma_1\delta(x-x')\delta(t-t')$.
The field $\phi$ is then conserved.  Writing
$\phi=\partial_{\onedir}h$ gives
\begin{equation}
  \gamma\partial_t h
  =K\partial_{\onedir}^2h+\frac{\lambda_c}{2}(\partial_{\onedir}h)^2+\eta_{\mathrm{KPZ}},
  \label{eq:kpz-limit}
\end{equation}
the one-dimensional Kardar--Parisi--Zhang (KPZ) equation
\cite{kardar1986,tauber2014}.
For $\Gamma_0\ne0$, the height noise acquires a $q^{-2}$ contribution
and the mapping no longer gives standard  local KPZ.   
Finally, by removing the advective terms $\lambda_c=\lambda_q=0$, one recovers a Langevin field theory which can sustain classical or quantum scaling at criticality depending whether strong diffusive noise ($\Gamma_1$) is included.

\subsubsection{Semi-classical and quantum limits}\label{sec:semiclassics}

For the sake of completeness, in this short sub-section we overview the nomenclature used above  when referring to the action as semi-classical or quantum (for a   comprehensive  textbook treatment, see Refs.~\cite{kamenev2023,sieberer2016}). 
To expose the classical limit, we normalize the  difference between the two Keldysh contour fields  as \(\phi_+-\phi_-=\hbar\tilde{\phi}\): this compensates the overall factor \(1/\hbar\) when the Keldysh action is exponentiated in the path integral, keeping the coupling to an external source finite as \(\hbar\to0\) (this is the correct limit, as the response formalism should apply   to   classical field theories as well).   Writing \(\phi_\pm=\phi\pm\hbar\tilde{\phi}/2\), a coherent quartic potential on the Keldysh path integral  gives
\[
\frac{V(\phi_+)-V(\phi_-)}{\hbar}
\sim{u}\phi^3\tilde{\phi}
+{\hbar^2u}\phi\tilde{\phi}^3.
\]
The term linear in the response field survives the classical limit, whereas the cubic-response vertex represents an explicit quantum correction. This applies also to other vertices, for instance $\lambda_q$ would carry a factor $\hbar$, while $\chi$ would scale as $\hbar^3$.

\subsection{Quantum convective fixed point}
\label{sec:synopsis-results}

\begin{table*}[t]
  \caption{Summary of the interacting   fixed points depicted in Fig.~\ref{fig:rg-cartoon}.}
  \label{tab:rg-outcomes}
  \begingroup%
  \small%
  \renewcommand{\arraystretch}{1.16}%
  \setlength{\tabcolsep}{6pt}%
  \begin{tabular*}{\textwidth}{@{\extracolsep{\fill}}lll@{}}
  \toprule
  \parbox[t]{0.16\textwidth}{\raggedright \textbf{Regime}} &
  \parbox[t]{0.39\textwidth}{\raggedright \textbf{Physical properties}} &
  \parbox[t]{0.36\textwidth}{\raggedright \textbf{RG signature}} \\
  \midrule
  \parbox[t]{0.16\textwidth}{\raggedright PD fixed point} &
  \parbox[t]{0.39\textwidth}{\raggedright Nonlinear noise suppresses the quartic potential and both convective couplings.} &
  \parbox[t]{0.36\textwidth}{\raggedright $\bar u_c^*=\bar\lambda_c^*=\bar\lambda_q^*=0$;
  non-Gaussian noise remains.} \\[3pt]
  \parbox[t]{0.16\textwidth}{\raggedright QC fixed point} &
  \parbox[t]{0.39\textwidth}{\raggedright Quartic nonlinearity, self-advection, and non-Gaussian noise coexist.} &
  \parbox[t]{0.36\textwidth}{\raggedright One relevant interaction direction;   extended critical crossover.} \\[3pt]
  \parbox[t]{0.16\textwidth}{\raggedright Strong coupling} &
  \parbox[t]{0.39\textwidth}{\raggedright Nonlinearities grow beyond perturbative control.} &
  \parbox[t]{0.36\textwidth}{\raggedright Runaway flow towards strong coupling.} \\
  \bottomrule
  \end{tabular*}%
  \endgroup%
\end{table*}

We now summarize at a qualitative level the outcome of our RG analysis. 
The RG flow shows that non-Gaussian noise can either suppress convection
or coexist with it at criticality (cf. Table~\ref{tab:rg-outcomes}).  We refer to these regimes as the
\emph{purely dissipative} fixed point (PD FP) and the \emph{quantum convective} fixed
point (QC FP), respectively.  The quantum convective fixed point acts as a separatrix between RG flows running
 toward PD or departing toward strong coupling.
 As we will see later on, in   practice the QC fixed point dominates the scaling behavior, since   the flat noise ($\Gamma_0$) may cut the RG flow before the crossover can be observed (see Sec.~\ref{sec:infrared-crossover}). 
Figure~\ref{fig:rg-cartoon} gives a schematic projection of this
competition, and Table~\ref{tab:rg-outcomes} summarizes its physical
content.  All the statements in this summary concern the
interaction sector with the relevant mass and flat-noise perturbations tuned to their critical values ($r\to0$ and $q_M\ll q \ll q_G$, or in other words large $\Gamma_1$).

We work to one loop at $d=2-\epsilon$, with $\epsilon>0$; bars denote
the dimensionless RG variables defined in Sec.~\ref{sec:rg-variables} (i.e. rescaled by appropriate momentum powers).
At PD, $\bar u_c^*=\bar\lambda_c^*=\bar\lambda_q^*=0$, whereas
$\bar\Gamma_2^*,\bar u_q^*,\bar\chi^*$ remain nonzero.  Convection,
relevant at the Gaussian theory, is irrelevant there with eigenvalue
$-\epsilon/2$ (negative eigenvalues denote attractive RG directions in our convention).  The interacting noise sector thus suppresses the
deterministic nonlinearities at long wavelengths.  The designation
{purely dissipative} specifies the suppression of convection and of the Landau-Ginzburg potential.

At the QC solution, the quartic nonlinearity, self-advection, and
nonlinear noise remain finite. In particular,  the  value of the rescaled operator $\tilde{u}^*_q$, encoding quantum fluctuations on top of the Landau-Ginzburg potential, is non-vanishing, hinting at a survival of quantum coherence in the scaling limit.
 The  balance between non-linearities and non-Gaussian noise result in one relevant
interaction direction: perturbations on opposite sides lead toward
PD or out of the perturbative regime (this makes the QC FP a saddle  in the manifold of flowing RG couplings).   This may indicate attraction towards a strong-coupling FP which should be explored with more advanced tools (cf. Sec.~\ref{sec:conclusions} for a list of options). The saddle behavior of the QC fixed point   appears analogous to that in the perturbative RG flow of
the KPZ equation in $d=2+\epsilon$ dimensions~\cite{freyTauber1994,canet2011}, although this similarity  would require further analysis in order to be substantiated. 

RG trajectories initialized at microscopic values of Eq.~\eqref{eq:minimal-active-model} near the stable manifold of QC can linger close to it for long RG times, before departing along the relevant direction towards strong coupling or PD.
Therefore, microscopic systems of active matter described by our critical field theory might  display QC scaling in spatial correlations and
relaxation over  broad time and scale windows.  {Its extent remains limited
by residual flat noise, which sets the crossover scale $q_M$ of Eq.~\eqref{eq:markov-crossover}; its renormalization will be discussed   in Sec.~\ref{sec:infrared-crossover}. Because flat noise is more relevant than the interaction instability of QC, this cutoff may be reached before the flow departs towards PD or strong coupling  (Sec.~\ref{sec:infrared-crossover}), leaving the QC fixed point as the only   observable scaling regime.}

\subsubsection{Critical exponents }
\label{sec:qc-properties}

Let us first compare the relative scaling of diffusive noise and spatial relaxation with the same ratio in  Langevin driven-diffusive systems. The relevant quantity is  $\kappa=(\Gamma_1/K)/(\Gamma_{1\perp}/K_\perp)$, the ratio of the two diffusive coefficients in the deterministic and noise sectors, as defined in Sec.~\ref{sec:rg-variables}. At the QC FP this ratio approaches $\kappa^*=1$, the same value reached in the Langevin driven-diffusive (or Burgers) limit of our equations, consistent with   the literature\newtext{~\cite{tauber2014}}.

The way this common value is reached contains information about the distinct universality classes
  which the QC and classical driven-diffusive (DD) fixed points belong to. For a RG coarse-graining factor $b>1$, we obtain
\begin{equation}
  \kappa(b)-1\sim b^{-\omega_\kappa},\qquad
  \omega_\kappa^{\mathrm{QC}}
  =\frac{5+6\sqrt{2}}{94}\epsilon+\order(\epsilon^2),
  \label{eq:qc-kappa-exponent}
\end{equation}
whereas $\omega_\kappa^{\mathrm{DD}}=\epsilon/6+
\order(\epsilon^2)$ at the same loop order~\cite{janssen1986field}. Thus the two regimes   share a limiting ratio while differing in the rate at which noise and relaxation approach it. Section~\ref{sec:rg-feedback} derives this rate from the convective flow and recovers the driven-diffusive value explicitly. The approach along irrelevant directions  discriminates between different universality classes~\cite{sieberer2014} and informs us that the QC fixed point is distinct from classical Langevin driven-diffusive dynamics or conventional Burgers hydrodynamics. 


Also static properties differ from their direct equilibrium counterparts. \auditedit{For the correlation length along the convective direction, $\xi_\parallel$, we find
\begin{equation}
  \nu_\parallel^{-1}=2-\zeta^*_c+\order(\epsilon^2),\qquad
  \zeta^*_c=\frac{10+12\sqrt{2}}{47}\epsilon,
  \label{eq:qc-nu}
\end{equation}
or $\nu_\parallel\simeq\tfrac12+0.14\epsilon+\order(\epsilon^2)$. At leading order in $\epsilon$ expansion, the transverse exponent keeps its Gaussian value, $\nu_\perp=1/2$. At this order the quartic and noise vertices do not shift the mass exponent: the correction comes entirely from the convective anomalous dimension $\zeta_c^*$ of the longitudinal coefficients (Sec.~\ref{sec:qc-fixed-point}). The quartic vertices instead enter  the scaling of the residual flat noise, and hence the crossover scale $q_M$ (Sec.~\ref{sec:infrared-crossover}).}

It is natural to compare with a
  thermal scalar $\phi^4$ theory. This is the closest non-linear quartic model that satisfies 
  detailed balance (contrary to our  critical theory) and that has $Z_2$ spontaneous symmetry breaking. At leading order in the epsilon expansion it has  
$\nu=1/2+\epsilon_{\rm eq}/12+\order(\epsilon_{\rm eq}^2)$~\cite{altlandSimons2023} with
$\epsilon_{\rm eq}=4-d$. Recall that the quantum power counting above involves $d+z$, with $z=2$ at Gaussian order. Thus the one-dimensional problem has effective dimension $d+z=1+2=3$ and lies one dimension below its quartic upper critical dimension. The corresponding equilibrium comparison is therefore a scalar $\phi^4$-theory in three spatial dimensions at finite temperature,   one dimension below $d_c=4$.  This different value of $\nu$ is another indirect indication of the breaking of detailed balance and therefore indication that the QC fixed point belongs to a distinct    universality class.

\subsection{Connection to microscopic models}
\label{sec:microscopic}

We now provide a microscopic example that motivates the operator content of the field theory    put forward in this work. The RG calculation determines the attractors that control the universal scaling behaviour at criticality, while microscopic accessibility remains a separate, and in general outstanding, problem for future work (see also Sec.~\ref{sec:outlook}). For instance, a certain microscopic model can provide the relevant operators necessary to the field theory analysis in this work, but it might display weak quantum correlations at criticality because of additional RG irrelevant non-linearities, or the specific initial conditions considered, or extra bath effects, that do not change the universality class of critical exponents, but might easily affect the microscopic open dynamics and their quantum mechanical content. 
  
Here, we consider a recent model of quantum Ising flocking encoded in Lindblad dynamics~\cite{khasseh2025}.
This
 combines an
alignment mechanism between two different spin species (labelled in the following by $\sigma=\uparrow, \downarrow$), and coherent conversion between them   (the classical counterparts are the active Ising models of Solon and Tailleur~\cite{solonTailleur2013,solonTailleur2015}, whose  
corresponding stochastic hydrodynamics are recovered from the dissipative quantum spin system of   Ref.~\cite{khasseh2025}  when its coherent drive $g$  is set to zero). With
$\hat n_{l\sigma}=\hat c^\dagger_{l\sigma}\hat c_{l\sigma}$,
the Lindblad equation of the model in~\cite{khasseh2025}  reads
\begin{equation}
 \partial_t\hat\varrho=-\ii[\hat H_g,\hat\varrho]
 +\sum_{X=M,A}\Gamma_X\sum_{l,\sigma}
 \mathcal D[\hat X_{l\sigma}]\hat\varrho,
 \label{eq:micro-lindblad}
\end{equation}
where $\mathcal D[L]\hat\varrho=L\hat\varrho L^\dagger
-\{L^\dagger L,\hat\varrho\}/2$, and
\begin{align}
 \hat M_{l\uparrow}&=\hat c^\dagger_{l\uparrow}\hat c_{l+1,\uparrow},
 &\hat M_{l\downarrow}&=\hat c^\dagger_{l+1,\downarrow}\hat c_{l\downarrow},
 \nonumber\\
 \hat A_{l\sigma}&=\hat c^\dagger_{l\sigma}\hat c_{l\bar\sigma}\hat P_l,
 &\hat H_g&=-g\sum_l(\hat c^\dagger_{l\uparrow}\hat c_{l\downarrow}
 +\mathrm{h.c.}).
 \label{eq:micro-channels}
\end{align}
Here $\bar\sigma$ denotes the opposite species.  Let us explain how these processes combine motion with Ising alignment. 

The dissipative motility operators $\hat M_{l\uparrow}$ and $\hat M_{l\downarrow}$ move an up particle one site to the left and a down particle one site to the right. The internal spin state therefore fixes the direction of motion. These jumps are irreversible: the reverse hop for the same species is absent. Hard-core exclusion blocks a hop when a site is already occupied by that species.

The alignment operator $\hat A_{l\sigma}$ instead converts the opposite species into $\sigma$ at the same site. Its diagonal factor $\hat P_l$ filters the conversion, so the rate depends on the local configuration through
$\hat P_l^\dagger\hat P_l=\exp[-2\mathcal K\hat m_l\hat{\mathcal M}_l]$,
where $\hat m_l=(\hat n_{l\uparrow}-\hat n_{l\downarrow})/2$ and $\hat{\mathcal M}_l$ is its neighborhood average. This fixes the normalization of the alignment strength $\mathcal K$. For positive $\mathcal K$ in an up-majority neighborhood, the factor suppresses an up-to-down flip and enhances a down-to-up flip. Alignment thus reinforces the local majority. Since the two states move in opposite directions, aligning the internal states also aligns their motion: this is the active-Ising mechanism of flocking~\cite{solonTailleur2013,scandolo2023,khasseh2025}.  
Finally, $\hat H_g$ coherently converts the two species into each other and it can create superpositions of   internal states, competing  with dissipative alignment. This is where quantum coherence enters the microscopic dynamics of~\eqref{eq:micro-lindblad}. 

We evaluate the coarse-grained magnetization equation on a uniform half-filled background, $\rho(x,t)=(n_\uparrow+n_\downarrow)/2=1/2$, and retain $m(x,t)$ as the dynamical field, subsequently identified with $\phi$. This choice is already discussed in~\cite{khasseh2025}: its density equation leaves $\rho=1/2$ invariant even for spatially varying magnetization.  Dynamical density fluctuations are outside this simplification; their coupling to magnetization must be included when determining the critical behavior of the full lattice model. However, as stated above (and further discussed at the end of this Section), there would be several additional effects to be considered if one were to demonstrate that the model in~\cite{khasseh2025} flows exactly into the RG fixed points discussed in this work. Rather, our purpose here is to show that there is a connection between recent attempts to build microscopic models of  quantum active matter and their potential RG descriptions.  
The actual connection between microscopic dynamics and RG fixed points is left to future work (Sec.~\ref{sec:outlook}). 

Within this framework and  retaining the magnetization noise generated by Lindblad jumps~\cite{khasseh2025}, we obtain
  
\begin{equation}
 \partial_t m+v\,m\partial_{\onedir}m
 =D_m\partial_{\onedir}^2m-r_m m-b_m m^3+\xi,
 \label{eq:micro-scalar}
\end{equation}
with  the coefficients at $g=0$ given by 
\begin{align}
 D_m&=\Gamma_M/2,\qquad v=2\Gamma_M,\nonumber\\
 r_m&=2\Gamma_A(1-\mathcal K\gamma_m),\nonumber\\
 b_m&=4\mathcal K\Gamma_A
             (3\mathcal K\gamma_m/2-1).
 \label{eq:micro-coefficients}
\end{align}
A perturbatively small value of $g$ would dress these couplings without qualitatively affecting our conclusions~\cite{khasseh2025}. 
Here $\gamma_m$ parametrizes the connected local variance,
$\langle\hat m_l^2\rangle-\langle\hat m_l\rangle^2\simeq\gamma_m\rho$,
and it is distinct from the friction $\gamma$. With   $\phi=m$,
Eq.~\eqref{eq:micro-scalar} matches Eq.~\eqref{eq:minimal-active-model}. 
Markovian noise comes from the Lindblad recycling terms of the jump operators in Eqs.~\eqref{eq:micro-channels} (cf. Appendix of~\cite{khasseh2025})
\begin{equation}
 \xi(x,t)=\sqrt{A_{\mathrm{loc}}(m)}\,\eta_{\mathrm{loc}}(x,t)
       -\partial_{\onedir}[\sqrt{A_M(m)}\,\eta_M(x,t)],
 \label{eq:micro-current-noise}
\end{equation}
with  
\begin{align}
 A_M(m)&=\Gamma_M(1/8-m^2/2),\nonumber\\
 A_{\mathrm{loc}}(m)&=2\gamma_m\Gamma_A +\order(m^2),\nonumber
 \label{eq:micro-noise-amplitudes}
\end{align}
where $\eta_{\mathrm{loc}}$ and $\eta_M$ are independent  white noises.   From Eq.~\eqref{eq:micro-current-noise} one can recognize both flat and diffusive  noise as well as the seed of a multiplicative noise term ($\Gamma_2$ in Sec.~\ref{sec:model});  it is also straightforward to extract  $\Gamma_0 \simeq\gamma_m\Gamma_A$ and 
  $\Gamma_1\simeq\Gamma_M/16$ which are two noise strengths relevant to our field theory construction.

Let us finally consider the two fine tunings  required to approach the critical regime of interest.  The     mass has to vanish (cf. expression in Eq.~\eqref{eq:micro-coefficients}) and diffusive noise has to dominate:

\begin{equation}
 \mathcal K\gamma_m\simeq1,
 \qquad \Gamma_A\ll\Gamma_M.
 \label{eq:micro-tuning}
\end{equation}
  The estimated crossover scale, Eq.~\eqref{eq:markov-crossover}, is then
\begin{equation}
 q_M^2\simeq16\gamma_m\frac{\Gamma_A}{\Gamma_M}.
 \label{eq:micro-window}
\end{equation}
Operators that are irrelevant near the fixed point can nevertheless change the  non-universal   values of relevant couplings when short-distance modes are eliminated. This would have impact for various aspects of our RG analysis, such as: which fixed point (PD, QC or strong coupling) is realized in concrete numerical simulations depending on initial conditions; more refined estimates of Eq.~\eqref{eq:micro-window} (since the one provided here is  just mean field); or the role of the coherent coupling $g$ in enhancing quantum fluctuations at criticality. 
In Sec.~\ref{sec:outlook} we discuss concrete options to address these items in forthcoming work.

\section{Fluctuation corrections to transport, Landau-Ginzburg potential and nonlinear noise}
\label{sec:fluctuation-corrections}

We now start a systematic presentation of the one-loop RG calculations that lead to the purely dissipative and quantum convective fixed points.
We outline the main steps of the derivation here and present the full calculations and the complete
set of diagrams in the appendices.   

\subsection{Dynamical field theory }
\label{sec:technical-theory}

We resort to Keldysh field theory for non-equilibrium systems~\cite{kamenevLevchenko2009,kamenev2023,sieberer2016,sieberer2025}, which is a natural pathway to perform RG on a dissipative dynamical system. Its classical response-functional limit connects to the Martin--Siggia--Rose and Janssen formulations~\cite{tauber2014}. 
 In addition to the physical field, $\phi(X)$ (with $X=(t,\bm x)$), the formalism introduces a 'response field', $\widetilde\phi(X)$.
To make the connection to linear response explicit, imagine adding a weak deterministic source $h(X)$ to the right-hand side of
Eq.~\eqref{eq:minimal-active-model}.
For the functional weight $\exp(-S)$, the source enters as
$S_h=S-\int_Xh\widetilde\phi$, giving
\begin{equation}
 \left.\frac{\delta\langle\phi(X)\rangle_h}{\delta h(X')}\right|_{h=0}
 =\langle\phi(X)\widetilde\phi(X')\rangle_c
 \equiv G^R(X,X'),
 \label{eq:technical-response-definition}
\end{equation}
which is the retarded Green's function.
An insertion of $\widetilde\phi$ therefore measures the response
to an applied source.  Causality requires this response to vanish
unless the source precedes the measured field.  Powers of
$\widetilde\phi$ also organize the noise cumulants: two response
fields encode the variance, and higher powers encode higher cumulants of the noise.

Since we expand around the upper critical dimension $d_c=2$ (cf. Sec.~\ref{sec:model}), we must allow for a possible breakdown of isotropy, since the advective term remains along the $\longdir$ direction. Relaxation and diffusive noise may then have
different longitudinal and transverse coefficients, as in
anisotropic   driven-diffusive theories~\cite{tauber2014}.
We retain $K$ and $\Gamma_1$ for the longitudinal spatial relaxation
and diffusive-noise coefficients, and denote their transverse
counterparts by $K_\perp$ and $\Gamma_{1\perp}$.  
\vthirtythreeedit{In this $d$-dimensional theory, $\partial_\parallel$ denotes the longitudinal derivative; it reduces to $\partial_x$ in one dimension.} For
$Q=(\omega,\bm q)$ and
$\int_Q=\int\dd\omega\,\dd^dq/(2\pi)^{d+1}$, the quadratic
kernels and action in Fourier space are
\begin{align}
 E_{\bm q}&=r+Kq_{\longdir}^2+K_\perp q_\perp^2,\nonumber\\
 D_{\bm q}&=\Gamma_0+\Gamma_1q_{\longdir}^2+\Gamma_{1\perp}q_\perp^2,
 \label{eq:technical-kernels}\\
 S_2&=\int_Q\left[\widetilde\phi(-Q)
 (E_{\bm q}-\ii\gamma\omega)\phi(Q)
 -D_{\bm q}\widetilde\phi(-Q)\widetilde\phi(Q)\right].
 \label{eq:technical-S2}
\end{align}
The transverse terms disappear in one dimension.  At one loop,
convective two-point corrections carry external longitudinal
momentum, while the quartic tadpoles are momentum independent.
Thus $K_\perp$ and $\Gamma_{1\perp}$ receive no correction at
this order.  The quartic sector allows momentum-dependent diagrams
at two loops, so this absence is not valid at all orders in the epsilon expansion.

The equation of motion and the noise cumulants fix the cubic and quartic
terms of the nonlinear action (cf. Sec.~\ref{sec:model}): 
\begin{align}
 S_3&=\int_X\left[
 \frac{\lambda_c}{2}(\partial_{\longdir}\widetilde\phi)\phi^2
 +\frac{\lambda_q}{2}(\partial_{\longdir}\phi)\widetilde\phi^2\right],
 \label{eq:technical-S3}\\
 S_4&=\int_X\left[+\frac{u_c}{6}\phi^3\widetilde\phi
 -\frac{\Gamma_2}{4}\phi^2\widetilde\phi^2
 +\frac{u_q}{6}\phi\widetilde\phi^3
 +\frac{\chi}{24}\widetilde\phi^4\right],
 \label{eq:technical-S4}
\end{align}
with $\int_X=\int\dd t\,\dd^dx$; the total action therefore reads
  $S=S_2+S_3+S_4$.  The two cubic vertices are
self-advection, $\lambda_c$, and corrections to the variance through  gradients of the field, $\lambda_q$; the
four quartic vertices distinguish the Landau--Ginzburg term ($u_c$) and its quantum counterpart ($u_q$) from multiplicative noise ($\Gamma_2$) and non-Gaussian noise ($\chi$). 
  Integration
by parts converts the first term in $S_3$ to
$-\lambda_c\widetilde\phi\phi\partial_{\longdir}\phi$, fixing its sign
relative to Eq.~\eqref{eq:minimal-active-model}.  All factorials
shown in the action are included to get cleaner Wick contractions at later stages of the calculation. 

\begin{figure*}[t]
 \centering
 \begingroup
 \setlength{\unitlength}{\textwidth}
 \begin{picture}(1,0.371429)
  \put(0,0){\includegraphics[width=\textwidth]{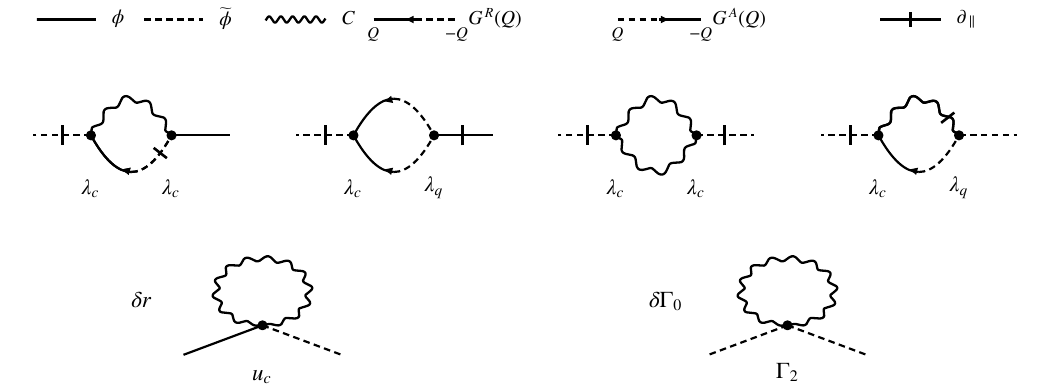}}
  \put(0.030,0.317571){\makebox(0,0)[l]{\small (a) Deterministic sector: $\widetilde\phi\phi$}}
  \put(0.530,0.317571){\makebox(0,0)[l]{\small (b) Noise sector: $\widetilde\phi\widetilde\phi$}}
  \put(0.030,0.154143){\makebox(0,0)[l]{\small (c) Retarded mass: $\widetilde\phi\phi$}}
  \put(0.530,0.154143){\makebox(0,0)[l]{\small (d) Flat noise: $\widetilde\phi\widetilde\phi$}}
 \end{picture}
 \endgroup
 \caption{One-loop corrections to the quadratic coefficients.
 Solid and dashed legs denote \(\phi\) and \(\widetilde\phi\),
 respectively.  A response propagator joins a dashed half-line to a
 continuous half-line, with its causal arrow pointing from
 \(\widetilde\phi\) to \(\phi\).    The symbols \(R,A\)  
 abbreviate retarded and advanced propagators.  Wavy lines
 denote \(C\); a  bar marks a differentiated leg.
 (a) The two bubbles correcting  \(K\); (b) the two bubbles correcting 
 \(\Gamma_1\). Their contributions proportional to \(\lambda_c^2\)
 and \(\lambda_c\lambda_q\) are given in
 Eq.~\eqref{eq:technical-kinetic-corrections}.
  (c),(d) Tadpoles proportional to \(u_c\)
 and \(\Gamma_2\) correct \(r\) and \(\Gamma_0\), respectively.}
 \label{fig:two-point}
\end{figure*}

Inverting $S_2$ gives the Gaussian propagators
\cite{sieberer2016,tauber2014,kamenev2023} 
\begin{align}
 G^R(Q)&=\langle\phi(Q)\widetilde\phi(-Q)\rangle_0
       =\frac{1}{E_{\bm q}-\ii\gamma\omega},\nonumber\\
 G^A(Q)&=\langle\widetilde\phi(Q)\phi(-Q)\rangle_0=G^R(-Q),\nonumber\\
 C(Q)&=\langle\phi(Q)\phi(-Q)\rangle_0
       =2D_{\bm q}G^R(Q)G^A(Q),
 \label{eq:technical-propagators}
\end{align}
where  $C$ is the conventional correlation function containing information about the statistical distribution of the field $\phi$.   We abbreviate
$G^R$ and $G^A$ by $R$ and $A$ in loop integrals.  In the figures,
a response line is dashed at its $\widetilde\phi$ end and continuous
at its $\phi$ end, with its arrow directed toward the physical field.
Figure~\ref{fig:two-point} specifies the ordered endpoints for
$G^R$ and $G^A$.  The arrows keep track of causal ordering:
parallel responses between two vertices are allowed, whereas a
directed response cycle vanishes.

The propagators make the power counting in Sec.~\ref{sec:model}
explicit.  At $r=\Gamma_0=0$, both quadratic kernels vanish as
$q^2$, and the equal-time correlation
$D_{\bm q}/(\gamma E_{\bm q})$ has no radial infrared divergence.
 With $z=2$, the canonical momentum dimensions are
\begin{align}
 [\phi]=[\widetilde\phi]&=d/2,\nonumber\\
 [\lambda_c]=[\lambda_q]&=1-d/2,\nonumber\\
 [u_c]=[\Gamma_2]=[u_q]=[\chi]&=2-d.
 \label{eq:technical-dimensions}
\end{align}
All six nonlinear vertices are marginal at $d=2$ as it should be at the upper critical dimension.
 An additional pair of spatial
derivatives lowers a quartic coupling dimension to $-d$, excluding
terms such as $\phi^2(\partial_{\longdir}\widetilde\phi)^2$ from the
relevant truncation near two dimensions.

We use a momentum-shell integration to perform RG~\cite{tauber2014,altlandSimons2023}.
Momenta are integrated over shells 
$\Lambda\ee^{-\delta\ell}<|\bm q|<\Lambda$, with $\delta\ell\to0^+$, and $\ell$ the running RG scale, while frequencies
are integrated without restriction~\cite{altlandSimons2023,kamenev2023}.  Expanding in small external
momenta and frequencies projects the result onto local operators.
For $S_{\rm int}=S_3+S_4$, the connected cumulant expansion gives
\begin{equation}
 \delta S=\langle S_{\rm int}\rangle_>
 -\frac12\langle S_{\rm int}^2\rangle_{>,c}
 +\frac16\langle S_{\rm int}^3\rangle_{>,c}-\cdots,
 \label{eq:technical-cumulants}
\end{equation}
where $>$ denotes the Gaussian shell average and $c$ its connected
part~\cite{altlandSimons2023,kamenev2023}.   We retain the local two-point coefficients, cubic terms linear
in external $p_{\longdir}$, and quartic terms at zero external momentum and
frequency.  Appendix~\ref{app:conventions} specifies the projections
and contraction rules.  These choices separate corrections  to mass and noise thresholds from corrections to transport and  
nonlinear vertices.

\subsection{Spectral mass, diffusion, and noise}
\label{sec:technical-kinetic}

The response and noise kernels contain two physically different
types of information.  Their constant parts control fluctuations corrections to the mass and to  flat-noise; their $p_{\longdir}^2$ terms control spatial
relaxation and diffusive noise.  At one loop, the first are corrected
by quartic tadpoles and the second by convective bubbles, as shown
in Fig.~\ref{fig:two-point}.

We define the self-energies $\Sigma_R$ and $\Sigma_N$ by
\begin{equation}
 \delta S_2=\int_P\left[
 \widetilde\phi(-P)\Sigma_R(P)\phi(P)
 -\Sigma_N(P)\widetilde\phi(-P)\widetilde\phi(P)\right].
 \label{eq:technical-self-energy-convention}
\end{equation}
Their constant terms give $\delta r$ and $\delta\Gamma_0$, and
their $p_{\longdir}^2$ coefficients give $\delta K$ and $\delta\Gamma_1$.

The local corrections have a common origin: the shell variance
$I_1=\int_>C(Q)$.  Through $u_c$, this variance changes the
linear response around zero mean field.  Through $\Gamma_2$,
it changes the flat-noise level and hence contributes to corrections to $q_M$.
The tadpoles in Fig.~\ref{fig:two-point}(c),(d) give
\begin{equation}
 \delta r=\frac{u_c}{2}I_1,\qquad
 \delta\Gamma_0=\frac{\Gamma_2}{4}I_1.
 \label{eq:technical-tadpoles}
\end{equation}
The first term shifts the ordering threshold; the second generates
flat noise from a field-dependent variance.     In particular, bare
$\Gamma_0=0$ is not preserved when the second tadpole is nonzero.
This is the fluctuation correction to the microscopic scale $q_M$.

Expanding the bubbles in
Fig.~\ref{fig:two-point}(a),(b) in powers of external momenta gives the corrections to the kinetic coefficients ($K$ and $\Gamma_1$) of the quadratic part of the action, $S_2$; namely, 

\begin{align}
 \delta K&=\frac{\lambda_c^2}{4}J_K
             -\frac{\lambda_c\lambda_q}{2}I_3,\nonumber\\
 \delta\Gamma_1&=\frac{\lambda_c^2}{4}I_4
             -\frac{\lambda_c\lambda_q}{4}J_\Gamma.
 \label{eq:technical-kinetic-corrections}
\end{align}
Here $I_3=\int_>RA$ and $I_4=\int_>C^2$ contain products of two
response or two correlation propagators.  The integrals $J_K$ and
$J_\Gamma$ include the external-momentum expansion of the propagators (cfr. Eqs.~\eqref{kin:loop-definitions} and~\eqref{kin:shell-values}
in Appendix~\ref{app:kinetic}).  The two different
right-hand sides of Eq.~\eqref{eq:technical-kinetic-corrections} illustrate how  diffusion and noise do not renormalize equally.  The full
action imposes no general fluctuation--dissipation relation between
the two corrections. By comparison, in equilibrium relaxational models A and B, detailed balance fixes the noise strength to damping through  temperature~\cite{hohenbergHalperin1977,tauber2014}.   

\subsection{Renormalization of the nonlinear couplings}
\label{sec:technical-interactions}

\begin{figure*}[t]
 \centering
 \begingroup
 \setlength{\unitlength}{\textwidth}
 \begin{picture}(1,0.211429)
  \put(0,0){\includegraphics[width=\textwidth]{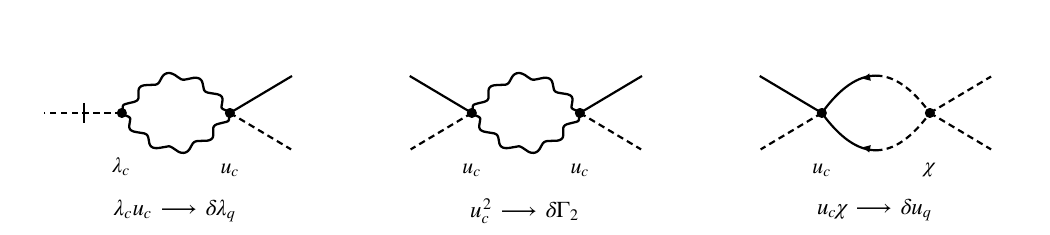}}
  \put(0.166667,0.190286){\makebox(0,0){\small (a) Quantum convective coupling}}
  \put(0.500000,0.190286){\makebox(0,0){\small (b) Multiplicative noise}}
  \put(0.833333,0.190286){\makebox(0,0){\small (c) Quantum quartic coupling}}
 \end{picture}
 \endgroup
 \caption{Representative corrections for non-linear couplings.  (a) Self-advection and the quartic nonlinearity contribute
 to the quantum convective coupling, renormalizing noise variance.  (b) Two quartic-nonlinearity vertices generate
 a multiplicative-noise contribution.   
 (c) The non-Gaussian-noise strength \(\chi\) contributes to the
 quantum quartic coupling \(u_q\) through \(u_c\).
 }
 \label{fig:interaction-feedback}
\end{figure*}

 The examples presented in the following and illustrated in
Fig.~\ref{fig:interaction-feedback}   show   the generation of
initially absent non-linear couplings, the reciprocal effects of nonlinear
noise on deterministic dynamics, and corrections to cubic vertices.  Appendices~\ref{app:cubic}
and~\ref{app:quartic} contain the complete derivation.

\emph{Cubic vertices}. 
Self-advection and the quartic nonlinearity generate corrections dependent on the gradient of the field in the
noise variance, even if $\lambda_q$  is zero at the microscopic level.  The  bubble
in Fig.~\ref{fig:interaction-feedback}(a) produces
$(\partial_{\longdir}\phi)\widetilde\phi^2$, with
\begin{equation}
 \left.\delta\lambda_q\right|_{\lambda_cu_c}
 =\frac{\lambda_cu_c}{2}I_4.
 \label{eq:technical-lambdaq-source}
\end{equation}
This adds $-\delta\lambda_q\partial_{\longdir}\phi$ to the Gaussian noise variance in
Eq.~\eqref{eq:nonlinear-noise-cumulants}.   

Self-advection, $\lambda_c$, is renormalized in turn by terms proportional to
$\lambda_cu_c$, $\lambda_c\Gamma_2$, and $\lambda_qu_c$.
The presence of a Landau-Ginzburg potential together with  terms  non-linear in the response field, breaks the Galilean protection for self-advection.
The complete corrections in Eqs.~\eqref{cub:counterterm-c}
and~\eqref{cub:counterterm-q} show that the flows of both convective couplings must be solved jointly with those of the local quartic
couplings.

\emph{Quartic vertices}. 
The amplitude dependence of the noise is generated by an analogous
mechanism.  Two $u_c$ vertices joined by two correlation functions, $C$, give the
multiplicative-noise contribution in Fig.~\ref{fig:interaction-feedback}(b):
\begin{equation}
 \left.\delta\Gamma_2\right|_{u_c^2}=u_c^2 I_4.
 \label{eq:technical-multiplicative-source}
\end{equation}
The quartic nonlinearity therefore makes the variance depend on
$\phi^2$, even when $\Gamma_2$ is initially absent.  Conversely,
a multiplicative-noise vertex paired with $u_c$ corrects the
quartic nonlinearity:
\begin{equation}
 \left.\delta u_c\right|_{u_c\Gamma_2}
 =\frac32u_c\Gamma_2 I_3.
 \label{eq:technical-quartic-feedback}
\end{equation}
 
 The feedback extends to the higher noise cumulants.  In
Fig.~\ref{fig:interaction-feedback}(c), $u_c$ allows the
non-Gaussian-noise strength $\chi$ to correct the quantum quartic
coupling:
\begin{equation}
 \left.\delta u_q\right|_{u_c\chi}=-\frac32u_c\chi I_3.
 \label{eq:technical-chi-feedback}
\end{equation}
The fourth noise cumulant thus corrects the quantum counterpart of the Landau--Ginzburg potential.  

In principle, one-loop cubic triangles, mixed cubic--quartic triangles, and cubic boxes can also contribute through the last terms in Eq.~\eqref{con:cumulant-sectors}. Directed percolation provides a familiar example of a cubic response-field theory in which a triangle corrects a three-point vertex~\cite{tauber2014,janssenTauber2005}. These topologies must therefore be checked even in a one-loop calculation.
 In Appendix~\ref{app:cancellations}  we show that a combination of  
causality, momentum factors, and/or cancellation after frequency integration prevent these diagrams from contributing at one loop in the truncation adopted here.

\section{Structure of the renormalization group flow}
\label{sec:rg-flow}

\subsection{Dimensionless couplings and beta functions}\label{sec:rg-variables}\label{sec:rg-feedback}

We now summarize the whole set of beta functions needed to determine the fixed points and their stability, as anticipated in Sec.~\ref{sec:synopsis-results}. 
As is customary in one-loop RG shell integration~\cite{tauber2014,altlandSimons2023}, we denote by $b>1$  the coarse-graining factor, and define
\begin{equation}
 \ell=\ln b,\qquad \Lambda_\ell=\Lambda\ee^{-\ell},\qquad
 \beta_g\equiv\frac{\dd g}{\dd\ell}.
 \label{eq:rg-time}
\end{equation}
Increasing $\ell$ probes longer wavelengths, starting from the ultra-violet ($\Lambda$).  A  
perturbation of eigenvalue $y$ therefore grows as $\ee^{y\ell}$,
with $y<0$ denoting infrared attraction.   The coefficients
$K_\perp,\Gamma_{1\perp}$ and $\gamma$ are held fixed because
their one-loop corrections vanish, while   the longitudinal coefficients
$K$ and $\Gamma_1$ may run, retaining the anisotropy generated by
self-advection~\cite{tauber2014}, as derived in Appendix~\ref{kin:momentum-expansion}.

We will need in the following both the ratio of diffusive noise to spatial relaxation, and the flow of either of these two couplings:  
\begin{equation}
 c=\frac{K}{K_\perp},\qquad
  \kappa= 
 \frac{\Gamma_1/K}{\Gamma_{1\perp}/K_\perp}.
 \label{eq:rg-anisotropy}
\end{equation}
The presence of $K_\perp$ and $\Gamma_{1\perp}$ makes these expressions cumbersome; however, at one-loop order they do not flow, simplifying considerably the RG algebra. 
 
 The dimensionless quartic couplings are
\begin{align}
 \bar u_c&=\frac{a_d\Lambda_\ell^{-\epsilon}\Gamma_{1\perp}u_c}
 {\gamma K_\perp^2\sqrt c},&
 \bar\Gamma_2&=\frac{a_d\Lambda_\ell^{-\epsilon}\Gamma_2}
 {\gamma K_\perp\sqrt c},\nonumber\\
 \bar u_q&=\frac{a_d\Lambda_\ell^{-\epsilon}u_q}
 {\gamma\Gamma_{1\perp}\sqrt c},&
 \bar\chi&=\frac{a_d\Lambda_\ell^{-\epsilon}K_\perp\chi}
 {\gamma\Gamma_{1\perp}^2\sqrt c},
 \label{eq:rg-quartic-variables}
\end{align}
while for the dimensionless convective couplings we define 
\begin{align}
 \bar\lambda_c&=\lambda_c\Lambda_\ell^{-\epsilon/2}
 \left(\frac{a_d\Gamma_{1\perp}}{\gamma K_\perp^3}\right)^{1/2}
 c^{-3/4}\kappa^{-1/2},\nonumber\\
 \bar\lambda_q&=\lambda_q\Lambda_\ell^{-\epsilon/2}
 \left(\frac{a_d}{\gamma\Gamma_{1\perp}K_\perp}\right)^{1/2}
 c^{-3/4}\kappa^{-1/2},
 \label{eq:rg-convective-variables}
\end{align}
with 
 $a_d=S_d/(2\pi)^d$, where $S_d$ is the area of the unit
sphere in $d$ dimensions. 
The explicit introduction of cutoff powers, $\Lambda_\ell$,
implements canonical rescaling, giving dimensions $\epsilon$ for
the quartics and $\epsilon/2$ for convection.  The remaining
factors absorb the inert  anisotropic kinetic
coefficients ($K_\perp$ and $\Gamma_{1\perp}$, as well as the damping coefficient $\gamma$).  

Corrections to kinetic coefficients of the Gaussian action, evaluated in Eqs.~\eqref{kin:explicit-shifts}--\eqref{kin:kappa-flow}, combine to give the following two beta functions
\begin{align}
 \zeta_c\equiv\frac{\dd\ln c}{\dd\ell}
 &=\frac{\bar\lambda_c^2\kappa(\kappa+3)}{16}
 -\frac{\bar\lambda_c\bar\lambda_q\kappa}{4},\nonumber\\
 \frac{\dd\ln\kappa}{\dd\ell}
 &=\frac{\bar\lambda_c^2(\kappa-1)(\kappa-3)}{32}
 +\frac{\bar\lambda_c\bar\lambda_q(\kappa-1)}{16}.
 \label{eq:rg-kappa}
\end{align}

The quartic flows contain $\epsilon_{\mathrm{eff}}\equiv\epsilon-\frac{\zeta_c}{2}$, while the cubic flows also contain the running of $\kappa$ through their normalization in Eq.~\eqref{eq:rg-convective-variables}. Combining these terms with the vertex corrections in Appendices~\ref{cub:flows}  gives the beta functions at leading order in $\epsilon$:
 \begin{align}
 \frac{\dd\bar\lambda_c}{\dd\ell}
 ={}&\bar\lambda_c\left[
 \frac{\epsilon}{2}-\frac{5\kappa+7}{16}\bar u_c
 +\frac{\bar\Gamma_2}{4}
 -\frac{4\kappa^2+5\kappa+3}{64}\bar\lambda_c^2
 \right.\nonumber\\[-1pt]
 &\left.\hspace{13mm}
 +\frac{5\kappa+1}{32}\bar\lambda_c\bar\lambda_q\right]
 +\frac{\bar u_c\bar\lambda_q}{4},
 \label{eq:rg-lambda-c}\\
 \frac{\dd\bar\lambda_q}{\dd\ell}
 ={}&\bar\lambda_q\left[
 \frac{\epsilon}{2}-\frac{3\kappa+1}{16}\bar u_c
 +\frac{\bar\Gamma_2}{4}
 -\frac{4\kappa^2+5\kappa+3}{64}\bar\lambda_c^2
 \right.\nonumber\\[-1pt]
 &\left.\hspace{13mm}
 +\frac{5\kappa+1}{32}\bar\lambda_c\bar\lambda_q\right]
 \nonumber\\
 &+\bar\lambda_c\left[
 \frac{3\kappa^2+2\kappa+3}{16}\bar u_c
 -\frac{5\kappa+7}{16}\bar\Gamma_2+\frac{\bar u_q}{4}
 \right],
 \label{eq:rg-lambda-q}\\\displaybreak[1]
 \frac{\dd\bar u_c}{\dd\ell}
 ={}&\bar u_c\left[
 \epsilon_{\mathrm{eff}}-\frac34(\kappa+1)\bar u_c
 +\frac34\bar\Gamma_2\right],
 \label{eq:rg-uc}\\
 \frac{\dd\bar\Gamma_2}{\dd\ell}
 ={}&\bar\Gamma_2\left[
 \epsilon_{\mathrm{eff}}-\frac54(\kappa+1)\bar u_c
 +\frac14\bar\Gamma_2\right]\nonumber\\
 &+\frac{3\kappa^2+2\kappa+3}{8}\bar u_c^2
 +\bar u_c\bar u_q,
 \label{eq:rg-Gamma2}\\
 \frac{\dd\bar u_q}{\dd\ell}
 ={}&\bar u_q\left[
 \epsilon_{\mathrm{eff}}-\frac34(\kappa+1)\bar u_c
 +\frac34\bar\Gamma_2\right]\nonumber\\
 &+\frac{3(3\kappa^2+2\kappa+3)}{16}\bar u_c\bar\Gamma_2
 -\frac{3(\kappa+1)}4\bar\Gamma_2^2
 -\frac34\bar u_c\bar\chi,
 \label{eq:rg-uq}\\
 \frac{\dd\bar\chi}{\dd\ell}
 ={}&\bar\chi\left[\epsilon_{\mathrm{eff}}
 +\frac32\bar\Gamma_2\right]
 -\frac{3(3\kappa^2+2\kappa+3)}{16}\bar\Gamma_2^2
 \nonumber\\
 &+\frac{3(\kappa+1)}2\bar\Gamma_2\bar u_q.
 \label{eq:rg-chi}
\end{align}
Equations~\eqref{eq:rg-kappa}--\eqref{eq:rg-chi} close on
\begin{equation}
 \bm g=(\kappa,\bar\lambda_c,\bar\lambda_q,
 \bar u_c,\bar\Gamma_2,\bar u_q,\bar\chi),
 \label{eq:rg-vector}
\end{equation}
and a fixed point satisfies $\bm\beta(\bm g^*)=0$. These values feed into the beta functions of the mass, $r$, and of the flat-noise level, $\Gamma_0$. At this order in the $\epsilon$ expansion the interaction flow is autonomous, while the flow of   $r$ and $\Gamma_0$ are driven by the interactions. In other words, one can solve first for the flow of non-linearities and convection, and then substitute their values in the beta functions of the masses (in the following, we will informally refer to both $r$ and $\Gamma_0$ as 'masses', since $\Gamma_0$ plays the role of a 'gap' in the quadratic noise sector as much as the mass  $r$ does in the retarded sector). The final bracket of
Eq.~\eqref{eq:rg-lambda-q} can generate $\bar\lambda_q$ from
self-advection and local interactions, as illustrated by
Eq.~\eqref{eq:technical-lambdaq-source}.  Setting
$\bar\lambda_q=0$ alone is therefore not a closed reduction of the RG flow, and this justifies its inclusion in the ansatz for the Keldysh action, Eq.~\eqref{eq:technical-S3}.
A couple of remarks on the structure of the beta functions are in order:

(i) 
The degeneracy in the RG flow is a feature of the leading-order $\epsilon$-expansion, yet it has important
mathematical consequences for the search for RG fixed points.  First of all, we can write 
\begin{equation}
 \mathcal B_4=\epsilon_{\mathrm{eff}}
 -\frac34(\kappa+1)\bar u_c+\frac34\bar\Gamma_2,
 \qquad \beta_{\bar u_c}=\bar u_c\mathcal B_4,
 \label{eq:rg-shared-bracket}
\end{equation}
so a fixed point lies either on $\bar u_c=0$ or on
$\mathcal B_4=0$.  More generally, the   beta functions in Eqs.~\eqref{eq:rg-uc}--\eqref{eq:rg-chi}
obey the   identity
\begin{align}
 &\bar u_c\beta_{\bar\chi}
 +\big[(\kappa+1)\bar u_c+\bar\Gamma_2\big]\beta_{\bar u_q}
 \nonumber\\
 &\quad-\frac32(\kappa+1)\bar\Gamma_2\beta_{\bar\Gamma_2}
 =\mathcal B_4\mathcal P_4,
 \label{eq:rg-degeneracy}\\
 \mathcal P_4={}&\bar u_c\bar\chi
 +\big[(\kappa+1)\bar u_c+\bar\Gamma_2\big]\bar u_q
 -\frac32(\kappa+1)\bar\Gamma_2^2.
 \label{eq:rg-degeneracy-polynomial}
\end{align}
It holds for arbitrary values of all seven variables, including
the convective dependence hidden in $\epsilon_{\mathrm{eff}}$.
On the branch $\bar u_c\ne0$, the conditions
$\mathcal B_4=\beta_{\bar\Gamma_2}=\beta_{\bar u_q}=0$
automatically imply $\beta_{\bar\chi}=0$.  Only three of the
four quartic fixed-point equations are therefore independent.  At fixed
$\kappa$ and $\epsilon_{\mathrm{eff}}$, their regular solutions
form a family parametrized by $\bar u_c$.  The branch
$\bar u_c=0$ must   be treated separately.

(ii) The flow of $\kappa$ measures the relative renormalization of
noise and relaxation.  With the transverse coefficients fixed,
\begin{equation}
 \frac{\dd\ln\kappa}{\dd\ell}
 =\frac{\dd\ln\Gamma_1}{\dd\ell}
 -\frac{\dd\ln K}{\dd\ell},
 \label{eq:rg-kappa-anomalous}
\end{equation}
and it has no canonical contribution.  Thus $\kappa$ is a dimensionless
ratio, while its logarithmic derivative is the difference of two
anomalous rates.  At one loop these rates depend directly on
$\bar\lambda_c$ and $\bar\lambda_q$ (cf. Fig.~\ref{fig:two-point}); the quartics affect them indirectly 
through the convective flow.  Equation~\eqref{eq:rg-kappa} preserves
$\kappa=1$.  Linearization about a fixed point on this surface gives
\begin{equation}
 \frac{\dd\,\delta\kappa}{\dd\ell}
 =\frac{\bar\lambda_c^*(\bar\lambda_q^*-\bar\lambda_c^*)}{16}
 \delta\kappa.
 \label{eq:rg-kappa-linear}
\end{equation}
The coefficient equals $-\zeta_c^*/4$.  For $\zeta_c^*>0$,
the ratio is attractive. To recover the driven-diffusive limit of Sec.~\ref{sec:limiting-cases}, set $\bar\lambda_q=\bar u_c=\bar\Gamma_2=\bar u_q=\bar\chi=0$ and $\kappa=1$. This potential-free convective theory then has the usual Galilean protection~\cite{tauber2014}. In our dimensionless variables, Eq.~\eqref{eq:rg-lambda-c} reduces to
\begin{equation}
 \beta_{\bar\lambda_c}=\bar\lambda_c
 \left(\frac{\epsilon}{2}-\frac{3\bar\lambda_c^2}{16}\right),
 \label{eq:rg-burgers-fixed-point}
\end{equation}
 with fixed point value for self-advection: $(\bar\lambda_c^{*})^2=\frac{8\epsilon}{3}$.
Inserting this value into Eq.~\eqref{eq:rg-kappa-linear} gives
$\dd\delta\kappa/\dd\ell=-\epsilon\delta\kappa/6$ and therefore
$\omega_\kappa^{\mathrm{DD}}=\epsilon/6+\order(\epsilon^2)$.
This exponent describes the infrared approach to $\kappa=1$ in the driven-diffusive limit~\cite{tauber2014,janssen1986field}.

\subsection{Spectral mass and flat Markovian noise  }
\label{sec:rg-masses}

Once the RG flow for non-linear couplings  is known, it is straightforward to solve for   the spectral mass, $r$, and the flat noise level, $\Gamma_0$, since they do not feed back into the flow of the convective or quartic couplings. 
  We rescale them as follows:
\begin{equation}
 \bar r=\frac{r}{K_\perp\sqrt{c}\Lambda_\ell^2},\qquad
 \bar\Gamma_0=\frac{\Gamma_0}{\sqrt{c}\Gamma_{1\perp}\Lambda_\ell^2}.
 \label{eq:rg-mass-variables}
\end{equation}

Using the tadpoles in Eq.~\eqref{eq:technical-tadpoles} and projecting on the manifold $\kappa=1$ (achieved at the QC fixed point), we find at leading order in epsilon expansion (i.e. for small $\bar r$ and $\bar\Gamma_0$)
\begin{align}
 \frac{\dd\bar r}{\dd\ell}
 &=\left(2-\frac{\zeta_c}{2}\right)\bar r+\frac12\bar u_c\,(1+\bar\Gamma_0-\bar r),\\
 \frac{\dd\bar\Gamma_0}{\dd\ell}
 &=\left(2-\frac{\zeta_c}{2}\right)\bar\Gamma_0+\frac14\bar\Gamma_2\,(1+\bar\Gamma_0-\bar r)
 \label{eq:rg-mass-flow}
\end{align}
which yields the fixed point values $\bar r_{\mathrm{crit}}=-\bar u_c^*/4$ and
$\bar\Gamma_{0,\mathrm{crit}}=-\bar\Gamma_2^*/8$.
\auditedit{Linearizing about this fixed point gives two relevant directions: the spectral mass and the flat noise measured relative to it, $\bar\Gamma_0-\bar r$. Their exponents are derived in Sec.~\ref{sec:qc-fixed-point}, Eq.~\eqref{eq:qc-mass-tau}.}

 \section{Fixed points and quantum convective scaling regime}
 \label{sec:results}
 
 We will now overview the manifold of fixed points supported by the beta functions derived in the previous Section, below the upper critical dimension, $d_c=2$. %

As usual, the Gaussian fixed point is retrieved by setting   all  couplings   to zero.
Linearizing Eqs.~\eqref{eq:rg-lambda-c}--\eqref{eq:rg-chi} about this
solution gives eigenvalues $\epsilon/2$ (positive eigenvalues are repulsive in our convention) for the two convective couplings
and $\epsilon$ for the four quartic couplings. Thus, below $d_c=2$, the
Gaussian theory is unstable even after tuning the two relevant mass
perturbations discussed in Sec.~\ref{sec:rg-masses}.  

By solving the RG beta functions for an interacting (non-Gaussian) fixed point, we find, as the two most stable solutions, a purely dissipative fixed point (with rescaled
values of $u_c$ and convective couplings equal to zero) and the quantum convective fixed point discussed in Sec.~\newedit{\ref{sec:synopsis-results}}, which supports all nonlinearities.

\subsection{Purely dissipative fixed point}
\label{sec:pd-fixed-point}

Setting \(\bar\lambda_c=\bar\lambda_q=\bar u_c=0\) gives \(\zeta_c=0\), \(\epsilon_{\mathrm{eff}}=\epsilon\), and \(\beta_\kappa=0\). At the retained one-loop order, the purely dissipative solution therefore forms a family parametrized by \(\kappa>0\):
\begin{equation}
\begin{aligned}
\bar\Gamma_2^*&=-4\epsilon,\qquad
\bar u_q^*=-6(\kappa+1)\epsilon,\\
\bar\chi^*&=\frac{3}{5}(9\kappa^2+22\kappa+9)\epsilon.
\end{aligned}
\label{eq:pd-values}
\end{equation}
The interaction-sector eigenvalues are
\[
\epsilon\left\{-\frac{1}{2},-\frac{1}{2},
-1,-2,-2,-5,0\right\}.
\]
Thus the family has six attractive directions and one marginal direction tangent to it (connected to the redundancy in $\kappa$), with the two relevant mass perturbations understood to be already tuned to criticality.   In the remainder of this subsection we specialize to the invariant slice \(\kappa=1\), which contains QC and the trajectories considered below. On this slice, \((\bar\Gamma_2^*,\bar u_q^*,\bar\chi^*)=(-4,-12,24)\epsilon\).

 \begin{figure*}[!t]
 \centering
 \includegraphics[width=\textwidth]{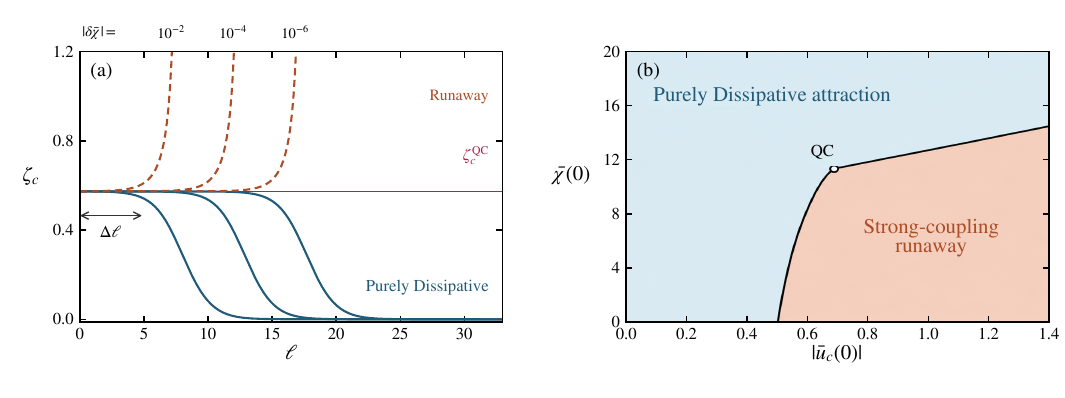}
 \caption{{QC instability and basins of attraction for  $\epsilon=\kappa=1$.
 Blue denotes flow toward PD and orange runaway.
 (a) All interactions start at QC$(s_0)$ except
 $\bar\chi(0)=\bar\chi^*+\delta\bar\chi$, with
 $|\delta\bar\chi|=10^{-2},10^{-4},10^{-6}$
 from left to right. Solid blue: positive $\delta\bar\chi$ shifts; dashed orange: negative $\delta\bar\chi$ 
 shifts. \vthirtytwoedit{The displayed RG plateau     $\Delta\ell=4.858$  is  for $\delta\bar\chi=10^{-2}$.}
 (b)  Final attractors  for initial values of $(|\bar u_c(0)|,\bar\chi(0))$  with the other four
 non-linear couplings set  at their  QC$(s_0)$ values; $s_0$ is chosen such that $\bar{\lambda}_q=0$.}}
 \label{fig:qc-crossover}
 \label{fig:attractors}
\end{figure*}

The name purely dissipative (PD) reflects the disappearance of self-advection,
its quantum partner, and the classical quartic drift. The values in Eq.~\eqref{eq:pd-values} allow us to write the effective action into a  compact form. {Unwinding the rescaling in
Eq.~\eqref{eq:rg-quartic-variables} allows us to express the dimensionful
fixed-point couplings in terms of the common coefficients}
\begin{equation}
 a=\frac{\Gamma_{1\perp}}{K_\perp},\qquad
 g_{\rm PD}=\frac{\epsilon\gamma K_\perp\sqrt c\,
 \Lambda_\ell^\epsilon}{a_d},
 \label{eq:pd-normalization}
\end{equation}
as $\Gamma_2^*=-4g_{\rm PD}$, $u_q^*=-12a g_{\rm PD}$, and
$\chi^*=24a^2g_{\rm PD}$. Substituting these coefficients into
Eq.~\eqref{eq:technical-S4}, while $S_3^*=0$, gives
\begin{align}
 S_{\mathrm{nl}}^{\rm PD}
 &=g_{\rm PD}\int_X\left(
 \phi^2\widetilde\phi^2-2a\phi\widetilde\phi^3
 +a^2\widetilde\phi^4\right)\nonumber\\
 &=g_{\rm PD}\int_X\widetilde\phi^2
 (\phi-a\widetilde\phi)^2.
 \label{eq:pd-square}
\end{align}
We now introduce the linear transformation  
\begin{equation}
 \psi=a^{-1/2}\phi-a^{1/2}\widetilde\phi,
 \qquad \widetilde\psi=a^{1/2}\widetilde\phi.
 \label{eq:pd-fields}
\end{equation}
Under this transformation, the bulk action becomes, up to a temporal boundary term,
\begin{align}
S_{\rm PD}=\int_X\Bigl[&
\widetilde\psi\bigl(\gamma\partial_t-K\partial_{\longdir}^2
-K_\perp\nabla_\perp^2+r\bigr)\psi\nonumber\\
&-\left(\frac{\Gamma_0}{a}-r\right)\widetilde\psi^2
+g_{\rm PD}\psi^2\widetilde\psi^2\Bigr],
\label{eq:pd-action}
\end{align}
with the coefficients evaluated at PD. At \(\kappa=1\), the relations \(\Gamma_1=aK\) and \(\Gamma_{1\perp}=aK_\perp\) cancel the quadratic gradient-noise terms generated by the field transformation. The local quadratic noise term and the nonlinear noise term remain.

The transformation preserves the causal Keldysh structure for arbitrary \(a>0\):
\[
\begin{gathered}
S_{\rm PD}[\psi,0]=0,\qquad
\langle\widetilde\psi\widetilde\psi\rangle=0,\\
\langle\psi\widetilde\psi\rangle
=\langle\phi\widetilde\phi\rangle=G^R.
\end{gathered}
\]
 Correlation functions of the original order parameter are recovered using \(\phi=\sqrt a(\psi+\widetilde\psi)\).

\subsection{Quantum convective fixed point}
\label{sec:qc-fixed-point}

At the QC fixed point, self-advection coexists with the quartic
Landau-Ginzburg potential and higher order noise terms. The RG identity in
Eq.~\eqref{eq:rg-degeneracy} leaves a free parameter in its coordinates.
Writing $\bar u_c^*=-\epsilon s$, with $s>0$, and defining
\begin{equation}
 \alpha\equiv\frac{56-8\sqrt2}{47},\qquad
 s_0\equiv\frac{28\sqrt2-8}{47},
 \label{eq:qc-constants}
\end{equation}
we obtain the family
\begin{align}
 \bar u_c^*&=-\epsilon s,\qquad
 \bar\Gamma_2^*=-\epsilon(2s+\alpha),\nonumber\\
 \bar u_q^*&=-\epsilon\left(3s+3\alpha+\frac{\alpha^2}{2s}\right),\nonumber\\
 \bar\chi^*&=\epsilon\left(4s+6\alpha+\frac{2\alpha^2}{s}\right),\nonumber\\
 \bar\lambda_c^*&=\sqrt{\epsilon(2+\sqrt2)s},\qquad
 \bar\lambda_q^*=\bar\lambda_c^*\left(1-\frac{s_0}{s}\right),
 \label{eq:qc-family}
\end{align}
\newchange{with $\kappa^*=1$. Direct substitution into
Eqs.~\eqref{eq:rg-kappa}--\eqref{eq:rg-chi} gives zero for all seven
beta functions at every $s>0$.} 
The $s$-dependence of this family of fixed points has no physical consequences, since the eigenvalues of the stability matrix are not affected by the value  of $s$.
Notice also that the  value of the rescaled operator $\tilde{u}^*_q$, encoding quantum fluctuations on top of the Landau-Ginzburg potential, is non-vanishing, hinting at a survival of quantum coherence jointly with convective operators ($\tilde{\lambda}^*_{c,q}\neq0$) in the scaling limit. This occurrence motivates the name of quantum convective (QC) for the fixed point found at this level of our analysis.

The cancellation of $s$ in the physical exponents can  be checked
directly. Equation~\eqref{eq:qc-family} gives
\begin{align}
 \bar\lambda_c^*(\bar\lambda_c^*-\bar\lambda_q^*)
 &=(\bar\lambda_c^*)^2\frac{s_0}{s}
 =\epsilon(2+\sqrt2)s_0=4\zeta_c^*,\nonumber\\
 \zeta_c^*&=\frac{10+12\sqrt2}{47}\epsilon.
 \label{eq:qc-invariant-rates}
\end{align}
Thus the growth of $(\bar\lambda_c^*)^2$ with $s$ is compensated by
the factor $s_0/s$ in $1-\bar\lambda_q^*/\bar\lambda_c^*$.
Equation~\eqref{eq:rg-kappa-linear} then gives
$\omega_\kappa=\zeta_c^*/4$. {For the masses it is convenient to use $\bar r$ and
$\tau\equiv\bar\Gamma_0-\bar r$, the flat noise measured relative to the
mass. In dimensionful form $\tau\propto\Upsilon\equiv\Gamma_0-ar$, with
$a=\Gamma_{1\perp}/K_\perp$ (incidentally, at $\kappa=1$, $\Upsilon/a$ is the   Gaussian
noise of the rotated fields of Eq.~\eqref{eq:pd-fields}). The two mass
equations of Sec.~\ref{sec:rg-masses} become
\begin{align}
 \frac{\dd\tau}{\dd\ell}&=\Big(2-\frac{\zeta_c}{2}
   +\frac{\bar\Gamma_2-2\bar u_c}{4}\Big)\tau+\frac{\bar\Gamma_2-2\bar u_c}{4},
 \nonumber\\
 \frac{\dd\bar r}{\dd\ell}&=\Big(2-\frac{\zeta_c}{2}\Big)\bar r
   +\frac{\bar u_c}{2}(1+\tau),
 \label{eq:qc-mass-tau}
\end{align}
so $\tau$ evolves on its own and drives $\bar r$. At QC the shared bracket in
Eq.~\eqref{eq:rg-shared-bracket} vanishes, so
$\bar\Gamma_2^*-2\bar u_c^*=-4\epsilon_{\mathrm{eff}}^*/3$, with
$\epsilon_{\mathrm{eff}}^*=\epsilon-\zeta_c^*/2=(42-6\sqrt2)\epsilon/47$.
The two relevant eigenvalues are therefore
$2-\zeta_c^*/2-\epsilon_{\mathrm{eff}}^*/3$ for $\tau$ and $2-\zeta_c^*/2$ for
$\bar r$ at fixed $\tau$.  

 \begin{figure*}[t]
 \centering
 \includegraphics[width=\textwidth]{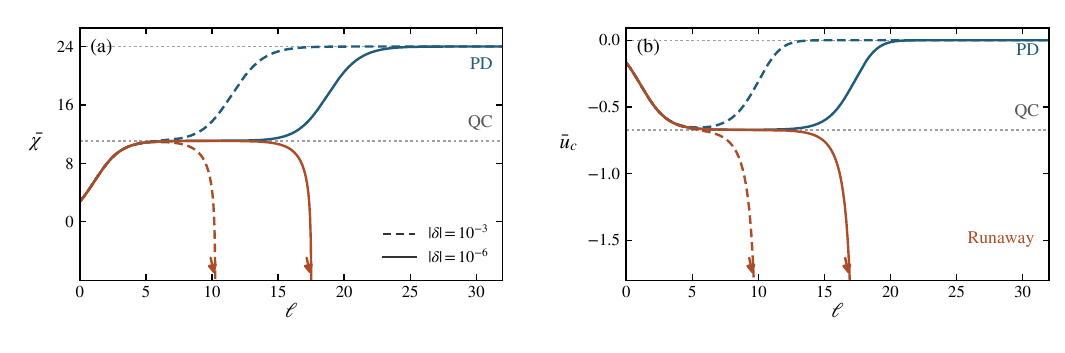}
 \caption{{Approach to and departure from QC at
 $\epsilon=\kappa=1$. Initial conditions are given by
 Eq.~\eqref{eq:qc-incoming-seeds}, with
 $\delta=\pm10^{-3}$ (dashed) and $\pm10^{-6}$ (solid).
 Positive $\delta$: blue, toward PD; negative $\delta$: orange, runaway.
 Panels show (a) $\bar\chi$ and (b) $\bar u_c$; horizontal guides mark
 QC$(s_0)$ and PD. }}
 \label{fig:qc-trajectories}
\end{figure*}

The barred masses are measured in units of the geometric means
$\sqrt{KK_\perp}\,\Lambda_\ell^2$ and $\sqrt{\Gamma_1\Gamma_{1\perp}}\,\Lambda_\ell^2$ (this can be easily seen from Eq.~\eqref{eq:rg-mass-variables} by noting that at $\kappa=1$, we have $c=K/K_\perp=\Gamma_1/\Gamma_{1\perp}$).
Since $K\propto c\propto q^{-\zeta_c^*}$ while
$K_\perp,\Gamma_{1\perp},\gamma$ do not run, the longitudinal dynamical
exponent is $z_\parallel=2-\zeta_c^*$ ($z_\perp=2$). At fixed $\tau$ the
dimensionful mass deviation does not run, so comparing it with $Kq^2$ gives
$\nu_\parallel=1/z_\parallel$, while $\nu_\perp=1/2$ in $d=2-\epsilon$. The
excess flat noise instead scales as
$\Upsilon(q)\propto q^{\epsilon_{\mathrm{eff}}^*/3}$, which sets $q_M$
(Sec.~\ref{sec:infrared-crossover}). In summary:
\begin{align}
 \omega_\kappa&=\frac{5+6\sqrt2}{94}\epsilon+\order(\epsilon^2),\nonumber
 \label{eq:qc-exponents-recalled}
\end{align}
as anticipated in Sec.~\ref{sec:qc-properties}.}

We now examine the stability of QC fixed point within the interaction sector.
At this order, the two masses do not feed back into the interaction
flow, while the interactions drive their evolution. We assume that
their two relevant perturbations are tuned to the critical manifold.
The remaining six-coupling flow on $\kappa=1$ has four attractive
directions, the   zero eigenvalue  mirroring the redundancy in $s$, and one
unstable direction with eigenvalue
\begin{equation}
 y_+=\alpha\epsilon=\frac{56-8\sqrt2}{47}\epsilon.
 \label{eq:qc-unstable-exponent}
\end{equation}
Allowing $\kappa$ to vary adds the attractive direction with eigenvalue
$-\omega_\kappa$. QC is therefore a saddle   after the masses have
been tuned. A small initial projection onto its unstable direction
\newchange{allows a long interval of flow near QC before departure, as shown
in Fig.~\ref{fig:qc-trajectories} and Sec.~\ref{sec:qc-basins}.
Section~\ref{sec:infrared-crossover}   examines the cutoff imposed
by residual flat noise, which eventually classicalizes the RG flow.}

Non-Gaussian noise provides a direct way to probe this instability.
We choose $s=s_0$, for which $\bar\lambda_q^*=0$. In the coordinate
order $(\bar\lambda_c,\bar\lambda_q,\bar u_c,\bar\Gamma_2,
\bar u_q,\bar\chi)$, the normalized unstable eigenvector at
$\epsilon=1$ is
\begin{equation}
 \bm v_+\simeq(-0.1103,0,0.1635,-0.0370,-0.4381,0.8762).
 \label{eq:qc-unstable-vector}
\end{equation}
Its largest component is along $\bar\chi$, accounting for about $77\%$
of the squared norm in these coordinates.  A displacement of $\bar\chi$ alone has a
nonzero projection onto the unstable mode, whose amplitude grows
proportionally to $\delta\bar\chi\,\exp(y_+\ell)$ in the linear regime.

{\figedit{Figure~\ref{fig:qc-crossover}(a)}} shows the resulting departure from QC.
We initialize the interactions at the value $s_0$ for which $\bar{\lambda_q}^*=0$, displace
only $\bar\chi(0)=\bar\chi^*+\delta\bar\chi$, and integrate all six
beta functions in Eqs.~\eqref{eq:rg-lambda-c}--\eqref{eq:rg-chi}
with $\epsilon=\kappa=1$. In particular, $\bar\lambda_q$ is allowed
to evolve from its initial value of zero. We monitor $\zeta_c(\ell)$
from Eq.~\eqref{eq:rg-kappa}: it has the same value at every QC
representative and vanishes at PD, so the plateau tracks how long the RG flow stays 
near QC.

\newchange{Positive displacements, $\delta\bar\chi>0$, approach PD,
whereas negative displacements, $\delta\bar\chi<0$, produce runaway
growth beyond perturbative control.}  Both signs of the
noise perturbation therefore disrupt the QC balance, but lead to
different outcomes. Decreasing $|\delta\bar\chi|$ prolongs the common
plateau because the unstable mode starts with a smaller amplitude.
Since the coarse-graining RG length grows as $\exp(\ell)$, a long plateau allows QC
scaling to influence correlations and relaxation over a broad range
of scales, even though QC is a saddle.   \newedit{\figedit{Fig.~\ref{fig:attractors}(b)}}
extends this analysis to various initial conditions of the RG flow, showing the interplay
between non-Gaussian noise and the Landau-Ginzburg potential.

\subsubsection[Basins of attraction and approach to QC]{Basins of attraction and approach to QC}
\label{sec:qc-basins}

{\figedit{Figure~\ref{fig:attractors}(b)}} demonstrates the three infrared regimes anticipated in  Table~\ref{tab:rg-outcomes}  through numerical
solutions of the RG flow.
We select the representative of the QC family
with $\bar\lambda_q^*=0$ and vary the initial values of $\bar u_c$ and
$\bar\chi$, keeping the other four interactions initially at their QC
values. Each point in this plane therefore defines an initial condition
for the full six-coupling flow in
Eqs.~\eqref{eq:rg-lambda-c}--\eqref{eq:rg-chi}; in particular,
$\bar\lambda_q$ is allowed to evolve after initialization. Blue denotes
trajectories that eventually approach PD, while orange denotes runaway
growth of the interactions towards a breakdown of the perturbative scheme or towards a strong coupling fixed point (the closest analogue in this context is the strong coupling fixed point of KPZ~\cite{kardar1986,freyTauber1994,canet2010,canet2011}).
The colors follow the same convention as
{Fig.~\ref{fig:qc-crossover}(a)}. {As we discuss in Sec.~\ref{sec:infrared-crossover}, in practice, the QC fixed point controls the observable scaling of the system: residual flat Markovian noise, which grows relative to the diffusive noise, generically ends the diffusive scaling before the flow reaches PD or strong coupling, and washes out the quantum scaling responsible for the manifold of fixed points discussed in this work.}

   Trajectories close to the stable
manifold can spend a long RG interval near QC before departing toward
PD or strong coupling. This   is not recorded by the two colors,
which distinguish only the eventual outcomes.
Figure~\ref{fig:qc-trajectories} follows trajectories from initial conditions far from QC through their approach to the QC
fixed point and their subsequent departure. In the coordinate order of
Eq.~\eqref{eq:rg-vector}, the initial conditions for the RG trajectories in Fig.~\ref{fig:qc-trajectories} are
\begin{equation}
 \begin{gathered}
 \bm g(0;\delta)=\mathsf M_0\left[\bm g_{\rm QC}(s_0)+\delta\bm r_+\right],\\
 \mathsf M_0=\operatorname{diag}\left(1,\tfrac12,\tfrac12,
 \tfrac14,\tfrac14,\tfrac14,\tfrac14\right),
 \end{gathered}
 \label{eq:qc-incoming-seeds}
\end{equation}
where $\bm r_+=(0,\bm v_+/v_{+,\chi})$, with $\bm v_+$ the unstable mode
of Eq.~\eqref{eq:qc-unstable-vector}.  

The common plateaus in both $\bar\chi$ and $\bar u_c$ show the approach
to QC before the trajectories separate. Reducing $|\delta|$  lengthens this interval: the attractive
directions first bring the flow toward QC, while the smaller unstable
projection takes longer to grow. Positive $\delta$ eventually leads
to PD; negative $\delta$ leads out of perturbative control. The plots
therefore exhibit approach, intermediate relaxation towards the QC fixed point, and departure in a single
initial-value calculation.   We next examine how residual mass and
flat-noise perturbations can limit the scales over which this picture
applies.

\subsection{Infrared cutoff by residual flat noise}
\label{sec:infrared-crossover}

  The quantum scaling discussed in Secs.~\ref{sec:pd-fixed-point} and~\ref{sec:qc-fixed-point}  and the associated fixed points require the flat
Markov noise to vanish in the infrared.
 Any residual excess is a relevant perturbation that eventually dominates the diffusive noise and ends the diffusive regime. This mechanism is familiar in driven quantum systems, including noise-driven liquids, lossy quantum wires, and dissipative condensates~\cite{dallaTorre2012,froml2020,marinoDiehl2016,marinoDiehlPRL2016,marino2022}; its equilibrium counterpart is the finite-temperature cutoff of a quantum critical regime~\cite{sachdev2011}. Here, however, no temperature is protected by fluctuation--dissipation relations, and the flat noise is itself renormalized.

{
In the following  $q\equiv\Lambda_\ell=\Lambda\ee^{-\ell}$ denotes the running
momentum cutoff [Eq.~\eqref{eq:rg-time}]. To determine the noise
excess near QC, we evaluate the interaction couplings at their fixed-point
values and tune the   spectral mass   to zero.
Equation~\eqref{eq:qc-mass-tau} for
$\tau=\bar\Gamma_0-\bar r$ then reads
\[
 \begin{aligned}
 \partial_\ell\tau&=y_\tau\tau-\frac{\epsilon_{\mathrm{eff}}^*}{3},\\
 y_\tau&=2-\frac{\zeta_c^*}{2}
             -\frac{\epsilon_{\mathrm{eff}}^*}{3}.
 \end{aligned}
\]
The additive tadpole shifts the critical value to
$\tau_{\mathrm{crit}}=\epsilon_{\mathrm{eff}}^*/(3y_\tau)$.
The deviation $\widehat\tau=\tau-\tau_{\mathrm{crit}}$ therefore
obeys the homogeneous flow
$\partial_\ell\widehat\tau=y_\tau\widehat\tau$.

Using the mass normalizations in Eq.~\eqref{eq:rg-mass-variables},
with $a=\Gamma_{1\perp}/K_\perp$, the corresponding dimensionful
noise excess is
\[
 \begin{aligned}
 \Upsilon_{\mathrm{ex}}(q)
 &=\Gamma_{1\perp}\sqrt{c(q)}\,q^2\widehat\tau(q)\\
 &=\Gamma_0(q)-a r(q)
   -\Gamma_{1\perp}\sqrt{c(q)}\,q^2\tau_{\mathrm{crit}}.
 \end{aligned}
\]
The last term is the running critical background.
 Combining this with
the flow of $\widehat\tau$ gives
$\partial_\ell\Upsilon_{\mathrm{ex}}
=-\epsilon_{\mathrm{eff}}^*\Upsilon_{\mathrm{ex}}/3$,
hence $\Upsilon_{\mathrm{ex}}(q)\propto
q^{\epsilon_{\mathrm{eff}}^*/3}$.
Although this excess decreases, the longitudinal diffusive noise
contribution decreases faster:
$\Gamma_1(q)q^2\propto q^{2-\zeta_c^*}=q^{z_\parallel}$.
Their balance at $q_M$ therefore gives
}

\begin{equation}
 q_M\simeq{q_0}\left(\frac{\Upsilon_0}{\Gamma_1q_0^2}\right)^{\nu_M},
 \qquad
 \nu_M^{-1}=z_\parallel-\frac{\epsilon_{\mathrm{eff}}^*}{3},
 \label{eq:qM-renormalized}
\end{equation}
with $q_0$ a matching scale within the QC regime, such that 
$\Upsilon_0\equiv\Upsilon_{\mathrm{ex}}(q_0)>0$,
$\Gamma_1\equiv\Gamma_1(q_0)$, and
$\Upsilon_0/(\Gamma_1q_0^2)\ll1$.
For $\epsilon\to0$ and $r=0$ this reduces to Eq.~\eqref{eq:markov-crossover}  upon taking $q_0=\Lambda$.
At one loop $\nu_M=\tfrac12+\tfrac{12+5\sqrt2}{94}\epsilon+\order(\epsilon^2)$, and the unexpanded one-loop value at $\epsilon=1$ is $\nu_M\simeq0.84$. Fluctuations therefore widen the logarithmic extent $\ell_M\equiv\ln(q_0/q_M)$ of the quantum regime relative to the bare estimate  {at the same matched noise ratio}.
 {For a projection $\delta$ onto the unstable direction of Eq.~\eqref{eq:qc-unstable-vector} at $q_0$, departure from QC occurs after an RG interval $\ell_{\rm dep}\simeq y_+^{-1}\ln(1/|\delta|)$. Both $\ell_{\rm dep}$ and $\ell_M$ are measured from $q_0$.} Figure~\ref{fig:qc-crossover}(a) displays this empirically: curves whose $|\delta\bar\chi|$ differ by a factor $10^2$ exit $y_+^{-1}\ln10^2\simeq4.84$ apart. The flow toward PD or strong coupling can begin within the diffusive regime only if $\ell_{\rm dep}\lesssim\ell_M$, i.e.
\begin{equation}
 |\delta|\gtrsim\left(\frac{\Upsilon_0}{\Gamma_1q_0^2}\right)^{y_+\nu_M},
 \qquad
 y_+\nu_M=\frac{28-4\sqrt2}{47}\epsilon+\order(\epsilon^2).
 \label{eq:qM-vs-departure}
\end{equation}
{
We have $y_+\nu_M<1$, with an unexpanded one-loop estimate of $0.80$
at $\epsilon=1$. A long QC plateau requires small $|\delta|$;
observing the subsequent flow within the diffusive regime then
requires even stronger noise suppression,
$\Upsilon_0/(\Gamma_1q_0^2)\lesssim|\delta|^{1/(y_+\nu_M)}$.
For comparable small detunings,
$\Upsilon_0/(\Gamma_1q_0^2)\sim|\delta|$,
the flat-noise cutoff is reached, and consequently the diffusive quantum scaling ends,  while the couplings are still
near QC. This means that the drift
towards the PD fixed point or runaway to strong coupling are
rarely observable in practice.
}

 Below $q_M$ the power counting is semiclassical~\cite{sieberer2014,dallaTorre2013Dicke,maghrebiGorshkov2016}: with $z=2$, flat noise fixes the canonical dimensions of $\widetilde\phi$ and $\phi$ to $(d+2)/2$ and $(d-2)/2$. The quantum vertices $u_q$, $\chi$, and $\lambda_q$ then acquire the dimensions $-d$, $-d-2$, and $-d/2$ and become irrelevant, whereas $u_c$ and $\lambda_c$ become more relevant, with dimensions $4-d$ and $(4-d)/2$.   Understanding the semiclassical scaling regime below $q_M$ would require a separate RG treatment, and it is left for future work.

\section{Conclusions}\label{sec:conclusions}

A RG flow involving several nonlinearly coupled parameters may have no physical fixed points, or only fixed points with too many unstable directions to be accessible through realistic tuning. The fixed point structure identified here therefore provides an encouraging first indication that    active matter may emerge in microscopic models suitably tuned near criticality in a regime dominated by quantum fluctuations, at least over a transient window of RG scales. Calculations starting from microscopic models would also offer more direct access to entanglement and other signatures of quantum coherence, helping to establish the quantum character of the regimes identified in this work.
 
 \secviedit{Diffusive noise allows self-advection, the Landau--Ginzburg nonlinearity, and the quantum noise vertices of Eq.~\eqref{eq:minimal-active-model} to compete on equal footing below $d_c=2$.  The noise and the convection are the main non-equilibrium agents: at criticality, they promote the effects of RG relevant nonlinearities beyond the classical quartic term, $u_c$, by including kurtosis, skewness, and quantum fluctuations around the double well potential.  
 At one loop, the RG flow contains the PD fixed point  and the QC saddle, whose relevant direction separates the basin of PD from runaway flow (Table~\ref{tab:rg-outcomes} and Fig.~\ref{fig:attractors}). } 
 
 This structure suggests an analogy with a tricritical point, where an additional tuning parameter separates continuous and first-order transitions. In the usual Landau description, this parameter controls the sign of the quartic coupling.   The runaway flow admits two possible interpretations: approach to a strong-coupling fixed point, as in KPZ for $d=2+\epsilon$~\cite{canet2011,freyTauber1994}, or a fluctuation-induced first-order transition, the standard interpretation of a runaway without a stable fixed point~\cite{colemanWeinberg1973,halperin1974,bak1976}.
\secviedit{Out of equilibrium, tricritical points occur in absorbing-state transitions~\cite{grassberger2006}   and in driven-dissipative Ising and Rydberg systems~\cite{marcuzzi2014,overbeck2017}. In classical polar and active Ising flocks, discontinuous ordering is common and is driven by the coupling between density and order~\cite{gregoireChate2004,solonTailleur2013,solon2015,chate2020}, much as density or elastic fluctuations lower the effective quartic coupling in compressible magnets~\cite{larkinPikin1969}. This suggests looking for QC scaling near the boundary between continuous and discontinuous ordering of the coarse-grained magnetization in the full Lindblad dynamics of Eq.~\eqref{eq:micro-lindblad}. At the level of the RG, a natural way to test the first-order interpretation is to include sextic couplings such as $\phi^5\widetilde\phi$, which are marginal in $d=1$ under our quantum power counting. This would require extending nonperturbative methods, such as the functional renormalization group~\cite{berges2002frg,dupuis2021}, to retain all the relevant quantum operators in Eqs.~\eqref{eq:technical-S2}--\eqref{eq:technical-S4}.}

\subsection{Perspectives}\label{sec:outlook}

\emph{Beyond one loop ---} Extending the RG calculation could also help to clarify the fate of the flow beyond the present approximation and test features that may be specific to one loop, including the identities in Eqs.~\eqref{eq:rg-shared-bracket}--\eqref{eq:rg-degeneracy-polynomial} responsible for the QC manifold. The number of candidate diagrams would however drastically increase at two-loop order.   Even then, a two-loop result may offer only limited quantitative improvement at $\epsilon=1$. For model A, obtaining $z=2.14(2)$ in two dimensions may require up to five loops~\cite{adzhemyan2022}. For KPZ, the strong-coupling fixed point is inaccessible at any finite perturbative order~\cite{wiese1998,lassig1995,freyTauber1994}. The runaway in Fig.~\ref{fig:attractors}(b) could have a similar origin.

\secviedit{We could pursue a nonperturbative calculation of the flow. The functional renormalization group~\cite{wetterich1993,berges2002frg,dupuis2021} has already been formulated for Keldysh theories of driven open systems~\cite{sieberer2013,sieberer2014} and applied to polar active fluids~\cite{jentschLee2023}, using derivative expansions. Retaining the full momentum dependence of the two-point functions~\cite{blaizot2006,benitez2012} has proved useful for KPZ, where it gives access to the strong-coupling fixed point and its scaling functions~\cite{canet2010,canet2011,kloss2012,canet2025}, and for turbulence~\cite{canet2016}. However, to our knowledge, extending this treatment to Keldysh actions with all the quantum noise vertices in Eqs.~\eqref{eq:technical-S2}--\eqref{eq:technical-S4}, even without convective couplings, remains an open technical problem.}

\emph{Microscopic models ---} 
  A complementary research program is to identify microscopic regimes in which quantum correlations persist over the scales governed by the RG flow found here.  Full or modified versions of Eq.~\eqref{eq:micro-lindblad}, retaining density fluctuations and coherent conversion ($g\neq0$), provide starting points for this search. Varying the coherent coupling and dissipative rates would allow one to investigate how microscopic processes and initial conditions set the approach to the QC or PD scaling limit in observables of physical interest, such as magnetization or its correlation functions; how they get cut-off by flat Markovian noise, and to which extent neglected irrelevant RG operators in Sec.~\ref{sec:microscopic} impact microscopic non-universal dynamics beyond our predictions. In principle, the dynamics of~\eqref{eq:micro-lindblad} could  provide access to the strong coupling regime more efficiently than any viable RG extension of our work.   The dissipative truncated Wigner approximation (TWA)~\cite{polkovnikov2010,hosseinabadi2025,hartmann2026}  offers a convenient first exploration of parameter space  over large systems and long times. Its semiclassical truncation, however, omits higher response-field vertices and does not retain the full quantum fluctuation hierarchy; it therefore cannot by itself establish conclusively the quantum character of the scaling regime found. Quantum trajectory methods~\cite{plenioKnight1998,daley2014}, combined with matrix product states~\cite{schollwock2011,weimer2021} and benchmarked against direct evolution of the density matrix on smaller systems, could extend this investigation beyond TWA. Relevant diagnostics would include   logarithmic negativity  of the mixed state and   distribution of   entanglement entropies over   ensembles of quantum trajectories~\cite{dos2026kinetically}, which are inaccessible with field theory methods.

\secviedit{\emph{Quantum Toner--Tu hydrodynamics? ---} A further question is how the scalar theory studied here extends to a polarization vector field  coupled to a conserved density, as in Toner--Tu hydrodynamics~\cite{tonerTu1995,tonerTu1998,tonerTuRamaswamy2005,marchetti2013}. }
\secviedit{The difference from classical flocking is especially clear in the ordered phase. Classical flocks order in $d=2$ despite the Mermin--Wagner theorem: convective nonlinearities, relevant below $d=4$, suppress Goldstone fluctuations and produce anomalous hydrodynamics and giant number fluctuations~\cite{tonerTu1995,tonerTu1998,tonerTuRamaswamy2005}.   As in Sec.~\ref{sec:model}, diffusive noise raises the effective dimension to $d+z=d+2>2$, allowing order already at the Gaussian level in one-dimensional systems.
A quantum version of Toner-Tu hydrodynamics constructed along the lines proposed here would
then have different physical content from the original   model~\cite{tonerTu1995}.
The main question is then what new transport phenomena can arise from nonlinear quantum vertices and non-Gaussian noise fluctuations absent from standard Toner-Tu hydrodynamics. This is an exciting direction of research which we are currently pursuing. }

\begin{acknowledgments}

We thank the KITP for its hospitality during the program 'Many body quantum optics'
(October-December 2024), during which this work was conceived. We also acknowledge helpful conversations with S. Ramaswamy and U. T\"{a}uber. The following LLMs have been used for streamlining the writing of the manuscript, benchmarking and performing analytical and numerical calculations, coding, and preparing and restyling figures: Open AI's Chat-GPT 4, 5, 6; Anthropic's Fable 5.1 and Opus 5.5.   
The author remains solely responsible for the integrity, originality, and contents of this paper. This research was supported in part by
National Science Foundation (NSF) Grant No. PHY2309135 to the Kavli Institute for Theoretical Physics,  by   the CAS
Dean's office and by the Office of Research, Innovation and Economic Development at SUNY Buffalo.  
  \end{acknowledgments}

\newpage
\appendix

\section{One-loop RG calculations: conventions}
\label{app:conventions}

\subsection{Shell integration and local projections}
\label{con:shell-projections}

Our Fourier convention is
\begin{equation}
 \phi(X)=\int_Q\ee^{\ii\bm q\cdot\bm x-\ii\omega t}\phi(Q),
 \label{con:fourier}
\end{equation}
with $X=(t,\bm x)$, $Q=(\omega,\bm q)$, and
$\int_Q=\int\dd\omega\,\dd^dq/(2\pi)^{d+1}$; the same  
applies to $\widetilde\phi$.

For one RG step, separate the fields as
\begin{equation}
 \phi=\phi_<+\phi_>,\qquad
 \widetilde\phi=\widetilde\phi_<+\widetilde\phi_>,
 \label{con:field-split}
\end{equation}
where '$>$' contains the shell
$\Lambda\ee^{-\delta\ell}<|\bm q|<\Lambda$ and '$<$' the retained
momenta below it. The Gaussian shell average integrates each loop over
\begin{equation}
 \int_>F(Q)=\int_{\Lambda\ee^{-\delta\ell}<|\bm q|<\Lambda}
 \frac{\dd^dq}{(2\pi)^d}\int_{-\infty}^{\infty}
 \frac{\dd\omega}{2\pi}F(\omega,\bm q).
 \label{con:shell}
\end{equation}
In the following we also write this full integral as $\int_{>Q}$ and denote its
spatial part by $\int_{>\bm q}$.
With $S_{\rm int}=S_3+S_4$ from
Eqs.~\eqref{eq:technical-S3} and~\eqref{eq:technical-S4}, the
correction to the action reads 
\begin{equation}
 \delta S=-\ln\langle\ee^{-S_{\rm int}}\rangle_>
 -S_{\rm int}[\phi_<,\widetilde\phi_<],
 \label{con:shell-action}
\end{equation}
where the brackets   denote the average over the   Gaussian  eliminated fast
fields.  
The last term subtracts the interaction already present among the
retained fields. Expanding the logarithm gives the connected cumulants
in Eq.~\eqref{eq:technical-cumulants}.

A connected one-loop diagram with $n_3$ cubic and $n_4$ quartic
vertices has $n_3+n_4$ internal lines and therefore
$N_{\rm ext}=n_3+2n_4$ external fields. Selecting the terms with
two, three, or four external fields gives
\begin{align}
 \delta S^{(2)}&=\langle S_4\rangle_>
                  -\tfrac12\langle S_3^2\rangle_{>,c},\nonumber\\
 \delta S^{(3)}&=-\langle S_3S_4\rangle_{>,c}
                  +\tfrac16\langle S_3^3\rangle_{>,c},\nonumber\\
 \delta S^{(4)}&=-\tfrac12\langle S_4^2\rangle_{>,c}
                  +\tfrac12\langle S_3^2S_4\rangle_{>,c}
                  -\tfrac1{24}\langle S_3^4\rangle_{>,c}.
 \label{con:cumulant-sectors}
\end{align}
Each right-hand side retains the connected, one-loop contributions
that cannot be disconnected by cutting one internal line. The mixed
coefficients follow from the binomial expansion of $(S_3+S_4)^n$
in Eq.~\eqref{eq:technical-cumulants}.
The two-point bubbles are evaluated in Appendix~\ref{app:kinetic},
the cubic--quartic bubbles in Appendix~\ref{app:cubic}, and the
quartic bubbles in Appendix~\ref{app:quartic}. The remaining
triangles and boxes are treated in Appendix~\ref{app:cancellations}.

To obtain corrections to the local action, we expand each loop in
external momenta small compared with $\Lambda$, and then retain the
term proportional to the shell width $\delta\ell$.
For example, the coefficient of $p_{\longdir}^2$ in a two-point diagram
corrects a longitudinal gradient term in Eq.~\eqref{eq:technical-S2}.
We first choose the momentum $Q$ integrated over the shell in
Eq.~\eqref{con:shell}. The external momentum then enters the vertex
factors and the other propagators, such as $C(Q+P)$.
In the expansion we differentiate those propagators while keeping
the integration range of $Q$ fixed.  

At zero external frequency, the constant and $p_{\longdir}^2$ terms in
$\Sigma_R$ give $\delta r,\delta K$, and those in $\Sigma_N$ give
$\delta\Gamma_0,\delta\Gamma_1$, with signs fixed by
Eq.~\eqref{eq:technical-self-energy-convention}. For the cubic
corrections we retain one longitudinal derivative and match to
\begin{equation}
 \delta S_3=\frac12\int_X\left[
 \delta\lambda_c(\partial_{\longdir}\widetilde\phi)\phi^2
 +\delta\lambda_q(\partial_{\longdir}\phi)\widetilde\phi^2\right].
 \label{con:cubic-projections}
\end{equation}
Appendix~\ref{app:cubic} specifies the external momenta and derivative
placements. Quartic corrections are projected at zero external
momenta and frequencies onto the four monomials in
Eq.~\eqref{eq:technical-S4}.

The contractions follow from the Gaussian action in
Eq.~\eqref{eq:technical-S2}. In the field basis
$(\phi,\widetilde\phi)$, its inverse quadratic form is
\begin{equation}
 \begin{pmatrix}0&P^A\\P^R&-2D\end{pmatrix}^{-1}
 =\begin{pmatrix}2DRA&R\\A&0\end{pmatrix},\qquad
 P^{R,A}=E\mp\ii\gamma\omega.
 \label{con:gaussian-inverse}
\end{equation}
Here $E=E_{\bm q}$ and $D=D_{\bm q}$ are  in
Eq.~\eqref{eq:technical-kernels}, while $R=G^R$ and $A=G^A$.
Equation~\eqref{con:gaussian-inverse} therefore reproduces the
propagators in Eq.~\eqref{eq:technical-propagators}:
$\langle\phi\phi\rangle_0=C=2DRA$ and
$\langle\widetilde\phi\widetilde\phi\rangle_0=0$.  

The response arrows in Fig.~\ref{fig:two-point} point from an earlier
response field to a later physical field. Two arrows going from the
same vertex to the other therefore require the same time ordering,
and this contraction is allowed. Arrows going in opposite directions
would require each vertex to occur later than the other, so the
contraction vanishes. In frequency space this distinction follows
from $R(-Q)=A(Q)$ in Eq.~\eqref{eq:technical-propagators}.
At zero external frequency and momentum the two lines carry $Q$ and $-Q$:
parallel arrows give $R(Q)R(-Q)=R(Q)A(Q)$, whereas opposite arrows
give $R(Q)A(-Q)=R(Q)^2$ (or $A(Q)^2$).
The poles of $R^2$ lie in the same half-plane, so its frequency
integral vanishes; the same holds for $A^2$. The $RA$ integral
has poles in both half-planes and survives, as evaluated in
Eq.~\eqref{kin:frequency-first}. This is the usual causal
structure of the response functional~\cite{tauber2014,kamenevLevchenko2009}.

\subsection{Anisotropy}
\label{con:anisotropy-section}

We retain $K,\Gamma_1$ for the longitudinal coefficients and
$K_\perp,\Gamma_{1\perp}$ for the transverse ones, exactly as in
the main text. We use the ratios defined in Eq.~\eqref{eq:rg-anisotropy},
\begin{equation}
 c=\frac{K}{K_\perp},\qquad
 \kappa=\frac{\Gamma_1/K}{\Gamma_{1\perp}/K_\perp}.
 \label{con:anisotropy}
\end{equation}
At $r=\Gamma_0=0$, writing $q_{\longdir}=q\cos\theta$ and
$q_\perp=q\sin\theta$ gives
\begin{align}
 E_{\bm q}&=K_\perp q^2\mathcal E(\theta),&
 \mathcal E&=c\cos^2\theta+\sin^2\theta,\nonumber\\
 D_{\bm q}&=\Gamma_{1\perp}q^2\mathcal D(\theta),&
 \mathcal D&=\kappa c\cos^2\theta+\sin^2\theta.
 \label{con:angular-kernels}
\end{align}
For the leading $\epsilon$ expansion, gradient and vertex corrections
are evaluated at $r=\Gamma_0=0$, with angular averages taken at $d=2$.
The kernels in Eq.~\eqref{con:angular-kernels} will be used in
Appendices~\ref{kin:integrations} and~\ref{cub:integrations} to
perform those averages.

\raggedbottom

\section{Two-point functions}
\label{app:kinetic}

\subsection{Self-energies and   longitudinal momentum expansion}
\label{kin:momentum-expansion}

The coefficients of $p_{\longdir}^2$ in $\Sigma_R$ and $\Sigma_N$ give
$\delta K$ and $\delta\Gamma_1$ through
Eq.~\eqref{eq:technical-self-energy-convention}. We evaluate the
bubbles in Fig.~\ref{fig:two-point}(a),(b) to obtain the shifts
quoted in Eq.~\eqref{eq:technical-kinetic-corrections}.
We set the external frequency to zero and take $P=(0,p_{\longdir},\bm0_\perp)$. For the two
internal lines we use the arguments $Q$ and $Q+P$. Writing
$B(Q)=R(Q)+A(Q)$, the second cumulant
$-\langle S_3^2\rangle_{>,c}/2$ in
Eq.~\eqref{con:cumulant-sectors} gives
\begin{align}
 \Sigma_R(P)={}&\frac{\lambda_c^2p_{\longdir}}{4}\int_{>Q}
 \bigl[(p_{\longdir}+q_{\longdir})C(Q)B(Q+P)\nonumber\\
 &\hspace{35mm}-q_{\longdir}B(Q)C(Q+P)\bigr]\nonumber\\
 &-\frac{\lambda_c\lambda_qp_{\longdir}^2}{4}\int_{>Q}
 \bigl[R(Q)A(Q+P)+A(Q)R(Q+P)\bigr],
 \label{kin:self-response}\\
 \Sigma_N(P)={}&\frac{\lambda_c^2p_{\longdir}^2}{4}\int_{>Q}C(Q)C(Q+P)
 \nonumber\\
 &-\frac{\lambda_c\lambda_qp_{\longdir}}{4}\int_{>Q}
 \bigl[(p_{\longdir}+q_{\longdir})B(Q)C(Q+P)\nonumber\\
 &\hspace{35mm}-q_{\longdir}C(Q)B(Q+P)\bigr].
 \label{kin:self-noise}
\end{align}
The $\lambda_c^2$ bracket in Eq.~\eqref{kin:self-response} and
the mixed bracket in Eq.~\eqref{kin:self-noise} contain one
correlation line, $C$, and one response line, summed over
its two orientations in $B=R+A$. The differentiated internal field
can carry momentum $p_{\longdir}+q_{\longdir}$ or $-q_{\longdir}$, giving the two terms in
each bracket. 

The two remaining terms have different internal line content.
For the $\lambda_c\lambda_q$ term in
Eq.~\eqref{kin:self-response}, the external response field belongs
to the $\lambda_c$ vertex and the external physical field to the
$\lambda_q$ vertex. Both come with derivatives in
Eq.~\eqref{eq:technical-S3}, giving the overall factor $p_{\longdir}^2$.
The two internal physical fields at the $\lambda_c$ vertex pair
with the two internal response fields at the $\lambda_q$ vertex.
The resulting parallel response lines give $R(Q)A(Q+P)$.
At zero external frequency, changing the integration variable
$\omega\to-\omega$ interchanges $R$ and $A$, turning this
product into $A(Q)R(Q+P)$. The two products therefore have equal
frequency integrals. The last bracket in
Eq.~\eqref{kin:self-response} is their symmetric sum.

For the $\lambda_c^2$ term in Eq.~\eqref{kin:self-noise}, each
vertex instead supplies one external response field. Their two
external derivatives again give $p_{\longdir}^2$, while the four remaining
physical fields pair into two correlation lines, $C(Q)C(Q+P)$.
There are two ways to pair the physical fields between the vertices.
Multiplying this count by the cumulant factor $-1/2$ and the two
vertex coefficients $\lambda_c/2$ gives
$-\tfrac12(\lambda_c/2)^2\times2=-\lambda_c^2/4$ in the action.
Since the noise correction enters as $-\Sigma_N\widetilde\phi^2$
in Eq.~\eqref{eq:technical-self-energy-convention}, its contribution
to $\Sigma_N$ is positive, as written in Eq.~\eqref{kin:self-noise}.

The mixed response term also has two pairings. The two orderings
of its distinct vertices cancel the second-cumulant factor $1/2$,
leaving $-(\lambda_c/2)(\lambda_q/2)\times2
=-\lambda_c\lambda_q/2$. Writing the response product as the
average of its two equal frequency integrals supplies a further
factor $1/2$, giving the coefficient
$-\lambda_c\lambda_qp_{\longdir}^2/4$ of the symmetric bracket in
Eq.~\eqref{kin:self-response}.

We now expand $B(Q+P)=B+p_{\longdir}\partial_{q_{\longdir}}B+\order(p_{\longdir}^2)$,
and likewise for $C$, following the fixed-shell prescription in
Appendix~\ref{con:shell-projections}. The first bracket in
Eq.~\eqref{kin:self-response} becomes
\begin{equation}
 p_{\longdir}\bigl[CB+q_{\longdir}(C\partial_{q_{\longdir}}B-B\partial_{q_{\longdir}}C)\bigr]
 +\order(p_{\longdir}^2).
 \label{kin:expanded-bracket}
\end{equation}
The zeroth-order terms cancel. The bracket in the mixed noise term
of Eq.~\eqref{kin:self-noise} interchanges $B$ and $C$ in
Eq.~\eqref{kin:expanded-bracket}, reversing the derivative contribution.
The mixed response term in Eq.~\eqref{kin:self-response} and the
$\lambda_c^2$ noise term in Eq.~\eqref{kin:self-noise} already
carry $p_{\longdir}^2$. Their propagators can therefore be evaluated at
$P=0$, giving $RA+AR=2RA$ and $C^2$, respectively.
Collecting these four contributions gives
\begin{align}
 \delta K&=\frac{\lambda_c^2}{4}J_K
             -\frac{\lambda_c\lambda_q}{2}I_3,\nonumber\\
 \delta\Gamma_1&=\frac{\lambda_c^2}{4}I_4
             -\frac{\lambda_c\lambda_q}{4}J_\Gamma,
 \label{kin:gradient-shifts}\\
 I_3&=\int_{>Q}RA,\qquad I_4=\int_{>Q}C^2,\nonumber\\
 J_K&=\int_{>Q}\bigl[CB+q_{\longdir}(C\partial_{q_{\longdir}}B-B\partial_{q_{\longdir}}C)\bigr],
 \nonumber\\
 J_\Gamma&=\int_{>Q}\bigl[CB+q_{\longdir}(B\partial_{q_{\longdir}}C-C\partial_{q_{\longdir}}B)\bigr].
 \label{kin:loop-definitions}
\end{align}
 In the self-energies,
the terms just computed are
$\Sigma_R(P)=p_{\longdir}^2\delta K+\cdots$ and
$\Sigma_N(P)=p_{\longdir}^2\delta\Gamma_1+\cdots$, as follows from
Eqs.~\eqref{kin:self-response}--\eqref{kin:expanded-bracket}.
More generally, each of these bubbles contains a differentiated
external field along the $\longdir$ direction even before we set the external frequency
and transverse momentum to zero. It therefore vanishes at $p_{\longdir}=0$
with those other components retained. Such bubbles cannot produce
a term proportional only to $p_\perp^2$, to $\ii\Omega$, or to a
constant. They consequently do not correct $K_\perp$,
$\Gamma_{1\perp}$, $\gamma$, $r$, or $\Gamma_0$ in
Eq.~\eqref{eq:technical-S2}.

\subsection{Frequency integrals and angular averages}
\label{kin:integrations}

We now evaluate $I_3$, $I_4$, $J_K$, and $J_\Gamma$ in
Eq.~\eqref{kin:loop-definitions}, which determine the kinetic
corrections through Eq.~\eqref{kin:gradient-shifts}.
We first integrate over frequency and then over the momentum shell.
The common term $CB$ in $J_K$ and $J_\Gamma$ will also enter the
cubic corrections in Appendix~\ref{app:cubic}; we denote its integral by
\begin{equation}
 L_0=\int_{>Q}C(Q)B(Q).
 \label{kin:L0-definition}
\end{equation}

At fixed momentum, write $E=E_{\bm q}$, $D=D_{\bm q}$, and
$\Delta=E^2+\gamma^2\omega^2$. The Gaussian propagators in
Eq.~\eqref{eq:technical-propagators} give $RA=\Delta^{-1}$,
$C=2D/\Delta$, and $B=R+A=2E/\Delta$.
Thus $I_3$ requires the frequency integral of $\Delta^{-1}$,
while $I_4$ and $L_0$ require that of $\Delta^{-2}$.
For $E>0$, changing variables to $y=\gamma\omega/E$ in the first
of these integrals gives
\begin{equation}
 \int\frac{\dd\omega}{2\pi}\frac{1}{\Delta}
 =\frac{1}{2\pi\gamma E}\int_{-\infty}^{\infty}
 \frac{\dd y}{1+y^2}
 =\frac{1}{2\gamma E}.
 \label{kin:frequency-first}
\end{equation}
To obtain the squared denominator, differentiate
Eq.~\eqref{kin:frequency-first} with respect to $E$:
\begin{equation}
 \int\frac{\dd\omega}{2\pi}\frac{1}{\Delta^2}
 =-\frac{1}{2E}\partial_E\frac{1}{2\gamma E}
 =\frac{1}{4\gamma E^3}.
 \label{kin:frequency-integrals}
\end{equation}
Substituting Eqs.~\eqref{kin:frequency-first} and~\eqref{kin:frequency-integrals}
into the definitions in Eqs.~\eqref{kin:loop-definitions} and~\eqref{kin:L0-definition}
then gives
\begin{align}
 I_3&=\int_{>\bm q}\frac{1}{2\gamma E_{\bm q}},&
 I_4&=\int_{>\bm q}\frac{D_{\bm q}^2}{\gamma E_{\bm q}^3},
 \nonumber\\
 L_0&=\int_{>\bm q}\frac{D_{\bm q}}{\gamma E_{\bm q}^2}.
 \label{kin:frequency-reduced}
\end{align}

It remains to evaluate the derivative contributions to $J_K$ and
$J_\Gamma$ in Eq.~\eqref{kin:loop-definitions}. They contain the
same difference, $C\partial_{q_{\longdir}}B-B\partial_{q_{\longdir}}C$, with
opposite signs. Inserting $C=2D/\Delta$ and $B=2E/\Delta$ into
this difference cancels the derivatives of their common denominator.
The derivatives of the numerators follow from
Eq.~\eqref{eq:technical-kernels}:
$\partial_{q_{\longdir}}E=2Kq_{\longdir}$ and $\partial_{q_{\longdir}}D=2\Gamma_1q_{\longdir}$.
Consequently,
\begin{equation}
 C\partial_{q_{\longdir}}B-B\partial_{q_{\longdir}}C
 =\frac{8q_{\longdir}(KD_{\bm q}-\Gamma_1E_{\bm q})}{\Delta^2}.
 \label{kin:derivative-identity}
\end{equation}
Multiplying Eq.~\eqref{kin:derivative-identity} by the remaining
$q_{\longdir}$ in Eq.~\eqref{kin:loop-definitions}, performing the frequency
integral with Eq.~\eqref{kin:frequency-integrals}, and adding the
common $L_0$ term from Eq.~\eqref{kin:frequency-reduced} yields
\begin{align}
 J_K&=\int_{>\bm q}\left[
 \frac{D_{\bm q}}{\gamma E_{\bm q}^2}
 +\frac{2q_{\longdir}^2(KD_{\bm q}-\Gamma_1E_{\bm q})}
 {\gamma E_{\bm q}^3}\right],\nonumber\\
 J_\Gamma&=\int_{>\bm q}\left[
 \frac{D_{\bm q}}{\gamma E_{\bm q}^2}
 -\frac{2q_{\longdir}^2(KD_{\bm q}-\Gamma_1E_{\bm q})}
 {\gamma E_{\bm q}^3}\right].
 \label{kin:derivative-reduced}
\end{align}

We now perform the spatial integrals in
Eqs.~\eqref{kin:frequency-reduced} and~\eqref{kin:derivative-reduced}.
For the leading gradient corrections, set $r=\Gamma_0=0$ and use
$E_{\bm q}=K_\perp q^2\mathcal E(\theta)$ and
$D_{\bm q}=\Gamma_{1\perp}q^2\mathcal D(\theta)$ from
Eq.~\eqref{con:angular-kernels}. Each integrand is then proportional
to $q^{-2}$, so all five integrals share the radial factor
\begin{equation}
 \int_{\Lambda\ee^{-\delta\ell}}^\Lambda q^{d-3}\dd q
 =\Lambda^{d-2}\delta\ell+\order(\delta\ell^2).
 \label{kin:radial-shell}
\end{equation}
The remaining integrations are angular averages, which we evaluate
at $d=2$ for the leading
$\epsilon$ expansion. We write
$\langle f\rangle_\theta=(2\pi)^{-1}\int_0^{2\pi}f\,\dd\theta$
and $x=\cos^2\theta$, so that
$\mathcal E=1+(c-1)x$ and $\mathcal D=1+(\kappa c-1)x$.
The integral $I_3$ requires $\langle\mathcal E^{-1}\rangle_\theta$;
$L_0$ requires $\langle\mathcal D/\mathcal E^2\rangle_\theta$;
and $I_4$ requires $\langle\mathcal D^2/\mathcal E^3\rangle_\theta$.
For $J_K$ and $J_\Gamma$, Eq.~\eqref{kin:derivative-reduced}
adds or subtracts the average of
$2cx(1-\kappa)(1-x)/\mathcal E^3$, since
$c\mathcal D-\kappa c\mathcal E=c(1-\kappa)(1-x)$.

All these averages follow from one elementary integral. Define
$F(a,b)=\langle(a+b\cos^2\theta)^{-1}\rangle_\theta$.
Because the integrand depends only on $\cos^2\theta$, all four
quadrants give the same contribution. We can therefore integrate
from $0$ to $\pi/2$ and set $t=\tan\theta$ on that interval:
\begin{align}
 F(a,b)&=\frac{2}{\pi}\int_0^{\pi/2}
 \frac{\dd\theta}{a+b\cos^2\theta}\nonumber\\
 &=\frac{2}{\pi}\int_0^\infty\frac{\dd t}{at^2+a+b}
 =\frac{1}{\sqrt{a(a+b)}}.
 \label{kin:angular-generator}
\end{align}
Here $a>0$ and $a+b>0$. Differentiating the integrand in
Eq.~\eqref{kin:angular-generator} once or twice produces the
squared and cubed denominators needed above:
\begin{align}
 \left\langle\frac{1}{(a+bx)^2}\right\rangle_\theta&=-\partial_a F,&
 \left\langle\frac{x}{(a+bx)^2}\right\rangle_\theta&=-\partial_b F,
 \nonumber\\
 \left\langle\frac{1}{(a+bx)^3}\right\rangle_\theta&=\tfrac12\partial_a^2F,&
 \left\langle\frac{x}{(a+bx)^3}\right\rangle_\theta&=\tfrac12\partial_a\partial_bF,
 \nonumber\\
 \left\langle\frac{x^2}{(a+bx)^3}\right\rangle_\theta&=\tfrac12\partial_b^2F.
 \label{kin:angular-derivatives}
\end{align}
We take these derivatives with $a,b$ independent and only then set
$a=1$, $b=c-1$. For example, the last three relations give
\begin{align}
 \langle\mathcal E^{-3}\rangle_\theta
 &=\frac{3c^2+2c+3}{8c^{5/2}},\nonumber\\
 \langle x\mathcal E^{-3}\rangle_\theta
 &=\frac{c+3}{8c^{5/2}},\qquad
 \langle x^2\mathcal E^{-3}\rangle_\theta
 =\frac{3}{8c^{5/2}}.
 \label{kin:third-denominator-moments}
\end{align}
The difference of the last two averages, multiplied by
$2c(1-\kappa)$, gives $(1-\kappa)/(4\sqrt c)$: this is the
additional angular term in $J_K$ and its opposite in $J_\Gamma$,
as identified after Eq.~\eqref{kin:radial-shell}.
To evaluate $I_4$, expand its
numerator as
$\mathcal D^2=1+2(\kappa c-1)x+(\kappa c-1)^2x^2$
and use all three averages in Eq.~\eqref{kin:third-denominator-moments}.
Together with $F$ and its first derivatives in
Eqs.~\eqref{kin:angular-generator} and~\eqref{kin:angular-derivatives},
this gives the angular factors for $I_3,L_0,J_K,J_\Gamma,I_4$,
in that order:
\begin{align}
 \langle\mathcal E^{-1}\rangle_\theta&=\frac{1}{\sqrt c},\nonumber\\
 \left\langle\frac{\mathcal D}{\mathcal E^2}\right\rangle_\theta
 &=\frac{\kappa+1}{2\sqrt c},\nonumber\\
 \left\langle\frac{\mathcal D}{\mathcal E^2}
 +\frac{2cx(1-\kappa)(1-x)}{\mathcal E^3}\right\rangle_\theta
 &=\frac{\kappa+3}{4\sqrt c},\nonumber\\
 \left\langle\frac{\mathcal D}{\mathcal E^2}
 -\frac{2cx(1-\kappa)(1-x)}{\mathcal E^3}\right\rangle_\theta
 &=\frac{3\kappa+1}{4\sqrt c},\nonumber\\
 \left\langle\frac{\mathcal D^2}{\mathcal E^3}\right\rangle_\theta
 &=\frac{3\kappa^2+2\kappa+3}{8\sqrt c}.
 \label{kin:angular}
\end{align}

To assemble the shell integrals, multiply each angular factor in
Eq.~\eqref{kin:angular} by the radial integral in
Eq.~\eqref{kin:radial-shell} and its kinetic prefactor in
Eqs.~\eqref{kin:frequency-reduced} or~\eqref{kin:derivative-reduced}.
Using $a_d=S_d/(2\pi)^d$ as in Sec.~\ref{sec:rg-variables}, define
\begin{equation}
 \mathcal H_d=\frac{a_d\Lambda^{d-2}\delta\ell}{\gamma\sqrt c}.
 \label{kin:shell-factor}
\end{equation}
The five integrals become
\begin{align}
 I_3&=\frac{\mathcal H_d}{2K_\perp},\nonumber\\
 L_0&=\mathcal H_d\frac{\Gamma_{1\perp}}{K_\perp^2}
                    \frac{\kappa+1}{2},\nonumber\\
 J_K&=\mathcal H_d\frac{\Gamma_{1\perp}}{K_\perp^2}
                    \frac{\kappa+3}{4},\nonumber\\
 J_\Gamma&=\mathcal H_d\frac{\Gamma_{1\perp}}{K_\perp^2}
                    \frac{3\kappa+1}{4},\nonumber\\
 I_4&=\mathcal H_d\frac{\Gamma_{1\perp}^2}{K_\perp^3}
                    \frac{3\kappa^2+2\kappa+3}{8}.
 \label{kin:shell-values}
\end{align}
Inserting Eq.~\eqref{kin:shell-values} into
Eq.~\eqref{kin:gradient-shifts} now gives the corrections to the
longitudinal coefficients in Eq.~\eqref{eq:technical-S2}:
\begin{align}
 \delta K={}&\mathcal H_d\left[
 \frac{\lambda_c^2\Gamma_{1\perp}}{16K_\perp^2}(\kappa+3)
 -\frac{\lambda_c\lambda_q}{4K_\perp}\right],\nonumber\\
 \delta\Gamma_1={}&\mathcal H_d\left[
 \frac{\lambda_c^2\Gamma_{1\perp}^2}{32K_\perp^3}
 (3\kappa^2+2\kappa+3)\right.\nonumber\\
 &\left.\hspace{15mm}
 -\frac{\lambda_c\lambda_q\Gamma_{1\perp}}{16K_\perp^2}
 (3\kappa+1)\right].
 \label{kin:explicit-shifts}
\end{align}

To recover the flow quoted in Eq.~\eqref{eq:rg-kappa}, divide
$\delta K$ and $\delta\Gamma_1$ in Eq.~\eqref{kin:explicit-shifts}
by $K\delta\ell$ and $\Gamma_1\delta\ell$, respectively, and take
$\delta\ell\to0$. Substituting $K=cK_\perp$,
$\Gamma_1=\kappa c\Gamma_{1\perp}$ from
Eq.~\eqref{eq:rg-anisotropy} and the dimensionless couplings in
Eq.~\eqref{eq:rg-convective-variables} gives
\begin{align}
 \frac{\dd\ln c}{\dd\ell}
 &=\frac{\bar\lambda_c^2\kappa(\kappa+3)}{16}
   -\frac{\bar\lambda_c\bar\lambda_q\kappa}{4},\nonumber\\
 \frac{\dd\ln\Gamma_1}{\dd\ell}
 &=\frac{\bar\lambda_c^2(3\kappa^2+2\kappa+3)}{32}
   -\frac{\bar\lambda_c\bar\lambda_q(3\kappa+1)}{16}.
 \label{kin:logarithmic-flows}
\end{align}
Here $K_\perp$ and $\Gamma_{1\perp}$ are fixed, as explained in
Sec.~\ref{sec:rg-variables}, and the shell is evaluated at the
running cutoff $\Lambda_\ell$ in Eq.~\eqref{eq:rg-time}.
The definition of $\kappa$ in Eq.~\eqref{eq:rg-anisotropy}
therefore gives $\dd\ln\kappa/\dd\ell=
\dd\ln\Gamma_1/\dd\ell-\dd\ln c/\dd\ell$.
Subtracting the two lines of Eq.~\eqref{kin:logarithmic-flows} yields
\begin{equation}
 \frac{\dd\ln\kappa}{\dd\ell}
 =\frac{\bar\lambda_c^2(\kappa-1)(\kappa-3)}{32}
  +\frac{\bar\lambda_c\bar\lambda_q(\kappa-1)}{16},
 \label{kin:kappa-flow}
\end{equation}
which completes the derivation of Eq.~\eqref{eq:rg-kappa}. 

For brevity of the presentation, we omit the derivation of the beta functions for the   spectral mass   and for the flat noise strength, as they are standard.  The Keldysh
renormalization procedure is reviewed in Ref.~\cite{sieberer2016},
and analogous mass and noise tadpoles are discussed in
Ref.~\cite{marinoDiehl2016}.

\section{Convective-vertex corrections}
\label{app:cubic}

\begin{figure*}[!t]
 \centering
\begingroup
\setlength{\unitlength}{\textwidth}
\begin{picture}(1,0.564286)
  \put(0,0){\includegraphics[width=\textwidth]{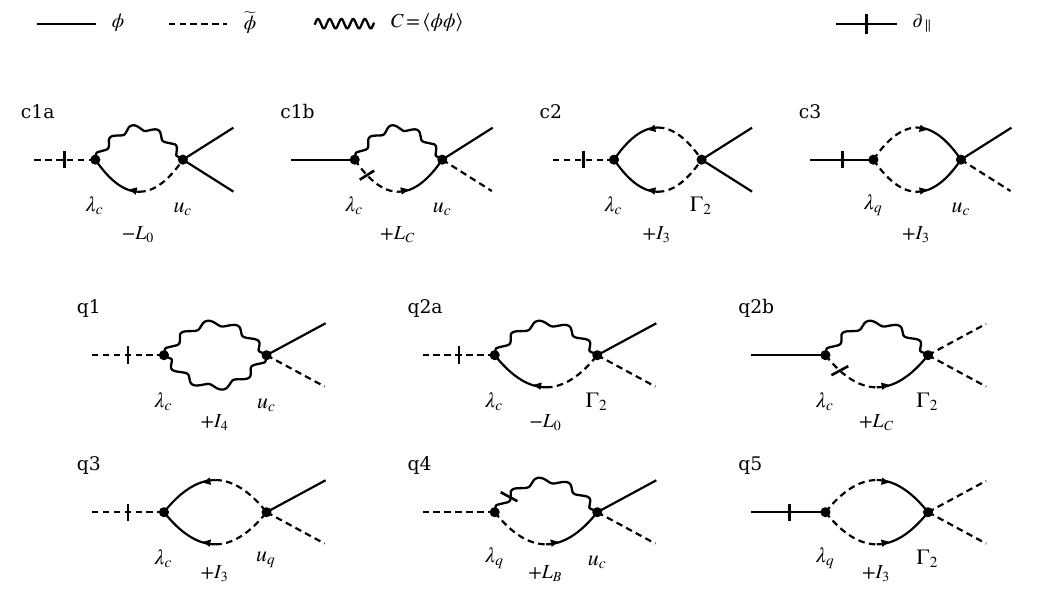}}
  \put(0.035,0.490929){\makebox(0,0)[l]{\small (a) Contributions to $\delta\lambda_c$}}
  \put(0.035,0.304714){\makebox(0,0)[l]{\small (b) Contributions to $\delta\lambda_q$}}
\end{picture}
\endgroup
 \caption{Nonzero cubic--quartic bubbles. The upper group corrects
 self-advection and the lower group corrects the quantum convective
 coupling. Diagram identifiers match Table~\ref{cub:table}; a marked
 endpoint carries the spatial derivative. }
 \label{fig:app-cubic}
\end{figure*}

We now calculate the loop corrections to self-advection, $\lambda_c$,
and to the field-gradient dependence of the noise variance, $\lambda_q$,
introduced in Eq.~\eqref{eq:nonlinear-noise-cumulants}.
The aim is to obtain the vertex contributions to the beta functions
in Eqs.~\eqref{eq:rg-lambda-c} and~\eqref{eq:rg-lambda-q}. We first evaluate the cubic--quartic
bubbles in Fig.~\ref{fig:app-cubic}, keeping the derivative placements
separate because they lead to different loop integrals.
We denote the two terms in Eq.~\eqref{eq:technical-S3} by $S_c,S_q$
and the four terms in Eq.~\eqref{eq:technical-S4} by
$S_{u_c},S_{\Gamma_2},S_{u_q},S_\chi$, for instance
$S_{u_c}=(u_c/6)\int_X\phi^3\widetilde\phi$.

\subsection{Projections and loop integrals}
\label{cub:projections-section}

The cubic correction must be matched to the two operators in
Eq.~\eqref{con:cubic-projections}. We set all other external frequencies
to zero and retain one power of longitudinal momentum. For
$P=(0,p_{\longdir},\bm0_\perp)$, the two projections are
\begin{align}
 \delta\mathcal V_{\widetilde\phi\phi\phi}(P,0,-P)
 &=\ii p_{\longdir}\delta\lambda_c+\order(p_{\longdir}^2),
 \label{cub:projection-c}\\
 \delta\mathcal V_{\phi\widetilde\phi\widetilde\phi}(P,0,-P)
 &=\ii p_{\longdir}\delta\lambda_q+\order(p_{\longdir}^2).
 \label{cub:projection-q}
\end{align}
The momentum arguments follow the field subscripts. Thus, in
Eq.~\eqref{cub:projection-c}, the response field carries $P$ and
the two physical fields carry $0$ and $-P$; in
Eq.~\eqref{cub:projection-q}, the physical field carries $P$.
In both cases the differentiated field carries $P$, giving
$\ii p_{\longdir}$ with the Fourier convention in Eq.~\eqref{con:fourier}.
The quantities $\delta\mathcal V$ are obtained by differentiating
$\delta S_3$ in Eq.~\eqref{con:cubic-projections} with respect to
its three fields. For the classical operator, the two derivatives
with respect to $\phi$ give a factor $2!$, so its coefficient
$\delta\lambda_c/2$ becomes $\delta\lambda_c$. The two response
fields give the same factor for the quantum operator.

Below we calculate $\delta S_3$ directly by Wick contraction.
We therefore match its operator coefficients to
$\delta\lambda_c/2$ and $\delta\lambda_q/2$ in
Eq.~\eqref{con:cubic-projections}. If a contraction puts the
derivative on another field, we first use
\begin{align}
 \int_X\widetilde\phi\phi\partial_{\longdir}\phi
 &=-\frac12\int_X(\partial_{\longdir}\widetilde\phi)\phi^2,
 \label{cub:parts-c}\\
 \int_X\phi\widetilde\phi\partial_{\longdir}\widetilde\phi
 &=-\frac12\int_X(\partial_{\longdir}\phi)\widetilde\phi^2,
 \label{cub:parts}
\end{align}
with periodic boundaries or fields decaying at infinity.

Selecting three external fields in Eq.~\eqref{con:cumulant-sectors}
gives the one-loop terms
\begin{equation}
 \delta S^{(3)}=-\langle S_3S_4\rangle_{>,c}
 +\frac16\langle S_3^3\rangle_{>,c}.
 \label{cub:cumulants}
\end{equation}
The first term joins one cubic and one quartic vertex by two internal
lines, leaving three external fields.  
These are the bubbles in Fig.~\ref{fig:app-cubic}.
The second term gives cubic triangles. Their contribution to
Eqs.~\eqref{cub:projection-c} and~\eqref{cub:projection-q} vanishes
after the routing sum and frequency integration, as shown by
Eq.~\eqref{can:total-derivative} in Appendix~\ref{can:cubic-triangles}.

For the bubbles, the next step is to identify the two internal
propagators and the location of the derivative. If the derivative is
external, it already supplies the required $\ii p_{\longdir}$, and the
propagators can be evaluated at $P=0$. Two parallel response lines
then give $I_3$, two correlation lines give $I_4$, and a
correlation--response pair gives an integral proportional to $L_0$.
These are the same integrals defined in
Eqs.~\eqref{kin:loop-definitions} and~\eqref{kin:L0-definition}.
With $B=R+A$, they read
\begin{equation}
 I_3=\int_{>Q}RA,\qquad I_4=\int_{>Q}C^2,\qquad
 L_0=\int_{>Q}CB.
 \label{cub:shared-integrals}
\end{equation}
Here two parallel responses give $RA$ by the time-ordering argument
in Appendix~\ref{con:shell-projections}.

If the differentiated field is internal, it supplies $\ii q_{\longdir}$.
The term at $P=0$ is odd in $q_{\longdir}$ and integrates to zero. The
retained external momentum must then come from the first-order
expansion of the other propagator. Expanding a correlation or a
response line introduces, respectively,
\begin{align}
 L_C&=\int_{>Q}q_{\longdir}B(Q)\partial_{q_{\longdir}}C(Q),\nonumber\\
 L_B&=\int_{>Q}q_{\longdir}C(Q)\partial_{q_{\longdir}}B(Q).
 \label{cub:derivative-integrals}
\end{align}
The subscripts identify the expanded line. As specified in the paragraph following
Eq.~\eqref{con:cumulant-sectors} in Appendix~\ref{con:shell-projections},
we choose $Q$ as the shell momentum before expanding in $P$.
For example, $C(Q-P)$ is differentiated at fixed $Q$, giving
$-p_{\longdir}\partial_{q_{\longdir}}C(Q)$, while the integration limits in
Eq.~\eqref{con:shell} remain fixed. Equation~\eqref{cub:expansion-example}
below applies this prescription to c1b.

An individual contraction contains one oriented response, $R$ or
$A$, rather than their sum $B$. To use
Eqs.~\eqref{cub:shared-integrals} and~\eqref{cub:derivative-integrals},
we therefore need its relation to $B$ under frequency integration.
The propagators in Eq.~\eqref{eq:technical-propagators} obey
$R=B/2+(R-A)/2$ and $A=B/2-(R-A)/2$. Here $B$ is even in
frequency and $R-A$ is odd. Since $C$ is even,
$\int_{>Q}C(R-A)=0$: this sets the difference of the $CR$ and
$CA$ integrals to zero, while their sum is $L_0$.
The same argument applies after a momentum derivative, which
preserves frequency parity. Thus
\begin{align}
 \int_{>Q}CR=\int_{>Q}CA&=\frac{L_0}{2},\nonumber\\
 \int_{>Q}q_{\longdir}A\partial_{q_{\longdir}}C&=\frac{L_C}{2},\qquad
 \int_{>Q}q_{\longdir}C\partial_{q_{\longdir}}R=\frac{L_B}{2}.
 \label{cub:oriented-integrals}
\end{align}

\subsection{Corrections to self-advection}
\label{cub:contractions}
\label{cub:classical-bubbles}

We first extract the coefficient of
$(\partial_{\longdir}\widetilde\phi)\phi^2/2$ in
Eq.~\eqref{con:cubic-projections}. The four diagrams c1a, c1b, c2,
and c3 in Fig.~\ref{fig:app-cubic}(a) contribute to this operator. Table~\ref{cub:table} summarizes the diagrams and their corrections.

\emph{Diagrams c1a and c1b.}
Both diagrams come from $-\langle S_cS_{u_c}\rangle_{>,c}$.
Using the action coefficients in
Eqs.~\eqref{eq:technical-S3} and~\eqref{eq:technical-S4}, their
common factor before counting contractions is
\begin{equation}
 -\frac{\lambda_c}{2}\frac{u_c}{6}
 =-\frac{\lambda_cu_c}{12}.
 \label{cub:bare-factor}
\end{equation}
For c1a, the differentiated response field at $S_c$ is external.
Its two internal physical fields connect to one physical and one
response field at $S_{u_c}$, leaving the two external physical
fields there. There are three choices of the internal physical
field at $S_{u_c}$ and two choices of its partner at $S_c$.
These six contractions produce one correlation and one response line:
\begin{equation}
 \delta S_{\mathrm{c1a}}
 =-\frac{\lambda_cu_c}{2}\int_{>Q}CR
       \int_X(\partial_{\longdir}\widetilde\phi)\phi^2.
 \label{cub:worked-action}
\end{equation}
Using Eq.~\eqref{cub:oriented-integrals} in
Eq.~\eqref{cub:worked-action} gives the action coefficient
$-\lambda_cu_cL_0/4$. Comparison with
Eq.~\eqref{con:cubic-projections} therefore yields
$\delta\lambda_c|_{\mathrm{c1a}}=-\lambda_cu_cL_0/2$.

For c1b, a physical field at $S_c$ is external, so its differentiated
response field is now internal. The external fields at $S_{u_c}$
are one physical and one response field. There are two choices
for the external physical field at $S_c$ and three at $S_{u_c}$.
After these choices, $S_c$ has one internal physical field and
its differentiated response field, while $S_{u_c}$ has two
internal physical fields. Either of the latter can pair with
the response field; the remaining pair forms the correlation
line. These two pairings give $2\times3\times2=12$ contractions.
Each carries the coefficient $-\lambda_cu_c/12$ in
Eq.~\eqref{cub:bare-factor}, so their sum has coefficient
$12(-\lambda_cu_c/12)=-\lambda_cu_c$ before the loop integral.

To extract the linear momentum dependence, put $\phi(-P)$ at
$S_c$ and $\widetilde\phi(P),\phi(0)$ at $S_{u_c}$, as in
Eq.~\eqref{cub:projection-c}. Assign momentum $Q$ to the internal
response field at $S_c$. Its derivative gives $\ii q_{\longdir}$, and
its contraction with a physical field at $S_{u_c}$ gives $A(Q)$.
Momentum conservation at $S_c$ leaves $P-Q$ on its internal
physical field; the correlation line therefore gives
$C(P-Q)=C(Q-P)$. The loop kernel is $\ii q_{\longdir}A(Q)C(Q-P)$.
Expanding its correlation line at fixed shell momentum gives
\begin{equation}
 C(Q-P)=C(Q)-p_{\longdir}\partial_{q_{\longdir}}C(Q)+\order(p_{\longdir}^2).
 \label{cub:expansion-example}
\end{equation}
The term $\ii q_{\longdir}AC$ integrates to zero because it is odd in
$q_{\longdir}$. The linear term is $-\ii p_{\longdir}q_{\longdir}A\partial_{q_{\longdir}}C$, whose
integral is $-\ii p_{\longdir}L_C/2$ by
Eq.~\eqref{cub:oriented-integrals}. Multiplication by the
coefficient $-\lambda_cu_c$ gives $+\ii p_{\longdir}\lambda_cu_cL_C/2$.
To write this as a local action term, the derivative on the
external $\phi(-P)$ supplies $-\ii p_{\longdir}$. Its operator coefficient
is consequently $-\lambda_cu_cL_C/2$:
\begin{align}
 \delta S_{\mathrm{c1b}}
 &=-\frac{\lambda_cu_cL_C}{2}
       \int_X\widetilde\phi\phi\partial_{\longdir}\phi\nonumber\\
 &=\frac{\lambda_cu_cL_C}{4}
       \int_X(\partial_{\longdir}\widetilde\phi)\phi^2,
 \label{cub:c1b-action}
\end{align}
where the second line uses Eq.~\eqref{cub:parts-c}.
Matching Eq.~\eqref{cub:c1b-action} to
Eq.~\eqref{con:cubic-projections} gives
$\delta\lambda_c|_{\mathrm{c1b}}=\lambda_cu_cL_C/2$.
Thus the two derivative placements of the same vertex pair combine
into $-\lambda_cu_c(L_0-L_C)/2$.

\emph{Diagrams c2 and c3.}
For c2, $S_c$ again supplies the external differentiated response
field. Its two physical fields now pair with the two response
fields at $S_{\Gamma_2}$, leaving two external physical fields at
that quartic vertex. The two pairings give parallel response lines
and hence $I_3$ in Eq.~\eqref{cub:shared-integrals}.
The factor $-(\lambda_c/2)(-\Gamma_2/4)\times2
=\lambda_c\Gamma_2/4$ therefore gives
\begin{equation}
 \delta S_{\mathrm{c2}}=\frac{\lambda_c\Gamma_2I_3}{4}
 \int_X(\partial_{\longdir}\widetilde\phi)\phi^2.
 \label{cub:c2-action}
\end{equation}
Equation~\eqref{con:cubic-projections} then gives
$\delta\lambda_c|_{\mathrm{c2}}=\lambda_c\Gamma_2I_3/2$.

For c3, the differentiated physical field at $S_q$ is external.
The two response fields at this vertex pair with two of the three
physical fields at $S_{u_c}$, again giving $I_3$.
There are three choices of the external physical field at $S_{u_c}$
and two pairings of the remaining fields, for a count of six.
The factor $-(\lambda_q/2)(u_c/6)\times6=-\lambda_qu_c/2$
then produces
\begin{align}
 \delta S_{\mathrm{c3}}
 &=-\frac{\lambda_qu_cI_3}{2}
 \int_X\widetilde\phi\phi\partial_{\longdir}\phi\nonumber\\
 &=\frac{\lambda_qu_cI_3}{4}
 \int_X(\partial_{\longdir}\widetilde\phi)\phi^2.
 \label{cub:c3-action}
\end{align}
The integration by parts in Eq.~\eqref{cub:parts-c} fixes the
positive sign of this correction:
$\delta\lambda_c|_{\mathrm{c3}}=\lambda_qu_cI_3/2$. Once again, all these corrections are summarized in Table~\ref{cub:table}. 

\subsection{Corrections to quantum convection}
\label{cub:quantum-bubbles}

We now extract the coefficient of
$(\partial_{\longdir}\phi)\widetilde\phi^2/2$ in
Eq.~\eqref{con:cubic-projections}. Figure~\ref{fig:app-cubic}(b)
contains the six contributions. The same vertex pair can correct
both convective couplings because different choices of external
fields leave different internal propagators.

\emph{Diagram q1.}
The pair $S_cS_{u_c}$ that produced c1a and c1b also gives q1.
Here each vertex leaves its response field external, while
$S_{u_c}$ also leaves one physical field external. The four
remaining physical fields form two correlation lines, giving
$I_4$ in Eq.~\eqref{cub:shared-integrals}.
There are three choices for the external physical field and two
pairings between the vertices. Each of these six contractions has
the same kernel $I_4$ and the coefficient $-\lambda_cu_c/12$
from Eq.~\eqref{cub:bare-factor}. Their sum therefore has
coefficient $6(-\lambda_cu_c/12)=-\lambda_cu_c/2$, giving
\begin{align}
 \delta S_{\mathrm{q1}}
 &=-\frac{\lambda_cu_cI_4}{2}
       \int_X\phi\widetilde\phi\partial_{\longdir}\widetilde\phi\nonumber\\
 &=\frac{\lambda_cu_cI_4}{4}
       \int_X(\partial_{\longdir}\phi)\widetilde\phi^2.
 \label{cub:q1-action}
\end{align}
The second line uses Eq.~\eqref{cub:parts} to put the derivative
on the physical field. Matching to Eq.~\eqref{con:cubic-projections}
gives $\delta\lambda_q|_{\mathrm{q1}}=\lambda_cu_cI_4/2$,
which is the source discussed in Eq.~\eqref{eq:technical-lambdaq-source}.

\emph{Diagrams q2a and q2b.}
These diagrams come from $-\langle S_cS_{\Gamma_2}\rangle_{>,c}$,
whose action prefactor is
$-(\lambda_c/2)(-\Gamma_2/4)=\lambda_c\Gamma_2/8$.
For q2a, $S_c$ supplies the external differentiated response field,
while $S_{\Gamma_2}$ supplies one external field of each type.
The internal lines are one $C$ and one response. The count is
$2\times2\times2=8$: choose the external physical field at
$S_{\Gamma_2}$, choose its external response field, and choose the
correlation-line partner at $S_c$. Combining this count with
$\int CR=L_0/2$ from Eq.~\eqref{cub:oriented-integrals} gives
\begin{align}
 \delta S_{\mathrm{q2a}}
 &=\frac{\lambda_c\Gamma_2L_0}{2}
 \int_X\phi\widetilde\phi\partial_{\longdir}\widetilde\phi\nonumber\\
 &=-\frac{\lambda_c\Gamma_2L_0}{4}
 \int_X(\partial_{\longdir}\phi)\widetilde\phi^2.
 \label{cub:q2a-action}
\end{align}
Using Eq.~\eqref{cub:parts} before the operator matching thus gives
$\delta\lambda_q|_{\mathrm{q2a}}=-\lambda_c\Gamma_2L_0/2$.

For q2b, the external field at $S_c$ is instead $\phi(P)$, and
both external response fields belong to $S_{\Gamma_2}$.
There are two choices of the external physical field at $S_c$
and two pairings of its internal fields with the two physical
fields at $S_{\Gamma_2}$, giving four contractions.
The derivative at $S_c$ is internal; the routing is
$\ii q_{\longdir}A(Q)C(Q+P)$. Expanding the correlation line as in
Eq.~\eqref{cub:expansion-example}, now with $+P$, leaves
$+\ii p_{\longdir}L_C/2$ after Eq.~\eqref{cub:oriented-integrals}.
The product of the prefactor, four contractions, and this integral
gives
\begin{equation}
 \delta S_{\mathrm{q2b}}=\frac{\lambda_c\Gamma_2L_C}{4}
 \int_X(\partial_{\longdir}\phi)\widetilde\phi^2.
 \label{cub:q2b-action}
\end{equation}
Therefore $\delta\lambda_q|_{\mathrm{q2b}}=\lambda_c\Gamma_2L_C/2$.
Together, Eqs.~\eqref{cub:q2a-action} and~\eqref{cub:q2b-action}
give the combination $\lambda_c\Gamma_2(L_C-L_0)/2$.

\emph{Diagram q4.}
This diagram comes from $-\langle S_qS_{u_c}\rangle_{>,c}$.
The differentiated physical field at $S_q$ is internal and
contracts with a physical field at $S_{u_c}$, giving $C$.
The other internal pair is a response line. Choose which of the
two response fields at $S_q$ is external, which of the three
physical fields at $S_{u_c}$ is external, and which of its two
remaining physical fields belongs to $C$. The count is
$2\times3\times2=12$, multiplying
$-(\lambda_q/2)(u_c/6)=-\lambda_qu_c/12$.

Put $\widetilde\phi(-P)$ at $S_q$ and
$\phi(P),\widetilde\phi(0)$ at $S_{u_c}$, as in
Eq.~\eqref{cub:projection-q}. With $Q$ on the correlation line,
the kernel is $\ii q_{\longdir}C(Q)R(Q-P)$. Its linear term is
$-\ii p_{\longdir}q_{\longdir}C\partial_{q_{\longdir}}R$, whose integral is
$-\ii p_{\longdir}L_B/2$ by Eq.~\eqref{cub:oriented-integrals}.
Multiplying by the contraction coefficient $-\lambda_qu_c$ gives
$+\ii p_{\longdir}\lambda_qu_cL_B/2$. The external response field at
$S_q$ carries $-P$, so its derivative supplies $-\ii p_{\longdir}$.
The corresponding local action term is therefore
\begin{align}
 \delta S_{\mathrm{q4}}
 &=-\frac{\lambda_qu_cL_B}{2}\int_X
 \phi\widetilde\phi\partial_{\longdir}\widetilde\phi\nonumber\\
 &=\frac{\lambda_qu_cL_B}{4}\int_X
 (\partial_{\longdir}\phi)\widetilde\phi^2.
 \label{cub:q4-action}
\end{align}
As in q1, Eq.~\eqref{cub:parts} moves the derivative onto the
physical field. The resulting contribution is
$\delta\lambda_q|_{\mathrm{q4}}=\lambda_qu_cL_B/2$.

\emph{Diagrams q3 and q5.}
The remaining two diagrams use $I_3$ in Eq.~\eqref{cub:shared-integrals}.
For q3, $S_c$ supplies the external differentiated response field.
Its two physical fields pair with two of the three response fields
at $S_{u_q}$. There are three choices of the external response
field at $S_{u_q}$ and two internal pairings, giving six
contractions and two parallel response lines. Their coefficient is
$-(\lambda_c/2)(u_q/6)\times6=-\lambda_cu_q/2$, so
\begin{align}
 \delta S_{\mathrm{q3}}
 &=-\frac{\lambda_cu_qI_3}{2}
 \int_X\phi\widetilde\phi\partial_{\longdir}\widetilde\phi\nonumber\\
 &=\frac{\lambda_cu_qI_3}{4}
 \int_X(\partial_{\longdir}\phi)\widetilde\phi^2.
 \label{cub:q3-action}
\end{align}
The same integration by parts as in Eq.~\eqref{cub:q1-action}
therefore gives $\delta\lambda_q|_{\mathrm{q3}}=\lambda_cu_qI_3/2$.
For q5, $S_q$ supplies the external differentiated physical field.
Its two response fields pair with the two physical fields at
$S_{\Gamma_2}$, leaving both response fields of that quartic vertex
external. The two pairings give
$-(\lambda_q/2)(-\Gamma_2/4)\times2=\lambda_q\Gamma_2/4$
and hence
\begin{equation}
 \delta S_{\mathrm{q5}}=\frac{\lambda_q\Gamma_2I_3}{4}
 \int_X(\partial_{\longdir}\phi)\widetilde\phi^2.
 \label{cub:q5-action}
\end{equation}
This operator already has the required derivative placement, so
$\delta\lambda_q|_{\mathrm{q5}}=\lambda_q\Gamma_2I_3/2$.

Table~\ref{cub:table} collects the ten nonzero corrections derived
above, with the same identifiers as Fig.~\ref{fig:app-cubic}.

\begin{table}[!t]
\centering
 \caption{Cubic--quartic bubbles in Fig.~\ref{fig:app-cubic}.
 Rows c correct $\lambda_c$ and rows q correct $\lambda_q$.
 The Wick combinatorial factor (second column) multiplies the cumulant and action coefficients}
 \label{cub:table}
 \begin{ruledtabular}
 \begin{tabular}{lrl}
 Diagram & $N$ & Coupling correction\\
 c1a & 6 & $-\frac12\lambda_cu_cL_0$\\
 c1b & 12 & $\frac12\lambda_cu_cL_C$\\
 c2 & 2 & $\frac12\lambda_c\Gamma_2I_3$\\
 c3 & 6 & $\frac12\lambda_qu_cI_3$\\[2pt]
 q1 & 6 & $\frac12\lambda_cu_cI_4$\\
 q2a & 8 & $-\frac12\lambda_c\Gamma_2L_0$\\
 q2b & 4 & $\frac12\lambda_c\Gamma_2L_C$\\
 q3 & 6 & $\frac12\lambda_cu_qI_3$\\
 q4 & 12 & $\frac12\lambda_qu_cL_B$\\
 q5 & 2 & $\frac12\lambda_q\Gamma_2I_3$
 \end{tabular}
 \end{ruledtabular}
\end{table}

Summing the c rows and q rows of Table~\ref{cub:table}, respectively,
gives the dimensionful shell increments
\begin{align}
 \delta\lambda_c={}&-\frac{\lambda_cu_c}{2}(L_0-L_C)
 +\frac{\lambda_c\Gamma_2+\lambda_qu_c}{2}I_3,
 \label{cub:counterterm-c}\\
 \delta\lambda_q={}&\frac{\lambda_cu_c}{2}I_4
 +\frac{\lambda_c\Gamma_2}{2}(L_C-L_0)\nonumber\\
 &+\frac{\lambda_cu_q+\lambda_q\Gamma_2}{2}I_3
 +\frac{\lambda_qu_c}{2}L_B.
 \label{cub:counterterm-q}
\end{align}
\subsection{Frequency and angular integration}
\label{cub:integrations}

The shell increments in Eqs.~\eqref{cub:counterterm-c}
and~\eqref{cub:counterterm-q} use $I_3,I_4,L_0$, already evaluated
in Eq.~\eqref{kin:shell-values}. To complete those increments,
we now evaluate $L_C$ and $L_B$ in
Eq.~\eqref{cub:derivative-integrals}. The frequency and angular
integrals follow the steps of Appendix~\ref{kin:integrations}.
Using $C=2D/\Delta$ and $B=2E/\Delta$, with
$\Delta=E^2+\gamma^2\omega^2$, $E=E_{\bm q}$ and $D=D_{\bm q}$,
we differentiate the two integrands in Eq.~\eqref{cub:derivative-integrals}:
\begin{align}
 q_{\longdir}B\partial_{q_{\longdir}}C
 &=\frac{8\Gamma_1Eq_{\longdir}^2}{\Delta^2}
   -\frac{16KDE^2q_{\longdir}^2}{\Delta^3},\nonumber\\
 q_{\longdir}C\partial_{q_{\longdir}}B
 &=\frac{8KDq_{\longdir}^2}{\Delta^2}
   -\frac{16KDE^2q_{\longdir}^2}{\Delta^3}.
 \label{cub:expanded-derivatives}
\end{align}
Here $\partial_{q_{\longdir}}E=2Kq_{\longdir}$ and
$\partial_{q_{\longdir}}D=2\Gamma_1q_{\longdir}$, from Eq.~\eqref{eq:technical-kernels}.
The squared denominator in Eq.~\eqref{cub:expanded-derivatives}
is integrated with Eq.~\eqref{kin:frequency-integrals}.
Differentiating that frequency integral once more gives the cubed denominator:
\begin{equation}
 \int\frac{\dd\omega}{2\pi}\Delta^{-3}
 =-\frac{1}{4E}\partial_E\frac{1}{4\gamma E^3}
 =\frac{3}{16\gamma E^5}.
 \label{cub:frequency-third}
\end{equation}
Applying Eqs.~\eqref{kin:frequency-integrals} and~\eqref{cub:frequency-third}
to the two lines of Eq.~\eqref{cub:expanded-derivatives} yields
\begin{align}
 L_C&=\int_{>\bm q}\frac{q_{\longdir}^2}{\gamma}
 \left(\frac{2\Gamma_1}{E_{\bm q}^2}
       -\frac{3KD_{\bm q}}{E_{\bm q}^3}\right),\nonumber\\
 L_B&=-\int_{>\bm q}
       \frac{KD_{\bm q}q_{\longdir}^2}{\gamma E_{\bm q}^3}.
 \label{cub:frequency-reduction}
\end{align}
The two terms in $L_C$ arise from the numerator and denominator
derivatives of $C$.   Comparing the definitions in
Eqs.~\eqref{kin:loop-definitions}, \eqref{kin:L0-definition},
and~\eqref{cub:derivative-integrals} also gives
\begin{equation}
 J_K=L_0+L_B-L_C,\qquad J_\Gamma=L_0+L_C-L_B,
 \label{cub:kinetic-identities}
\end{equation}
linking the convective and kinetic calculations. Substituting
Eq.~\eqref{cub:angular-integrals} below into these identities
reproduces $J_K,J_\Gamma$ in Eq.~\eqref{kin:shell-values}.

To complete the spatial integrals in Eq.~\eqref{cub:frequency-reduction},
set $r=\Gamma_0=0$ and insert the kernels in
Eq.~\eqref{con:angular-kernels}. As in Appendix~\ref{kin:integrations},
the radial dependence is $q^{-2}$, leaving the following two angular
averages. With $x=\cos^2\theta$, the first follows from
$-\partial_bF$ in Eq.~\eqref{kin:angular-derivatives} at $a=1,b=c-1$:
\begin{equation}
 \left\langle\frac{x}{\mathcal E^2}\right\rangle_\theta
 =\frac{1}{2c^{3/2}}.
 \label{cub:first-angular-moment}
\end{equation}
For the second, expand $\mathcal D=1+(\kappa c-1)x$ and use
the last two moments in Eq.~\eqref{kin:third-denominator-moments}:
\begin{align}
 \left\langle\frac{\mathcal D x}{\mathcal E^3}\right\rangle_\theta
 &=\left\langle\frac{x}{\mathcal E^3}\right\rangle_\theta
 +(\kappa c-1)\left\langle\frac{x^2}{\mathcal E^3}\right\rangle_\theta
 \nonumber\\
 &=\frac{3\kappa+1}{8c^{3/2}}.
 \label{cub:second-angular-moment}
\end{align}
Substituting Eqs.~\eqref{cub:first-angular-moment} and~\eqref{cub:second-angular-moment}
into Eq.~\eqref{cub:frequency-reduction}, and using the radial
factor in Eq.~\eqref{kin:radial-shell}, gives $L_C$ and $L_B$ below.
We include $L_0$ from Eq.~\eqref{kin:shell-values} because all three
enter the vertex corrections together. With $\mathcal H_d$ defined
in Eq.~\eqref{kin:shell-factor},
\begin{align}
 L_0&=\mathcal H_d\frac{\Gamma_{1\perp}}{K_\perp^2}
                  \frac{\kappa+1}{2},\nonumber\\
 L_C&=-\mathcal H_d\frac{\Gamma_{1\perp}}{K_\perp^2}
                  \frac{\kappa+3}{8},\nonumber\\
 L_B&=-\mathcal H_d\frac{\Gamma_{1\perp}}{K_\perp^2}
                  \frac{3\kappa+1}{8}.
 \label{cub:angular-integrals}
\end{align}
For $L_C$ in Eq.~\eqref{cub:frequency-reduction}, the two angular
contributions combine as
$\kappa-3(3\kappa+1)/8=-(\kappa+3)/8$.
The combination $L_0-L_C$ entering
Eqs.~\eqref{cub:counterterm-c} and~\eqref{cub:counterterm-q}
therefore has angular coefficient
$(\kappa+1)/2+(\kappa+3)/8=(5\kappa+7)/8$.
The values of $I_3$ and $I_4$ are collected in
Eq.~\eqref{kin:shell-values}. These averages are evaluated at
$d=2$ for the leading $\epsilon$ expansion.

\subsection{Beta functions}
\label{cub:flows}
The increments in Eqs.~\eqref{cub:counterterm-c} and~\eqref{cub:counterterm-q}
correct the dimensionful couplings. Differentiating the definition
of $\bar\lambda_c$ in Eq.~\eqref{eq:rg-convective-variables} gives
\begin{align}
 \beta_{\bar\lambda_c}={}&\bar\lambda_c\left[
 \frac{\epsilon}{2}-\frac34\zeta_c
 -\frac12\frac{\dd\ln\kappa}{\dd\ell}\right]\nonumber\\
 &+\Lambda_\ell^{-\epsilon/2}
 \left(\frac{a_d\Gamma_{1\perp}}{\gamma K_\perp^3}\right)^{1/2}
 c^{-3/4}\kappa^{-1/2}
 \lim_{\delta\ell\to0}\frac{\delta\lambda_c}{\delta\ell}.
 \label{cub:flow-conversion}
\end{align}
The first line combines canonical scaling and kinetic anomalous dimensions; the
second converts the shell increment to dimensionless form.
The same product rule applies to $\bar\lambda_q$, using its
normalization in Eq.~\eqref{eq:rg-convective-variables}.
The transverse coefficients and $\gamma$ are fixed. 
Using the kinetic flows in Eq.~\eqref{eq:rg-kappa}, the common
bracket in Eq.~\eqref{cub:flow-conversion} becomes
\begin{equation}
 \frac{\epsilon}{2}
 -\frac{4\kappa^2+5\kappa+3}{64}\bar\lambda_c^2
 +\frac{5\kappa+1}{32}\bar\lambda_c\bar\lambda_q.
 \label{cub:kinetic-normalization}
\end{equation}

For the second line of Eq.~\eqref{cub:flow-conversion}, insert
Eq.~\eqref{cub:counterterm-c} with the integrals in
Eqs.~\eqref{kin:shell-values} and~\eqref{cub:angular-integrals}, and express the quartics through
Eq.~\eqref{eq:rg-quartic-variables}. For example,
$-\lambda_cu_c(L_0-L_C)/2$ becomes
$-(5\kappa+7)\bar\lambda_c\bar u_c/16$ in the beta function.
The other two terms give
$(\bar\lambda_c\bar\Gamma_2+\bar\lambda_q\bar u_c)/4$.
Together with Eq.~\eqref{cub:kinetic-normalization}, these
reproduce Eq.~\eqref{eq:rg-lambda-c}. Applying the same steps to
Eq.~\eqref{cub:counterterm-q} gives Eq.~\eqref{eq:rg-lambda-q}.
The cubic powers in these beta functions therefore come from
kinetic normalization (i.e. anomalous dimensions in the first line of Eq.~\eqref{cub:flow-conversion}), not from diagrams directly proportional to the corresponding cubic products of couplings. 

\section{Local quartic vertices}
\label{app:quartic}

\begin{figure*}[!t]
\centering
\includegraphics[width=\textwidth]{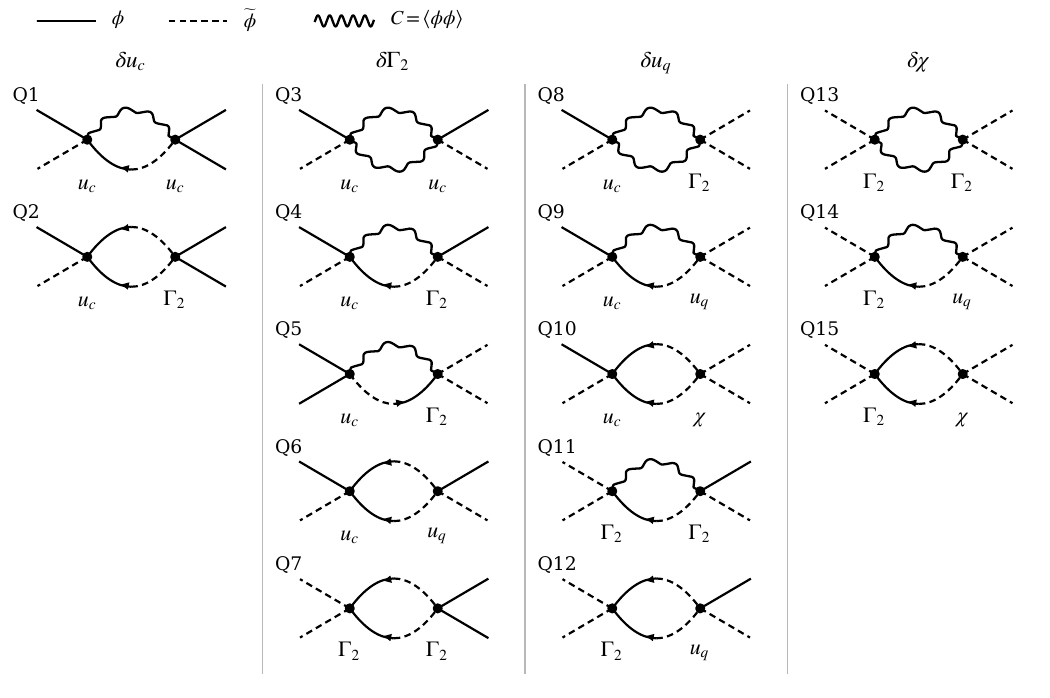}
\caption{Local quartic fish diagrams, grouped by the corrected
coupling. Vertex labels refer to Eq.~\eqref{eq:technical-S4}.
Physical external legs are solid and response external legs dashed;
wavy lines denote $C$, and mixed dashed--solid lines denote
responses, with arrows from response to physical endpoints.
Q1--Q2 contribute to Eq.~\eqref{qua:delta-uc}, Q3--Q7 to
Eq.~\eqref{qua:delta-Gamma}, Q8--Q12 to Eq.~\eqref{qua:delta-uq},
and Q13--Q15 to Eq.~\eqref{qua:delta-chi}.
}
\label{fig:app-quartic}
\end{figure*}

We now calculate the corrections to $u_c,\Gamma_2,u_q$, and
$\chi$ in Eq.~\eqref{eq:technical-S4}. These couplings multiply
operators without derivatives, so we set all external momenta
and frequencies to zero in loop corrections. The calculation follows the usual
response-functional contraction rules~\cite{tauber2014,marinoDiehl2016}.
Figure~\ref{fig:app-quartic} groups the surviving diagrams by the
coupling they correct; we use this grouping below to connect the
loop contributions to the beta functions in
Eqs.~\eqref{eq:rg-uc}--\eqref{eq:rg-chi}.

The second cumulant, the first term in $\delta S^{(4)}$ of
Eq.~\eqref{con:cumulant-sectors}, joins
two quartic vertices by two internal lines, leaving four external
fields:
\begin{equation}
 \delta S_{4,\mathrm{fish}}
 =-\frac12\langle S_4^2\rangle_{>,c}\big|_{\mathrm{local},4}.
 \label{qua:cumulant}
\end{equation}
The other quartic contributions in that equation contain convective
vertices. Their triangles vanish by the frequency identity in
Eq.~\eqref{can:total-derivative}, and their boxes by causality, as
shown in Appendices~\ref{can:quartic-triangles} and~\ref{can:quartic-boxes},
respectively. We extract the surviving coupling corrections by matching
Eq.~\eqref{qua:cumulant} to the normalization of
Eq.~\eqref{eq:technical-S4}:
\begin{align}
 \delta S_4=\int_X\bigg[
 &\frac{\delta u_c}{6}\phi^3\widetilde\phi
 -\frac{\delta\Gamma_2}{4}\phi^2\widetilde\phi^2\nonumber\\
 &+\frac{\delta u_q}{6}\phi\widetilde\phi^3
 +\frac{\delta\chi}{24}\widetilde\phi^4\bigg].
 \label{qua:coefficients}
\end{align}

The two internal momenta are $Q$ and $-Q$. Parallel response
arrows therefore give $R(Q)R(-Q)=R(Q)A(Q)$, whereas opposite
arrows give $R(Q)A(-Q)=R(Q)^2$ or its advanced counterpart.
The latter form directed response cycles and have vanishing
frequency integrals, by the argument in Appendix~\ref{con:shell-projections}~\cite{tauber2014}. Thus the surviving line
pairs reduce to the integrals in Eq.~\eqref{cub:shared-integrals}:
\begin{equation}
 \int_{>Q}RA=I_3,\qquad
 \int_{>Q}CR=\int_{>Q}CA=\frac{L_0}{2},\qquad
 \int_{>Q}C^2=I_4.
 \label{qua:integrals}
\end{equation}
The equality of the mixed integrals follows from
Eq.~\eqref{cub:oriented-integrals}. Their frequency and angular
evaluations are already given in Appendix~\ref{kin:integrations},
ending in Eq.~\eqref{kin:shell-values}.

The first column of Fig.~\ref{fig:app-quartic} contains the two
contributions to $u_c$. In Q1, one correlation line and one
response line join two $u_c$ vertices. For a fixed location of the internal response
field, its physical partner has three choices. The correlation
line then has three choices at the vertex carrying the internal
response field and two at the other vertex. Each response
location therefore gives $3\times3\times2=18$ contractions;
together they give $18CR+18CA$. With the vertex   in
Eq.~\eqref{qua:cumulant}, the coefficient of
$\phi^3\widetilde\phi$ is
$-\tfrac12(u_c/6)^2\,18L_0=-u_c^2L_0/4$.
Matching to Eq.~\eqref{qua:coefficients} gives
$-3u_c^2L_0/2$. Q2 instead has two parallel response lines between
$u_c$ and $\Gamma_2$, giving the second term, already quoted in Eq.~\eqref{eq:technical-quartic-feedback}:
\begin{equation}
 \delta u_c=-\frac32u_c^2L_0+\frac32u_c\Gamma_2I_3.
 \label{qua:delta-uc}
\end{equation}

For $\Gamma_2$, Q3 joins two $u_c$ vertices by correlation lines
and gives $u_c^2I_4$, the source of multiplicative noise in
Eq.~\eqref{eq:technical-multiplicative-source}.
Q4 and Q5 each have one correlation and one response line, but
leave different external fields at the two vertices. Their
contributions are $-2u_c\Gamma_2L_0$ and
$-u_c\Gamma_2L_0/2$. Finally, Q6 and Q7 contain two parallel
responses and contribute $2u_cu_qI_3$ and $\Gamma_2^2I_3/2$.
Their sum is
\begin{equation}
 \delta\Gamma_2=u_c^2I_4-\frac52u_c\Gamma_2L_0
                +2u_cu_qI_3+\frac12\Gamma_2^2I_3.
 \label{qua:delta-Gamma}
\end{equation}

Following a similar logic, we find
\begin{align}
 \delta u_q={}&\frac32u_c\Gamma_2I_4-\frac32u_cu_qL_0
              -\frac32u_c\chi I_3\nonumber\\
             &-\frac32\Gamma_2^2L_0+\frac32\Gamma_2u_qI_3,
 \label{qua:delta-uq}
\end{align}
and
\begin{equation}
 \delta\chi=-\frac32\Gamma_2^2I_4
            +3\Gamma_2u_qL_0+3\Gamma_2\chi I_3,
 \label{qua:delta-chi}
\end{equation}
which correspond to diagrams Q8-Q12 and Q13-Q15 in Fig.~\ref{fig:app-quartic}, respectively. Conversion of these results into beta functions follows the same procedure as for cubic vertices in
Appendix~\ref{cub:flows}.

\section{Vanishing one-loop triangle and box corrections}
\label{app:cancellations}

\begin{figure*}[!t]
\centering
 \begingroup
\setlength{\unitlength}{\textwidth}
\begin{picture}(1,0.507143)
 \put(0,0){\includegraphics[width=\textwidth]{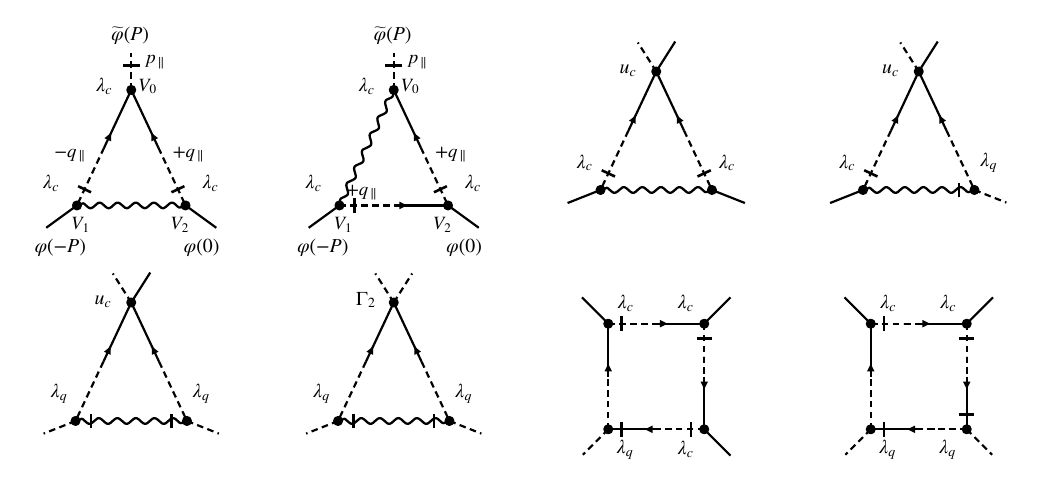}}
 \put(0.125,0.462857){\makebox(0,0){\small (a)}}
 \put(0.375,0.462857){\makebox(0,0){\small (b)}}
 \put(0.625,0.462857){\makebox(0,0){\small (c)}}
 \put(0.875,0.462857){\makebox(0,0){\small (d)}}
 \put(0.125,0.201429){\makebox(0,0){\small (e)}}
 \put(0.375,0.201429){\makebox(0,0){\small (f)}}
 \put(0.625,0.201429){\makebox(0,0){\small (g)}}
 \put(0.875,0.201429){\makebox(0,0){\small (h)}}
\end{picture}
\endgroup
\caption{(a),(b) Cubic triangles used in the   calculation;
their internal momentum labels are evaluated at $P=0$ after
extracting the external factor $p_{\longdir}$.
Examples of mixed cubic--quartic triangles that could in principle
correct (c) $u_c$, (d) $\Gamma_2$, (e) $u_q$, and (f) $\chi$.
Examples of cubic boxes that could correct (g) $u_c$ and (h) $\Gamma_2$.}
\label{fig:cancellations}
\end{figure*}

We complete the vertex calculation by evaluating the remaining
one-loop terms in Eq.~\eqref{con:cumulant-sectors}: the so-called cubic
triangles, built from three cubic vertices, which can renormalize
self-advection or its quantum counterpart; triangles containing
two cubic vertices and one quartic vertex, which can renormalize
quartic nonlinearities; and boxes built from four cubic vertices,
which can also contribute to quartic couplings.
Figure~\ref{fig:cancellations} shows the cubic contractions we
will evaluate explicitly, followed by examples of corrections
to the quartic operators in Eq.~\eqref{eq:technical-S4}. The fixed-point power counting associated with Eq.~\eqref{eq:technical-dimensions}
gives $\bar\lambda_{c,q}=\order(\sqrt\epsilon)$ and quartic
couplings of order $\epsilon$, \revision{as follows from the beta functions in Sec.~\ref{sec:rg-variables}}.%
 Cubic triangles are therefore
$\order(\epsilon^{3/2})$, like the cubic--quartic bubbles in
Appendix~\ref{app:cubic}. Mixed quartic triangles and cubic boxes
are $\order(\epsilon^2)$, like the fish diagrams in
Fig.~\ref{fig:app-quartic}. These topologies can thus contribute
at the retained order in their respective beta functions and must
be inspected individually.
 
\subsection{Cubic triangles}
\label{can:cubic-triangles}

The cubic triangles correcting self-advection  come from $\langle S_3^3\rangle_{>,c}/6$; associated diagrams are portrayed in Fig.~\ref{fig:cancellations}.

Consider Fig.~\ref{fig:cancellations}(a). The upper vertex
$V_0$ carries $\widetilde\phi(P)$, while the lower vertices
$V_1,V_2$ carry $\phi(-P),\phi(0)$, respectively.
The $S_c$ interaction in Eq.~\eqref{eq:technical-S3}
differentiates the response field. Its derivative is  
external at $V_0$ and internal at $V_1,V_2$, as marked in the
figure. After removing the external fields, $V_0$ has two
physical legs left; each lower vertex has one physical and
one response leg inside the loop. In panel (a), the two lower physical legs
form $C$, and the response legs connect to $V_0$.

We assign momentum $-Q$ to the response field at $V_1$ on
its connection to $V_0$. Momentum conservation fixes the
response field at $V_2$ to $Q+P$. The two response
contractions are then $R(Q)$ and $A(Q+P)$, and the lower
correlation line is $C(Q+P)$, by
Eq.~\eqref{eq:technical-propagators}. Using the Fourier
convention in Eq.~\eqref{con:fourier}, the three marked
derivatives in Fig.~\ref{fig:cancellations}(a) give the loop factor
\begin{align}
 &(\ii p_{\longdir})(-\ii q_{\longdir})\ii(q_{\longdir}+p_{\longdir})
 R(Q)A(Q+P)C(Q+P)\nonumber\\
 &\qquad=\ii p_{\longdir}q_{\longdir}^2C(Q)R(Q)A(Q)+\order(p_{\longdir}^2).
 \label{can:worked-triangle}
\end{align}
Here we have omitted the couplings and the Wick weight,
which are common to the three contractions. Since the
external derivative already supplies $p_{\longdir}$, shifts of the
internal factors contribute only at $\order(p_{\longdir}^2)$.
This is why the internal momentum labels in the figure,
and the second line of Eq.~\eqref{can:worked-triangle}, are
evaluated at $P=0$.

Now compare Fig.~\ref{fig:cancellations}(b) with panel (a).
Moving $C$ to $V_0V_1$ puts the differentiated response field
at $V_1$ on the lower edge. Its momentum changes from $-Q$
to $Q+P$, while the derivative at $V_2$ still carries $Q+P$.
The two responses now give $A(Q+P)^2$.
Thus, at the retained order, panel (b) contributes
$(\ii p_{\longdir})(\ii q_{\longdir})^2CA^2=-\ii p_{\longdir}q_{\longdir}^2CA^2$.
Reflecting panel (b) puts $C$ on $V_0V_2$ and interchanges
$R$ and $A$ at $P=0$, giving $-\ii p_{\longdir}q_{\longdir}^2CR^2$.
The three contractions have equal Wick weights: each has
the same external-field choices and permits the two internal
physical legs at $V_0$ to be interchanged. Adding panel (a),
panel (b), and its reflection therefore gives
\begin{equation}
 \ii p_{\longdir}q_{\longdir}^2C(Q)\bigl[R(Q)A(Q)-A(Q)^2-R(Q)^2\bigr].
 \label{can:routing-sum}
\end{equation}
The individual diagrams are nonzero; it is this sum that
vanishes after frequency integration. Removing $\ii p_{\longdir}$
as prescribed by Eq.~\eqref{cub:projection-c}, the
correction to $\lambda_c$ is proportional to
\begin{equation}
 I_\triangle=\int_{>Q}q_{\longdir}^2C(Q)
             [R(Q)A(Q)-R(Q)^2-A(Q)^2].
 \label{can:triangle-integral}
\end{equation}

To evaluate Eq.~\eqref{can:triangle-integral}, set
$\Delta=E_{\bm q}^2+\gamma^2\omega^2$ at fixed shell
momentum, where $E_{\bm q}>0$.
Substituting $C=2D_{\bm q}RA$ from
Eq.~\eqref{eq:technical-propagators} gives
\begin{align}
 C(RA-R^2-A^2)
 &=2D_{\bm q}\frac{-E_{\bm q}^2+3\gamma^2\omega^2}{\Delta^3}
 \nonumber\\
 &=-2D_{\bm q}\,\partial_\omega
       \left(\frac{\omega}{\Delta^2}\right).
 \label{can:total-derivative}
\end{align}
The noise kernel in Eq.~\eqref{eq:technical-kernels} is
frequency independent, and Eq.~\eqref{con:shell} integrates
over the full frequency axis. Since $\omega/\Delta^2$
vanishes at both limits, the frequency integral is zero at
each shell momentum, and hence $I_\triangle=0$.
Analogously one can show that triangle corrections to $\lambda_q$ vanish.

\subsection{Quartic triangles}
\label{can:quartic-triangles}

The quartic triangles in Eq.~\eqref{con:cumulant-sectors}
come from $\langle S_3^2S_4\rangle_{>,c}/2$.
All external momenta are zero under the local projection of
Eq.~\eqref{qua:coefficients}, so each   derivative from cubic terms 
must act internally.  

The quartic vertex now takes the place of $V_0$ in the cubic
calculation of Appendix~\ref{can:cubic-triangles}.
Panels (c)--(f) of Fig.~\ref{fig:cancellations} show possible
corrections to $u_c,\Gamma_2,u_q,\chi$, respectively, with
$C$ joining the two cubic vertices. Moving $C$ to the other
two edges gives the three response products derived in
Eq.~\eqref{can:routing-sum}. With two $S_c$ vertices, the internal
derivative factors  give
$q_{\longdir}^2C(RA-R^2-A^2)$. Each replacement of $S_c$ by $S_q$ moves its
derivative to the field of opposite momentum. Thus every allowed sector is proportional to
$+I_\triangle$ or $-I_\triangle$, and vanishes by
Eq.~\eqref{can:total-derivative}. We therefore conclude that the quartic corrections in
Eqs.~\eqref{qua:delta-uc}--\eqref{qua:delta-chi} receive no triangle contribution.

\subsection{Cubic boxes}
\label{can:quartic-boxes}

The box term $-\langle S_3^4\rangle_{>,c}/24$ in
Eq.~\eqref{con:cumulant-sectors} has one external leg at each
cubic vertex. In principle, it could renormalize a quartic coupling. At zero external momentum, the derivative at each vertex must again act internally, leaving one
internal physical field and one internal response field at every vertex.
The box therefore contains only response lines, with one
incoming and one outgoing arrow at each vertex.
Panels (g) and (h) of Fig.~\ref{fig:cancellations} illustrate
the possible corrections to $u_c$ and $\Gamma_2$.
Their time ordering would
require $t_1>t_2>t_3>t_4>t_1$, which is impossible. This is the
same causal exclusion used for two-line cycles in
Appendix~\ref{con:shell-projections}~\cite{tauber2014,kamenevLevchenko2009}.

\subsection{Dependence on the noise kernel}
\label{can:scope}

The triangle cancellation uses the frequency-independent
numerator of the Gaussian correlation function. To see its
role explicitly, keep $R,A$ unchanged and replace the noise
kernel by $D(\bm q,\omega)=D_{\bm q}+\eta|\omega|$, with $\eta>0$,
so that $C=2D(\bm q,\omega)/\Delta$. Such a frequency-dependent
noise contribution occurs at zero temperature for an Ohmic
bath~\cite{altlandSimons2023,kamenevLevchenko2009}. With $z=2$,
it scales as $q^2$, like the diffusive noise in
Eq.~\eqref{eq:technical-kernels}.
Using $\partial_\omega D=\eta\,\mathrm{sgn}\omega$, integration by parts in
Eq.~\eqref{can:total-derivative} now gives
\begin{align}
 \int\frac{\dd\omega}{2\pi}C(RA-R^2-A^2)
 &=2\int\frac{\dd\omega}{2\pi}
 (\partial_\omega D)\frac{\omega}{\Delta^2}\nonumber\\
 &=\frac{2\eta}{\pi}\int_0^\infty\frac{\omega\,\dd\omega}{\Delta^2}
 =\frac{\eta}{\pi\gamma^2E_{\bm q}^2}.
 \label{can:colored-noise}
\end{align}
The common triangle integral in Eq.~\eqref{can:triangle-integral}
would  then be nonzero. The boxes in Appendix~\ref{can:quartic-boxes}
remain zero: they contain no correlation line, so this change
of the noise kernel does not affect their causal argument.

These final remarks are meant to stress that the cancellations discussed in this Section are specific to the noise used in  our model and are not expected to hold in general.

\par
\raggedbottom
\makeatletter
\@booleantrue\raggedcolumn@sw
\makeatother
 
\manuscriptbibliography{references}
\color{black}\hypersetup{urlcolor=blue!55!black,linkcolor=blue!45!black,citecolor=blue!55!black}


\begin{thebibliography}{104}%
\makeatletter
\providecommand \@ifxundefined [1]{%
 \@ifx{#1\undefined}
}%
\providecommand \@ifnum [1]{%
 \ifnum #1\expandafter \@firstoftwo
 \else \expandafter \@secondoftwo
 \fi
}%
\providecommand \@ifx [1]{%
 \ifx #1\expandafter \@firstoftwo
 \else \expandafter \@secondoftwo
 \fi
}%
\providecommand \natexlab [1]{#1}%
\providecommand \enquote  [1]{``#1''}%
\providecommand \bibnamefont  [1]{#1}%
\providecommand \bibfnamefont [1]{#1}%
\providecommand \citenamefont [1]{#1}%
\providecommand \href@noop [0]{\@secondoftwo}%
\providecommand \href [0]{\begingroup \@sanitize@url \@href}%
\providecommand \@href[1]{\@@startlink{#1}\@@href}%
\providecommand \@@href[1]{\endgroup#1\@@endlink}%
\providecommand \@sanitize@url [0]{\catcode `\\12\catcode `\$12\catcode
  `\&12\catcode `\#12\catcode `\^12\catcode `\_12\catcode `\%12\relax}%
\providecommand \@@startlink[1]{}%
\providecommand \@@endlink[0]{}%
\providecommand \url  [0]{\begingroup\@sanitize@url \@url }%
\providecommand \@url [1]{\endgroup\@href {#1}{\urlprefix }}%
\providecommand \urlprefix  [0]{URL }%
\providecommand \Eprint [0]{\href }%
\providecommand \doibase [0]{http://dx.doi.org/}%
\providecommand \selectlanguage [0]{\@gobble}%
\providecommand \bibinfo  [0]{\@secondoftwo}%
\providecommand \bibfield  [0]{\@secondoftwo}%
\providecommand \translation [1]{[#1]}%
\providecommand \BibitemOpen [0]{}%
\providecommand \bibitemStop [0]{}%
\providecommand \bibitemNoStop [0]{.\EOS\space}%
\providecommand \EOS [0]{\spacefactor3000\relax}%
\providecommand \BibitemShut  [1]{\csname bibitem#1\endcsname}%
\let\auto@bib@innerbib\@empty
\bibitem [{\citenamefont {Marchetti}\ \emph
  {et~al.}(2013{\natexlab{a}})\citenamefont {Marchetti}, \citenamefont
  {Joanny}, \citenamefont {Ramaswamy}, \citenamefont {Liverpool}, \citenamefont
  {Prost}, \citenamefont {Rao},\ and\ \citenamefont {Simha}}]{marchetti2013}%
  \BibitemOpen
  \bibfield  {author} {\bibinfo {author} {\bibfnamefont {M.~Cristina}\
  \bibnamefont {Marchetti}}, \bibinfo {author} {\bibfnamefont
  {Jean-Fran{\c{c}}ois}\ \bibnamefont {Joanny}}, \bibinfo {author}
  {\bibfnamefont {Sriram}\ \bibnamefont {Ramaswamy}}, \bibinfo {author}
  {\bibfnamefont {T.~B.}\ \bibnamefont {Liverpool}}, \bibinfo {author}
  {\bibfnamefont {Jacques}\ \bibnamefont {Prost}}, \bibinfo {author}
  {\bibfnamefont {Madan}\ \bibnamefont {Rao}}, \ and\ \bibinfo {author}
  {\bibfnamefont {R.~Aditi}\ \bibnamefont {Simha}},\ }\bibfield  {title}
  {\enquote {\bibinfo {title} {Hydrodynamics of soft active matter},}\ }\href
  {\doibase 10.1103/RevModPhys.85.1143} {\bibfield  {journal} {\bibinfo
  {journal} {Reviews of Modern Physics}\ }\textbf {\bibinfo {volume} {85}},\
  \bibinfo {pages} {1143--1189} (\bibinfo {year}
  {2013}{\natexlab{a}})}\BibitemShut {NoStop}%
\bibitem [{\citenamefont {Khasseh}\ \emph {et~al.}(2025)\citenamefont
  {Khasseh}, \citenamefont {Wald}, \citenamefont {Moessner}, \citenamefont
  {Weber},\ and\ \citenamefont {Heyl}}]{khasseh2025}%
  \BibitemOpen
  \bibfield  {author} {\bibinfo {author} {\bibfnamefont {Reyhaneh}\
  \bibnamefont {Khasseh}}, \bibinfo {author} {\bibfnamefont {Sascha}\
  \bibnamefont {Wald}}, \bibinfo {author} {\bibfnamefont {Roderich}\
  \bibnamefont {Moessner}}, \bibinfo {author} {\bibfnamefont {Christoph~A.}\
  \bibnamefont {Weber}}, \ and\ \bibinfo {author} {\bibfnamefont {Markus}\
  \bibnamefont {Heyl}},\ }\bibfield  {title} {\enquote {\bibinfo {title}
  {Active quantum flocks},}\ }\href {\doibase 10.1103/rd46-hr3q} {\bibfield
  {journal} {\bibinfo  {journal} {Physical Review Letters}\ }\textbf {\bibinfo
  {volume} {135}},\ \bibinfo {pages} {248302} (\bibinfo {year}
  {2025})}\BibitemShut {NoStop}%
\bibitem [{\citenamefont {Radhakrishnan}\ \emph {et~al.}(2026)\citenamefont
  {Radhakrishnan}, \citenamefont {Khasseh}, \citenamefont {Schmidt},\ and\
  \citenamefont {Heyl}}]{radhakrishnan2026}%
  \BibitemOpen
  \bibfield  {author} {\bibinfo {author} {\bibfnamefont {Byjesh~N.}\
  \bibnamefont {Radhakrishnan}}, \bibinfo {author} {\bibfnamefont {Reyhaneh}\
  \bibnamefont {Khasseh}}, \bibinfo {author} {\bibfnamefont {Thomas}\
  \bibnamefont {Schmidt}}, \ and\ \bibinfo {author} {\bibfnamefont {Markus}\
  \bibnamefont {Heyl}},\ }\href@noop {} {\enquote {\bibinfo {title}
  {Superfluidity in active quantum flocks},}\ } (\bibinfo {year} {2026}),\
  \Eprint {http://arxiv.org/abs/2609.06515} {arXiv:2609.06515} \BibitemShut
  {NoStop}%
\bibitem [{\citenamefont {Adachi}\ \emph {et~al.}(2022)\citenamefont {Adachi},
  \citenamefont {Takasan},\ and\ \citenamefont {Kawaguchi}}]{adachi2022}%
  \BibitemOpen
  \bibfield  {author} {\bibinfo {author} {\bibfnamefont {Kyosuke}\ \bibnamefont
  {Adachi}}, \bibinfo {author} {\bibfnamefont {Kazuaki}\ \bibnamefont
  {Takasan}}, \ and\ \bibinfo {author} {\bibfnamefont {Kyogo}\ \bibnamefont
  {Kawaguchi}},\ }\bibfield  {title} {\enquote {\bibinfo {title}
  {Activity-induced phase transition in a quantum many-body system},}\ }\href
  {\doibase 10.1103/PhysRevResearch.4.013194} {\bibfield  {journal} {\bibinfo
  {journal} {Physical Review Research}\ }\textbf {\bibinfo {volume} {4}},\
  \bibinfo {pages} {013194} (\bibinfo {year} {2022})},\ \Eprint
  {http://arxiv.org/abs/2008.00996} {arXiv:2008.00996} \BibitemShut {NoStop}%
\bibitem [{\citenamefont {Takasan}\ \emph {et~al.}(2024)\citenamefont
  {Takasan}, \citenamefont {Adachi},\ and\ \citenamefont
  {Kawaguchi}}]{takasan2024}%
  \BibitemOpen
  \bibfield  {author} {\bibinfo {author} {\bibfnamefont {Kazuaki}\ \bibnamefont
  {Takasan}}, \bibinfo {author} {\bibfnamefont {Kyosuke}\ \bibnamefont
  {Adachi}}, \ and\ \bibinfo {author} {\bibfnamefont {Kyogo}\ \bibnamefont
  {Kawaguchi}},\ }\bibfield  {title} {\enquote {\bibinfo {title}
  {Activity-induced ferromagnetism in one-dimensional quantum many-body
  systems},}\ }\href {\doibase 10.1103/PhysRevResearch.6.023096} {\bibfield
  {journal} {\bibinfo  {journal} {Physical Review Research}\ }\textbf {\bibinfo
  {volume} {6}},\ \bibinfo {pages} {023096} (\bibinfo {year}
  {2024})}\BibitemShut {NoStop}%
\bibitem [{\citenamefont {Antonov}\ \emph {et~al.}(2025)\citenamefont
  {Antonov}, \citenamefont {Zheng}, \citenamefont {Liebchen},\ and\
  \citenamefont {L{\"o}wen}}]{zheng2025}%
  \BibitemOpen
  \bibfield  {author} {\bibinfo {author} {\bibfnamefont {Alexander~P.}\
  \bibnamefont {Antonov}}, \bibinfo {author} {\bibfnamefont {Yuanjian}\
  \bibnamefont {Zheng}}, \bibinfo {author} {\bibfnamefont {Benno}\ \bibnamefont
  {Liebchen}}, \ and\ \bibinfo {author} {\bibfnamefont {Hartmut}\ \bibnamefont
  {L{\"o}wen}},\ }\bibfield  {title} {\enquote {\bibinfo {title} {Engineering
  active motion in quantum matter},}\ }\href {\doibase 10.1103/z3gm-32jn}
  {\bibfield  {journal} {\bibinfo  {journal} {Physical Review Research}\
  }\textbf {\bibinfo {volume} {7}},\ \bibinfo {pages} {033008} (\bibinfo {year}
  {2025})},\ \Eprint {http://arxiv.org/abs/2305.16131} {arXiv:2305.16131}
  \BibitemShut {NoStop}%
\bibitem [{\citenamefont {Penner}\ \emph {et~al.}(2025)\citenamefont {Penner},
  \citenamefont {Viotti}, \citenamefont {Fazio}, \citenamefont {Arrachea},\
  and\ \citenamefont {von Oppen}}]{penner2025}%
  \BibitemOpen
  \bibfield  {author} {\bibinfo {author} {\bibfnamefont {Alexander-Georg}\
  \bibnamefont {Penner}}, \bibinfo {author} {\bibfnamefont {Ludmila}\
  \bibnamefont {Viotti}}, \bibinfo {author} {\bibfnamefont {Rosario}\
  \bibnamefont {Fazio}}, \bibinfo {author} {\bibfnamefont {Liliana}\
  \bibnamefont {Arrachea}}, \ and\ \bibinfo {author} {\bibfnamefont {Felix}\
  \bibnamefont {von Oppen}},\ }\bibfield  {title} {\enquote {\bibinfo {title}
  {Heat-to-motion conversion for quantum active matter},}\ }\href {\doibase
  10.1103/r6tm-nx19} {\bibfield  {journal} {\bibinfo  {journal} {Physical
  Review B}\ }\textbf {\bibinfo {volume} {112}},\ \bibinfo {pages} {L180303}
  (\bibinfo {year} {2025})},\ \Eprint {http://arxiv.org/abs/2503.07751}
  {arXiv:2503.07751} \BibitemShut {NoStop}%
\bibitem [{\citenamefont {Hardt}\ \emph {et~al.}(2025)\citenamefont {Hardt},
  \citenamefont {Doostani}, \citenamefont {Diehl}, \citenamefont {Del~Ser},\
  and\ \citenamefont {Rosch}}]{hardt2025propelling}%
  \BibitemOpen
  \bibfield  {author} {\bibinfo {author} {\bibfnamefont {Dennis}\ \bibnamefont
  {Hardt}}, \bibinfo {author} {\bibfnamefont {Reza}\ \bibnamefont {Doostani}},
  \bibinfo {author} {\bibfnamefont {Sebastian}\ \bibnamefont {Diehl}}, \bibinfo
  {author} {\bibfnamefont {Nina}\ \bibnamefont {Del~Ser}}, \ and\ \bibinfo
  {author} {\bibfnamefont {Achim}\ \bibnamefont {Rosch}},\ }\bibfield  {title}
  {\enquote {\bibinfo {title} {Propelling ferrimagnetic domain walls by
  dynamical frustration},}\ }\href@noop {} {\bibfield  {journal} {\bibinfo
  {journal} {Nature Communications}\ }\textbf {\bibinfo {volume} {16}},\
  \bibinfo {pages} {3817} (\bibinfo {year} {2025})}\BibitemShut {NoStop}%
\bibitem [{\citenamefont {Steiner}\ \emph {et~al.}(2026)\citenamefont
  {Steiner}, \citenamefont {von Oppen},\ and\ \citenamefont
  {Egger}}]{steiner2026}%
  \BibitemOpen
  \bibfield  {author} {\bibinfo {author} {\bibfnamefont {Jacob~F.}\
  \bibnamefont {Steiner}}, \bibinfo {author} {\bibfnamefont {Felix}\
  \bibnamefont {von Oppen}}, \ and\ \bibinfo {author} {\bibfnamefont
  {Reinhold}\ \bibnamefont {Egger}},\ }\bibfield  {title} {\enquote {\bibinfo
  {title} {Active quantum matter from monitored pure-state dynamics},}\ }\href
  {\doibase 10.1103/zhdk-kgzx} {\bibfield  {journal} {\bibinfo  {journal}
  {Physical Review Letters}\ }\textbf {\bibinfo {volume} {137}},\ \bibinfo
  {pages} {120403} (\bibinfo {year} {2026})},\ \Eprint
  {http://arxiv.org/abs/2603.12629} {arXiv:2603.12629 [quant-ph]} \BibitemShut
  {NoStop}%
\bibitem [{\citenamefont {Metelmann}\ and\ \citenamefont
  {Clerk}(2015)}]{metelmannClerk2015}%
  \BibitemOpen
  \bibfield  {author} {\bibinfo {author} {\bibfnamefont {A.}~\bibnamefont
  {Metelmann}}\ and\ \bibinfo {author} {\bibfnamefont {A.~A.}\ \bibnamefont
  {Clerk}},\ }\bibfield  {title} {\enquote {\bibinfo {title} {Nonreciprocal
  photon transmission and amplification via reservoir engineering},}\ }\href
  {\doibase 10.1103/PhysRevX.5.021025} {\bibfield  {journal} {\bibinfo
  {journal} {Physical Review X}\ }\textbf {\bibinfo {volume} {5}},\ \bibinfo
  {pages} {021025} (\bibinfo {year} {2015})},\ \Eprint
  {http://arxiv.org/abs/1502.07274} {arXiv:1502.07274 [quant-ph]} \BibitemShut
  {NoStop}%
\bibitem [{\citenamefont {Belyansky}\ \emph {et~al.}(2025)\citenamefont
  {Belyansky}, \citenamefont {Weis}, \citenamefont {Hanai}, \citenamefont
  {Littlewood},\ and\ \citenamefont {Clerk}}]{belyansky2025}%
  \BibitemOpen
  \bibfield  {author} {\bibinfo {author} {\bibfnamefont {Ron}\ \bibnamefont
  {Belyansky}}, \bibinfo {author} {\bibfnamefont {Cheyne}\ \bibnamefont
  {Weis}}, \bibinfo {author} {\bibfnamefont {Ryo}\ \bibnamefont {Hanai}},
  \bibinfo {author} {\bibfnamefont {Peter~B.}\ \bibnamefont {Littlewood}}, \
  and\ \bibinfo {author} {\bibfnamefont {Aashish~A.}\ \bibnamefont {Clerk}},\
  }\bibfield  {title} {\enquote {\bibinfo {title} {Phase transitions in
  nonreciprocal driven-dissipative condensates},}\ }\href {\doibase
  10.1103/gphr-d1bc} {\bibfield  {journal} {\bibinfo  {journal} {Physical
  Review Letters}\ }\textbf {\bibinfo {volume} {135}},\ \bibinfo {pages}
  {123401} (\bibinfo {year} {2025})},\ \Eprint
  {http://arxiv.org/abs/2502.05267} {arXiv:2502.05267 [quant-ph]} \BibitemShut
  {NoStop}%
\bibitem [{\citenamefont {Nadolny}\ \emph {et~al.}(2025)\citenamefont
  {Nadolny}, \citenamefont {Bruder},\ and\ \citenamefont
  {Brunelli}}]{nadolny2025}%
  \BibitemOpen
  \bibfield  {author} {\bibinfo {author} {\bibfnamefont {Tobias}\ \bibnamefont
  {Nadolny}}, \bibinfo {author} {\bibfnamefont {Christoph}\ \bibnamefont
  {Bruder}}, \ and\ \bibinfo {author} {\bibfnamefont {Matteo}\ \bibnamefont
  {Brunelli}},\ }\bibfield  {title} {\enquote {\bibinfo {title} {Nonreciprocal
  synchronization of active quantum spins},}\ }\href {\doibase
  10.1103/PhysRevX.15.011010} {\bibfield  {journal} {\bibinfo  {journal}
  {Physical Review X}\ }\textbf {\bibinfo {volume} {15}},\ \bibinfo {pages}
  {011010} (\bibinfo {year} {2025})},\ \Eprint
  {http://arxiv.org/abs/2406.03357} {arXiv:2406.03357} \BibitemShut {NoStop}%
\bibitem [{\citenamefont {Hanai}\ \emph {et~al.}(2025)\citenamefont {Hanai},
  \citenamefont {Ootsuki},\ and\ \citenamefont
  {Tazai}}]{hanai2025photoinduced}%
  \BibitemOpen
  \bibfield  {author} {\bibinfo {author} {\bibfnamefont {Ryo}\ \bibnamefont
  {Hanai}}, \bibinfo {author} {\bibfnamefont {Daiki}\ \bibnamefont {Ootsuki}},
  \ and\ \bibinfo {author} {\bibfnamefont {Rina}\ \bibnamefont {Tazai}},\
  }\bibfield  {title} {\enquote {\bibinfo {title} {Photoinduced non-reciprocal
  magnetism},}\ }\href@noop {} {\bibfield  {journal} {\bibinfo  {journal}
  {Nature communications}\ }\textbf {\bibinfo {volume} {16}},\ \bibinfo {pages}
  {8195} (\bibinfo {year} {2025})}\BibitemShut {NoStop}%
\bibitem [{\citenamefont {Hohenberg}\ and\ \citenamefont
  {Halperin}(1977)}]{hohenbergHalperin1977}%
  \BibitemOpen
  \bibfield  {author} {\bibinfo {author} {\bibfnamefont {P.~C.}\ \bibnamefont
  {Hohenberg}}\ and\ \bibinfo {author} {\bibfnamefont {B.~I.}\ \bibnamefont
  {Halperin}},\ }\bibfield  {title} {\enquote {\bibinfo {title} {Theory of
  dynamic critical phenomena},}\ }\href {\doibase 10.1103/RevModPhys.49.435}
  {\bibfield  {journal} {\bibinfo  {journal} {Reviews of Modern Physics}\
  }\textbf {\bibinfo {volume} {49}},\ \bibinfo {pages} {435--479} (\bibinfo
  {year} {1977})}\BibitemShut {NoStop}%
\bibitem [{\citenamefont {T{\"a}uber}\ \emph {et~al.}(2002)\citenamefont
  {T{\"a}uber}, \citenamefont {Akkineni},\ and\ \citenamefont
  {Santos}}]{tauberAkkineniSantos2002}%
  \BibitemOpen
  \bibfield  {author} {\bibinfo {author} {\bibfnamefont {Uwe~C.}\ \bibnamefont
  {T{\"a}uber}}, \bibinfo {author} {\bibfnamefont {Vamsi~K.}\ \bibnamefont
  {Akkineni}}, \ and\ \bibinfo {author} {\bibfnamefont {Jaimie~E.}\
  \bibnamefont {Santos}},\ }\bibfield  {title} {\enquote {\bibinfo {title}
  {Effects of violating detailed balance on critical dynamics},}\ }\href
  {\doibase 10.1103/PhysRevLett.88.045702} {\bibfield  {journal} {\bibinfo
  {journal} {Physical Review Letters}\ }\textbf {\bibinfo {volume} {88}},\
  \bibinfo {pages} {045702} (\bibinfo {year} {2002})},\ \Eprint
  {http://arxiv.org/abs/cond-mat/0109433} {arXiv:cond-mat/0109433} \BibitemShut
  {NoStop}%
\bibitem [{\citenamefont {Sieberer}\ \emph {et~al.}(2015)\citenamefont
  {Sieberer}, \citenamefont {Chiocchetta}, \citenamefont {Gambassi},
  \citenamefont {T{\"a}uber},\ and\ \citenamefont
  {Diehl}}]{sieberer2015equilibrium}%
  \BibitemOpen
  \bibfield  {author} {\bibinfo {author} {\bibfnamefont {L.~M.}\ \bibnamefont
  {Sieberer}}, \bibinfo {author} {\bibfnamefont {A.}~\bibnamefont
  {Chiocchetta}}, \bibinfo {author} {\bibfnamefont {A.}~\bibnamefont
  {Gambassi}}, \bibinfo {author} {\bibfnamefont {U.~C.}\ \bibnamefont
  {T{\"a}uber}}, \ and\ \bibinfo {author} {\bibfnamefont {S.}~\bibnamefont
  {Diehl}},\ }\bibfield  {title} {\enquote {\bibinfo {title} {Thermodynamic
  equilibrium as a symmetry of the {Schwinger--Keldysh} action},}\ }\href
  {\doibase 10.1103/PhysRevB.92.134307} {\bibfield  {journal} {\bibinfo
  {journal} {Physical Review B}\ }\textbf {\bibinfo {volume} {92}},\ \bibinfo
  {pages} {134307} (\bibinfo {year} {2015})},\ \Eprint
  {http://arxiv.org/abs/1505.00912} {arXiv:1505.00912} \BibitemShut {NoStop}%
\bibitem [{\citenamefont {T{\"a}uber}(2017)}]{tauber2017}%
  \BibitemOpen
  \bibfield  {author} {\bibinfo {author} {\bibfnamefont {Uwe~C.}\ \bibnamefont
  {T{\"a}uber}},\ }\bibfield  {title} {\enquote {\bibinfo {title} {Phase
  transitions and scaling in systems far from equilibrium},}\ }\href {\doibase
  10.1146/annurev-conmatphys-031016-025444} {\bibfield  {journal} {\bibinfo
  {journal} {Annual Review of Condensed Matter Physics}\ }\textbf {\bibinfo
  {volume} {8}},\ \bibinfo {pages} {185--210} (\bibinfo {year} {2017})},\
  \Eprint {http://arxiv.org/abs/1604.04487} {arXiv:1604.04487
  [cond-mat.stat-mech]} \BibitemShut {NoStop}%
\bibitem [{\citenamefont {Sieberer}\ \emph {et~al.}(2025)\citenamefont
  {Sieberer}, \citenamefont {Buchhold}, \citenamefont {Marino},\ and\
  \citenamefont {Diehl}}]{sieberer2025}%
  \BibitemOpen
  \bibfield  {author} {\bibinfo {author} {\bibfnamefont {Lukas~M.}\
  \bibnamefont {Sieberer}}, \bibinfo {author} {\bibfnamefont {Michael}\
  \bibnamefont {Buchhold}}, \bibinfo {author} {\bibfnamefont {Jamir}\
  \bibnamefont {Marino}}, \ and\ \bibinfo {author} {\bibfnamefont {Sebastian}\
  \bibnamefont {Diehl}},\ }\bibfield  {title} {\enquote {\bibinfo {title}
  {Universality in driven open quantum matter},}\ }\href {\doibase
  10.1103/RevModPhys.97.025004} {\bibfield  {journal} {\bibinfo  {journal}
  {Reviews of Modern Physics}\ }\textbf {\bibinfo {volume} {97}},\ \bibinfo
  {pages} {025004} (\bibinfo {year} {2025})},\ \Eprint
  {http://arxiv.org/abs/2312.03073} {arXiv:2312.03073 [cond-mat.stat-mech]}
  \BibitemShut {NoStop}%
\bibitem [{\citenamefont {Chat{\'e}}(2020)}]{chate2020}%
  \BibitemOpen
  \bibfield  {author} {\bibinfo {author} {\bibfnamefont {Hugues}\ \bibnamefont
  {Chat{\'e}}},\ }\bibfield  {title} {\enquote {\bibinfo {title} {Dry aligning
  dilute active matter},}\ }\href {\doibase
  10.1146/annurev-conmatphys-031119-050752} {\bibfield  {journal} {\bibinfo
  {journal} {Annual Review of Condensed Matter Physics}\ }\textbf {\bibinfo
  {volume} {11}},\ \bibinfo {pages} {189--212} (\bibinfo {year}
  {2020})}\BibitemShut {NoStop}%
\bibitem [{\citenamefont {Toner}\ and\ \citenamefont {Tu}(1995)}]{tonerTu1995}%
  \BibitemOpen
  \bibfield  {author} {\bibinfo {author} {\bibfnamefont {John}\ \bibnamefont
  {Toner}}\ and\ \bibinfo {author} {\bibfnamefont {Yuhai}\ \bibnamefont {Tu}},\
  }\bibfield  {title} {\enquote {\bibinfo {title} {Long-range order in a
  two-dimensional dynamical {XY} model: How birds fly together},}\ }\href
  {\doibase 10.1103/PhysRevLett.75.4326} {\bibfield  {journal} {\bibinfo
  {journal} {Physical Review Letters}\ }\textbf {\bibinfo {volume} {75}},\
  \bibinfo {pages} {4326--4329} (\bibinfo {year} {1995})}\BibitemShut {NoStop}%
\bibitem [{\citenamefont {Toner}\ and\ \citenamefont {Tu}(1998)}]{tonerTu1998}%
  \BibitemOpen
  \bibfield  {author} {\bibinfo {author} {\bibfnamefont {John}\ \bibnamefont
  {Toner}}\ and\ \bibinfo {author} {\bibfnamefont {Yuhai}\ \bibnamefont {Tu}},\
  }\bibfield  {title} {\enquote {\bibinfo {title} {Flocks, herds, and schools:
  A quantitative theory of flocking},}\ }\href {\doibase
  10.1103/PhysRevE.58.4828} {\bibfield  {journal} {\bibinfo  {journal}
  {Physical Review E}\ }\textbf {\bibinfo {volume} {58}},\ \bibinfo {pages}
  {4828--4858} (\bibinfo {year} {1998})}\BibitemShut {NoStop}%
\bibitem [{\citenamefont {Toner}\ \emph {et~al.}(2005)\citenamefont {Toner},
  \citenamefont {Tu},\ and\ \citenamefont {Ramaswamy}}]{tonerTuRamaswamy2005}%
  \BibitemOpen
  \bibfield  {author} {\bibinfo {author} {\bibfnamefont {John}\ \bibnamefont
  {Toner}}, \bibinfo {author} {\bibfnamefont {Yuhai}\ \bibnamefont {Tu}}, \
  and\ \bibinfo {author} {\bibfnamefont {Sriram}\ \bibnamefont {Ramaswamy}},\
  }\bibfield  {title} {\enquote {\bibinfo {title} {Hydrodynamics and phases of
  flocks},}\ }\href {\doibase 10.1016/j.aop.2005.04.011} {\bibfield  {journal}
  {\bibinfo  {journal} {Annals of Physics}\ }\textbf {\bibinfo {volume}
  {318}},\ \bibinfo {pages} {170--244} (\bibinfo {year} {2005})}\BibitemShut
  {NoStop}%
\bibitem [{\citenamefont {Marchetti}\ \emph
  {et~al.}(2013{\natexlab{b}})\citenamefont {Marchetti}, \citenamefont
  {Joanny}, \citenamefont {Ramaswamy}, \citenamefont {Liverpool}, \citenamefont
  {Prost}, \citenamefont {Rao},\ and\ \citenamefont
  {Simha}}]{RevModPhys.85.1143}%
  \BibitemOpen
  \bibfield  {author} {\bibinfo {author} {\bibfnamefont {M.~C.}\ \bibnamefont
  {Marchetti}}, \bibinfo {author} {\bibfnamefont {J.~F.}\ \bibnamefont
  {Joanny}}, \bibinfo {author} {\bibfnamefont {S.}~\bibnamefont {Ramaswamy}},
  \bibinfo {author} {\bibfnamefont {T.~B.}\ \bibnamefont {Liverpool}}, \bibinfo
  {author} {\bibfnamefont {J.}~\bibnamefont {Prost}}, \bibinfo {author}
  {\bibfnamefont {Madan}\ \bibnamefont {Rao}}, \ and\ \bibinfo {author}
  {\bibfnamefont {R.~Aditi}\ \bibnamefont {Simha}},\ }\bibfield  {title}
  {\enquote {\bibinfo {title} {Hydrodynamics of soft active matter},}\ }\href
  {\doibase 10.1103/RevModPhys.85.1143} {\bibfield  {journal} {\bibinfo
  {journal} {Rev. Mod. Phys.}\ }\textbf {\bibinfo {volume} {85}},\ \bibinfo
  {pages} {1143--1189} (\bibinfo {year} {2013}{\natexlab{b}})}\BibitemShut
  {NoStop}%
\bibitem [{\citenamefont {Ramaswamy}(2017)}]{ramaswamy2017active}%
  \BibitemOpen
  \bibfield  {author} {\bibinfo {author} {\bibfnamefont {Sriram}\ \bibnamefont
  {Ramaswamy}},\ }\bibfield  {title} {\enquote {\bibinfo {title} {Active
  matter},}\ }\href@noop {} {\bibfield  {journal} {\bibinfo  {journal} {Journal
  of Statistical Mechanics: Theory and Experiment}\ }\textbf {\bibinfo {volume}
  {2017}},\ \bibinfo {pages} {054002} (\bibinfo {year} {2017})}\BibitemShut
  {NoStop}%
\bibitem [{\citenamefont {Bowick}\ \emph {et~al.}(2022)\citenamefont {Bowick},
  \citenamefont {Fakhri}, \citenamefont {Marchetti},\ and\ \citenamefont
  {Ramaswamy}}]{bowick2022symmetry}%
  \BibitemOpen
  \bibfield  {author} {\bibinfo {author} {\bibfnamefont {Mark~J}\ \bibnamefont
  {Bowick}}, \bibinfo {author} {\bibfnamefont {Nikta}\ \bibnamefont {Fakhri}},
  \bibinfo {author} {\bibfnamefont {M~Cristina}\ \bibnamefont {Marchetti}}, \
  and\ \bibinfo {author} {\bibfnamefont {Sriram}\ \bibnamefont {Ramaswamy}},\
  }\bibfield  {title} {\enquote {\bibinfo {title} {Symmetry, thermodynamics,
  and topology in active matter},}\ }\href@noop {} {\bibfield  {journal}
  {\bibinfo  {journal} {Physical Review X}\ }\textbf {\bibinfo {volume} {12}},\
  \bibinfo {pages} {010501} (\bibinfo {year} {2022})}\BibitemShut {NoStop}%
\bibitem [{\citenamefont {Chen}\ \emph {et~al.}(2025)\citenamefont {Chen},
  \citenamefont {Jentsch}, \citenamefont {Lee}, \citenamefont {Maitra},
  \citenamefont {Ramaswamy},\ and\ \citenamefont
  {Toner}}]{chen2025inconvenient}%
  \BibitemOpen
  \bibfield  {author} {\bibinfo {author} {\bibfnamefont {Leiming}\ \bibnamefont
  {Chen}}, \bibinfo {author} {\bibfnamefont {Patrick}\ \bibnamefont {Jentsch}},
  \bibinfo {author} {\bibfnamefont {Chiu~Fan}\ \bibnamefont {Lee}}, \bibinfo
  {author} {\bibfnamefont {Ananyo}\ \bibnamefont {Maitra}}, \bibinfo {author}
  {\bibfnamefont {Sriram}\ \bibnamefont {Ramaswamy}}, \ and\ \bibinfo {author}
  {\bibfnamefont {John}\ \bibnamefont {Toner}},\ }\bibfield  {title} {\enquote
  {\bibinfo {title} {The inconvenient truth about flocks},}\ }\href@noop {}
  {\bibfield  {journal} {\bibinfo  {journal} {arXiv preprint arXiv:2503.17064}\
  } (\bibinfo {year} {2025})}\BibitemShut {NoStop}%
\bibitem [{\citenamefont {te~Vrugt}\ \emph {et~al.}(2025)\citenamefont
  {te~Vrugt}, \citenamefont {Liebchen},\ and\ \citenamefont
  {Cates}}]{teVrugtLiebchenCates2025}%
  \BibitemOpen
  \bibfield  {author} {\bibinfo {author} {\bibfnamefont {Michael}\ \bibnamefont
  {te~Vrugt}}, \bibinfo {author} {\bibfnamefont {Benno}\ \bibnamefont
  {Liebchen}}, \ and\ \bibinfo {author} {\bibfnamefont {Michael~E.}\
  \bibnamefont {Cates}},\ }\href {\doibase 10.48550/arXiv.2507.21621} {\enquote
  {\bibinfo {title} {What exactly is 'active matter'?}}\ } (\bibinfo {year}
  {2025}),\ \bibinfo {note} {arXiv:2507.21621},\ \Eprint
  {http://arxiv.org/abs/2507.21621} {arXiv:2507.21621 [cond-mat.soft]}
  \BibitemShut {NoStop}%
\bibitem [{\citenamefont {te~Vrugt}\ and\ \citenamefont
  {Wittkowski}(2025)}]{teVrugtWittkowski2025}%
  \BibitemOpen
  \bibfield  {author} {\bibinfo {author} {\bibfnamefont {Michael}\ \bibnamefont
  {te~Vrugt}}\ and\ \bibinfo {author} {\bibfnamefont {Raphael}\ \bibnamefont
  {Wittkowski}},\ }\bibfield  {title} {\enquote {\bibinfo {title} {Metareview:
  A survey of active matter reviews},}\ }\href {\doibase
  10.1140/epje/s10189-024-00466-z} {\bibfield  {journal} {\bibinfo  {journal}
  {The European Physical Journal E}\ }\textbf {\bibinfo {volume} {48}},\
  \bibinfo {pages} {12} (\bibinfo {year} {2025})}\BibitemShut {NoStop}%
\bibitem [{\citenamefont {Wittkowski}\ \emph {et~al.}(2014)\citenamefont
  {Wittkowski}, \citenamefont {Tiribocchi}, \citenamefont {Stenhammar},
  \citenamefont {Allen}, \citenamefont {Marenduzzo},\ and\ \citenamefont
  {Cates}}]{wittkowski2014}%
  \BibitemOpen
  \bibfield  {author} {\bibinfo {author} {\bibfnamefont {Raphael}\ \bibnamefont
  {Wittkowski}}, \bibinfo {author} {\bibfnamefont {Adriano}\ \bibnamefont
  {Tiribocchi}}, \bibinfo {author} {\bibfnamefont {Joakim}\ \bibnamefont
  {Stenhammar}}, \bibinfo {author} {\bibfnamefont {Rosalind~J.}\ \bibnamefont
  {Allen}}, \bibinfo {author} {\bibfnamefont {Davide}\ \bibnamefont
  {Marenduzzo}}, \ and\ \bibinfo {author} {\bibfnamefont {Michael~E.}\
  \bibnamefont {Cates}},\ }\bibfield  {title} {\enquote {\bibinfo {title}
  {Scalar {$\phi^4$} field theory for active-particle phase separation},}\
  }\href {\doibase 10.1038/ncomms5351} {\bibfield  {journal} {\bibinfo
  {journal} {Nature Communications}\ }\textbf {\bibinfo {volume} {5}},\
  \bibinfo {pages} {4351} (\bibinfo {year} {2014})}\BibitemShut {NoStop}%
\bibitem [{\citenamefont {Tjhung}\ \emph {et~al.}(2018)\citenamefont {Tjhung},
  \citenamefont {Nardini},\ and\ \citenamefont {Cates}}]{tjhung2018}%
  \BibitemOpen
  \bibfield  {author} {\bibinfo {author} {\bibfnamefont {Elsen}\ \bibnamefont
  {Tjhung}}, \bibinfo {author} {\bibfnamefont {Cesare}\ \bibnamefont
  {Nardini}}, \ and\ \bibinfo {author} {\bibfnamefont {Michael~E.}\
  \bibnamefont {Cates}},\ }\bibfield  {title} {\enquote {\bibinfo {title}
  {Cluster phases and bubbly phase separation in active fluids: Reversal of the
  ostwald process},}\ }\href {\doibase 10.1103/PhysRevX.8.031080} {\bibfield
  {journal} {\bibinfo  {journal} {Physical Review X}\ }\textbf {\bibinfo
  {volume} {8}},\ \bibinfo {pages} {031080} (\bibinfo {year}
  {2018})}\BibitemShut {NoStop}%
\bibitem [{\citenamefont {Polkovnikov}\ \emph {et~al.}(2011)\citenamefont
  {Polkovnikov}, \citenamefont {Sengupta}, \citenamefont {Silva},\ and\
  \citenamefont {Vengalattore}}]{polkovnikov2011}%
  \BibitemOpen
  \bibfield  {author} {\bibinfo {author} {\bibfnamefont {Anatoli}\ \bibnamefont
  {Polkovnikov}}, \bibinfo {author} {\bibfnamefont {Krishnendu}\ \bibnamefont
  {Sengupta}}, \bibinfo {author} {\bibfnamefont {Alessandro}\ \bibnamefont
  {Silva}}, \ and\ \bibinfo {author} {\bibfnamefont {Mukund}\ \bibnamefont
  {Vengalattore}},\ }\bibfield  {title} {\enquote {\bibinfo {title}
  {Colloquium: Nonequilibrium dynamics of closed interacting quantum
  systems},}\ }\href {\doibase 10.1103/RevModPhys.83.863} {\bibfield  {journal}
  {\bibinfo  {journal} {Reviews of Modern Physics}\ }\textbf {\bibinfo {volume}
  {83}},\ \bibinfo {pages} {863--883} (\bibinfo {year} {2011})}\BibitemShut
  {NoStop}%
\bibitem [{\citenamefont {Berges}\ \emph {et~al.}(2008)\citenamefont {Berges},
  \citenamefont {Rothkopf},\ and\ \citenamefont
  {Schmidt}}]{bergesRothkopf2008}%
  \BibitemOpen
  \bibfield  {author} {\bibinfo {author} {\bibfnamefont {J{\"u}rgen}\
  \bibnamefont {Berges}}, \bibinfo {author} {\bibfnamefont {Alexander}\
  \bibnamefont {Rothkopf}}, \ and\ \bibinfo {author} {\bibfnamefont {Jonas}\
  \bibnamefont {Schmidt}},\ }\bibfield  {title} {\enquote {\bibinfo {title}
  {Nonthermal fixed points: Effective weak coupling for strongly correlated
  systems far from equilibrium},}\ }\href {\doibase
  10.1103/PhysRevLett.101.041603} {\bibfield  {journal} {\bibinfo  {journal}
  {Physical Review Letters}\ }\textbf {\bibinfo {volume} {101}},\ \bibinfo
  {pages} {041603} (\bibinfo {year} {2008})}\BibitemShut {NoStop}%
\bibitem [{\citenamefont {Nowak}\ \emph {et~al.}(2011)\citenamefont {Nowak},
  \citenamefont {Sexty},\ and\ \citenamefont {Gasenzer}}]{nowak2011}%
  \BibitemOpen
  \bibfield  {author} {\bibinfo {author} {\bibfnamefont {Boris}\ \bibnamefont
  {Nowak}}, \bibinfo {author} {\bibfnamefont {D{\'e}nes}\ \bibnamefont
  {Sexty}}, \ and\ \bibinfo {author} {\bibfnamefont {Thomas}\ \bibnamefont
  {Gasenzer}},\ }\bibfield  {title} {\enquote {\bibinfo {title} {Superfluid
  turbulence: Nonthermal fixed point in an ultracold {Bose} gas},}\ }\href
  {\doibase 10.1103/PhysRevB.84.020506} {\bibfield  {journal} {\bibinfo
  {journal} {Physical Review B}\ }\textbf {\bibinfo {volume} {84}},\ \bibinfo
  {pages} {020506(R)} (\bibinfo {year} {2011})}\BibitemShut {NoStop}%
\bibitem [{\citenamefont {Pr{\"u}fer}\ \emph {et~al.}(2018)\citenamefont
  {Pr{\"u}fer}, \citenamefont {Kunkel}, \citenamefont {Strobel}, \citenamefont
  {Lannig}, \citenamefont {Linnemann}, \citenamefont {Schmied}, \citenamefont
  {Berges}, \citenamefont {Gasenzer},\ and\ \citenamefont
  {Oberthaler}}]{prufer2018}%
  \BibitemOpen
  \bibfield  {author} {\bibinfo {author} {\bibfnamefont {Maximilian}\
  \bibnamefont {Pr{\"u}fer}}, \bibinfo {author} {\bibfnamefont {Philipp}\
  \bibnamefont {Kunkel}}, \bibinfo {author} {\bibfnamefont {Helmut}\
  \bibnamefont {Strobel}}, \bibinfo {author} {\bibfnamefont {Stefan}\
  \bibnamefont {Lannig}}, \bibinfo {author} {\bibfnamefont {Daniel}\
  \bibnamefont {Linnemann}}, \bibinfo {author} {\bibfnamefont
  {Christian-Marcel}\ \bibnamefont {Schmied}}, \bibinfo {author} {\bibfnamefont
  {J{\"u}rgen}\ \bibnamefont {Berges}}, \bibinfo {author} {\bibfnamefont
  {Thomas}\ \bibnamefont {Gasenzer}}, \ and\ \bibinfo {author} {\bibfnamefont
  {Markus~K.}\ \bibnamefont {Oberthaler}},\ }\bibfield  {title} {\enquote
  {\bibinfo {title} {Observation of universal dynamics in a spinor {Bose} gas
  far from equilibrium},}\ }\href {\doibase 10.1038/s41586-018-0659-0}
  {\bibfield  {journal} {\bibinfo  {journal} {Nature}\ }\textbf {\bibinfo
  {volume} {563}},\ \bibinfo {pages} {217--220} (\bibinfo {year}
  {2018})}\BibitemShut {NoStop}%
\bibitem [{\citenamefont {Erne}\ \emph {et~al.}(2018)\citenamefont {Erne},
  \citenamefont {B{\"u}cker}, \citenamefont {Gasenzer}, \citenamefont
  {Berges},\ and\ \citenamefont {Schmiedmayer}}]{erne2018}%
  \BibitemOpen
  \bibfield  {author} {\bibinfo {author} {\bibfnamefont {Sebastian}\
  \bibnamefont {Erne}}, \bibinfo {author} {\bibfnamefont {Robert}\ \bibnamefont
  {B{\"u}cker}}, \bibinfo {author} {\bibfnamefont {Thomas}\ \bibnamefont
  {Gasenzer}}, \bibinfo {author} {\bibfnamefont {J{\"u}rgen}\ \bibnamefont
  {Berges}}, \ and\ \bibinfo {author} {\bibfnamefont {J{\"o}rg}\ \bibnamefont
  {Schmiedmayer}},\ }\bibfield  {title} {\enquote {\bibinfo {title} {Universal
  dynamics in an isolated one-dimensional {Bose} gas far from equilibrium},}\
  }\href {\doibase 10.1038/s41586-018-0667-0} {\bibfield  {journal} {\bibinfo
  {journal} {Nature}\ }\textbf {\bibinfo {volume} {563}},\ \bibinfo {pages}
  {225--229} (\bibinfo {year} {2018})}\BibitemShut {NoStop}%
\bibitem [{\citenamefont {Rodriguez-Nieva}\ \emph {et~al.}(2022)\citenamefont
  {Rodriguez-Nieva}, \citenamefont {Pi{\~n}eiro~Orioli},\ and\ \citenamefont
  {Marino}}]{rodriguez2022far}%
  \BibitemOpen
  \bibfield  {author} {\bibinfo {author} {\bibfnamefont {Joaquin~F}\
  \bibnamefont {Rodriguez-Nieva}}, \bibinfo {author} {\bibfnamefont {Asier}\
  \bibnamefont {Pi{\~n}eiro~Orioli}}, \ and\ \bibinfo {author} {\bibfnamefont
  {Jamir}\ \bibnamefont {Marino}},\ }\bibfield  {title} {\enquote {\bibinfo
  {title} {Far-from-equilibrium universality in the two-dimensional heisenberg
  model},}\ }\href@noop {} {\bibfield  {journal} {\bibinfo  {journal}
  {Proceedings of the National Academy of Sciences}\ }\textbf {\bibinfo
  {volume} {119}},\ \bibinfo {pages} {e2122599119} (\bibinfo {year}
  {2022})}\BibitemShut {NoStop}%
\bibitem [{\citenamefont {Marcuzzi}\ \emph {et~al.}(2016)\citenamefont
  {Marcuzzi}, \citenamefont {Marino}, \citenamefont {Gambassi},\ and\
  \citenamefont {Silva}}]{marcuzzi2016prethermalization}%
  \BibitemOpen
  \bibfield  {author} {\bibinfo {author} {\bibfnamefont {Matteo}\ \bibnamefont
  {Marcuzzi}}, \bibinfo {author} {\bibfnamefont {Jamir}\ \bibnamefont
  {Marino}}, \bibinfo {author} {\bibfnamefont {Andrea}\ \bibnamefont
  {Gambassi}}, \ and\ \bibinfo {author} {\bibfnamefont {Alessandro}\
  \bibnamefont {Silva}},\ }\bibfield  {title} {\enquote {\bibinfo {title}
  {Prethermalization from a low-density holstein-primakoff expansion},}\
  }\href@noop {} {\bibfield  {journal} {\bibinfo  {journal} {Physical Review
  B}\ }\textbf {\bibinfo {volume} {94}},\ \bibinfo {pages} {214304} (\bibinfo
  {year} {2016})}\BibitemShut {NoStop}%
\bibitem [{\citenamefont {Sieberer}\ \emph {et~al.}(2013)\citenamefont
  {Sieberer}, \citenamefont {Huber}, \citenamefont {Altman},\ and\
  \citenamefont {Diehl}}]{sieberer2013}%
  \BibitemOpen
  \bibfield  {author} {\bibinfo {author} {\bibfnamefont {L.~M.}\ \bibnamefont
  {Sieberer}}, \bibinfo {author} {\bibfnamefont {S.~D.}\ \bibnamefont {Huber}},
  \bibinfo {author} {\bibfnamefont {E.}~\bibnamefont {Altman}}, \ and\ \bibinfo
  {author} {\bibfnamefont {S.}~\bibnamefont {Diehl}},\ }\bibfield  {title}
  {\enquote {\bibinfo {title} {Dynamical critical phenomena in
  driven-dissipative systems},}\ }\href {\doibase
  10.1103/PhysRevLett.110.195301} {\bibfield  {journal} {\bibinfo  {journal}
  {Physical Review Letters}\ }\textbf {\bibinfo {volume} {110}},\ \bibinfo
  {pages} {195301} (\bibinfo {year} {2013})},\ \Eprint
  {http://arxiv.org/abs/1301.5854} {arXiv:1301.5854 [cond-mat.quant-gas]}
  \BibitemShut {NoStop}%
\bibitem [{\citenamefont {Altman}\ \emph {et~al.}(2015)\citenamefont {Altman},
  \citenamefont {Sieberer}, \citenamefont {Chen}, \citenamefont {Diehl},\ and\
  \citenamefont {Toner}}]{altman2015}%
  \BibitemOpen
  \bibfield  {author} {\bibinfo {author} {\bibfnamefont {Ehud}\ \bibnamefont
  {Altman}}, \bibinfo {author} {\bibfnamefont {Lukas~M.}\ \bibnamefont
  {Sieberer}}, \bibinfo {author} {\bibfnamefont {Leiming}\ \bibnamefont
  {Chen}}, \bibinfo {author} {\bibfnamefont {Sebastian}\ \bibnamefont {Diehl}},
  \ and\ \bibinfo {author} {\bibfnamefont {John}\ \bibnamefont {Toner}},\
  }\bibfield  {title} {\enquote {\bibinfo {title} {Two-dimensional
  superfluidity of exciton polaritons requires strong anisotropy},}\ }\href
  {\doibase 10.1103/PhysRevX.5.011017} {\bibfield  {journal} {\bibinfo
  {journal} {Physical Review X}\ }\textbf {\bibinfo {volume} {5}},\ \bibinfo
  {pages} {011017} (\bibinfo {year} {2015})}\BibitemShut {NoStop}%
\bibitem [{\citenamefont {Fontaine}\ \emph {et~al.}(2022)\citenamefont
  {Fontaine}, \citenamefont {Squizzato}, \citenamefont {Baboux}, \citenamefont
  {Amelio}, \citenamefont {Lema{\^i}tre}, \citenamefont {Morassi},
  \citenamefont {Sagnes}, \citenamefont {Le~Gratiet}, \citenamefont {Harouri},
  \citenamefont {Wouters}, \citenamefont {Carusotto}, \citenamefont {Amo},
  \citenamefont {Richard}, \citenamefont {Minguzzi}, \citenamefont {Canet},
  \citenamefont {Ravets},\ and\ \citenamefont {Bloch}}]{fontaine2022}%
  \BibitemOpen
  \bibfield  {author} {\bibinfo {author} {\bibfnamefont {Quentin}\ \bibnamefont
  {Fontaine}}, \bibinfo {author} {\bibfnamefont {Davide}\ \bibnamefont
  {Squizzato}}, \bibinfo {author} {\bibfnamefont {Florent}\ \bibnamefont
  {Baboux}}, \bibinfo {author} {\bibfnamefont {Ivan}\ \bibnamefont {Amelio}},
  \bibinfo {author} {\bibfnamefont {Aristide}\ \bibnamefont {Lema{\^i}tre}},
  \bibinfo {author} {\bibfnamefont {Martina}\ \bibnamefont {Morassi}}, \bibinfo
  {author} {\bibfnamefont {Isabelle}\ \bibnamefont {Sagnes}}, \bibinfo {author}
  {\bibfnamefont {Luc}\ \bibnamefont {Le~Gratiet}}, \bibinfo {author}
  {\bibfnamefont {Abdelmounaim}\ \bibnamefont {Harouri}}, \bibinfo {author}
  {\bibfnamefont {Michiel}\ \bibnamefont {Wouters}}, \bibinfo {author}
  {\bibfnamefont {Iacopo}\ \bibnamefont {Carusotto}}, \bibinfo {author}
  {\bibfnamefont {Alberto}\ \bibnamefont {Amo}}, \bibinfo {author}
  {\bibfnamefont {Maxime}\ \bibnamefont {Richard}}, \bibinfo {author}
  {\bibfnamefont {Anna}\ \bibnamefont {Minguzzi}}, \bibinfo {author}
  {\bibfnamefont {L{\'e}onie}\ \bibnamefont {Canet}}, \bibinfo {author}
  {\bibfnamefont {Sylvain}\ \bibnamefont {Ravets}}, \ and\ \bibinfo {author}
  {\bibfnamefont {Jacqueline}\ \bibnamefont {Bloch}},\ }\bibfield  {title}
  {\enquote {\bibinfo {title} {{Kardar--Parisi--Zhang} universality in a
  one-dimensional polariton condensate},}\ }\href {\doibase
  10.1038/s41586-022-05001-8} {\bibfield  {journal} {\bibinfo  {journal}
  {Nature}\ }\textbf {\bibinfo {volume} {608}},\ \bibinfo {pages} {687--691}
  (\bibinfo {year} {2022})}\BibitemShut {NoStop}%
\bibitem [{\citenamefont {Else}\ \emph {et~al.}(2016)\citenamefont {Else},
  \citenamefont {Bauer},\ and\ \citenamefont {Nayak}}]{else2016}%
  \BibitemOpen
  \bibfield  {author} {\bibinfo {author} {\bibfnamefont {Dominic~V.}\
  \bibnamefont {Else}}, \bibinfo {author} {\bibfnamefont {Bela}\ \bibnamefont
  {Bauer}}, \ and\ \bibinfo {author} {\bibfnamefont {Chetan}\ \bibnamefont
  {Nayak}},\ }\bibfield  {title} {\enquote {\bibinfo {title} {{Floquet} time
  crystals},}\ }\href {\doibase 10.1103/PhysRevLett.117.090402} {\bibfield
  {journal} {\bibinfo  {journal} {Physical Review Letters}\ }\textbf {\bibinfo
  {volume} {117}},\ \bibinfo {pages} {090402} (\bibinfo {year}
  {2016})}\BibitemShut {NoStop}%
\bibitem [{\citenamefont {Rudner}\ \emph {et~al.}(2013)\citenamefont {Rudner},
  \citenamefont {Lindner}, \citenamefont {Berg},\ and\ \citenamefont
  {Levin}}]{rudner2013}%
  \BibitemOpen
  \bibfield  {author} {\bibinfo {author} {\bibfnamefont {Mark~S.}\ \bibnamefont
  {Rudner}}, \bibinfo {author} {\bibfnamefont {Netanel~H.}\ \bibnamefont
  {Lindner}}, \bibinfo {author} {\bibfnamefont {Erez}\ \bibnamefont {Berg}}, \
  and\ \bibinfo {author} {\bibfnamefont {Michael}\ \bibnamefont {Levin}},\
  }\bibfield  {title} {\enquote {\bibinfo {title} {Anomalous edge states and
  the bulk-edge correspondence for periodically driven two-dimensional
  systems},}\ }\href {\doibase 10.1103/PhysRevX.3.031005} {\bibfield  {journal}
  {\bibinfo  {journal} {Physical Review X}\ }\textbf {\bibinfo {volume} {3}},\
  \bibinfo {pages} {031005} (\bibinfo {year} {2013})}\BibitemShut {NoStop}%
\bibitem [{\citenamefont {D'Alessio}\ \emph {et~al.}(2016)\citenamefont
  {D'Alessio}, \citenamefont {Kafri}, \citenamefont {Polkovnikov},\ and\
  \citenamefont {Rigol}}]{dalessio2016}%
  \BibitemOpen
  \bibfield  {author} {\bibinfo {author} {\bibfnamefont {Luca}\ \bibnamefont
  {D'Alessio}}, \bibinfo {author} {\bibfnamefont {Yariv}\ \bibnamefont
  {Kafri}}, \bibinfo {author} {\bibfnamefont {Anatoli}\ \bibnamefont
  {Polkovnikov}}, \ and\ \bibinfo {author} {\bibfnamefont {Marcos}\
  \bibnamefont {Rigol}},\ }\bibfield  {title} {\enquote {\bibinfo {title} {From
  quantum chaos and eigenstate thermalization to statistical mechanics and
  thermodynamics},}\ }\href {\doibase 10.1080/00018732.2016.1198134} {\bibfield
   {journal} {\bibinfo  {journal} {Advances in Physics}\ }\textbf {\bibinfo
  {volume} {65}},\ \bibinfo {pages} {239--362} (\bibinfo {year}
  {2016})}\BibitemShut {NoStop}%
\bibitem [{\citenamefont {Sieberer}\ \emph {et~al.}(2016)\citenamefont
  {Sieberer}, \citenamefont {Buchhold},\ and\ \citenamefont
  {Diehl}}]{sieberer2016}%
  \BibitemOpen
  \bibfield  {author} {\bibinfo {author} {\bibfnamefont {Lukas~M.}\
  \bibnamefont {Sieberer}}, \bibinfo {author} {\bibfnamefont {Michael}\
  \bibnamefont {Buchhold}}, \ and\ \bibinfo {author} {\bibfnamefont
  {Sebastian}\ \bibnamefont {Diehl}},\ }\bibfield  {title} {\enquote {\bibinfo
  {title} {Keldysh field theory for driven open quantum systems},}\ }\href
  {\doibase 10.1088/0034-4885/79/9/096001} {\bibfield  {journal} {\bibinfo
  {journal} {Reports on Progress in Physics}\ }\textbf {\bibinfo {volume}
  {79}},\ \bibinfo {pages} {096001} (\bibinfo {year} {2016})}\BibitemShut
  {NoStop}%
\bibitem [{\citenamefont {{Dalla Torre}}\ \emph {et~al.}(2013)\citenamefont
  {{Dalla Torre}}, \citenamefont {Diehl}, \citenamefont {Lukin}, \citenamefont
  {Sachdev},\ and\ \citenamefont {Strack}}]{dallaTorre2013Dicke}%
  \BibitemOpen
  \bibfield  {author} {\bibinfo {author} {\bibfnamefont {Emanuele~G.}\
  \bibnamefont {{Dalla Torre}}}, \bibinfo {author} {\bibfnamefont {Sebastian}\
  \bibnamefont {Diehl}}, \bibinfo {author} {\bibfnamefont {Mikhail~D.}\
  \bibnamefont {Lukin}}, \bibinfo {author} {\bibfnamefont {Subir}\ \bibnamefont
  {Sachdev}}, \ and\ \bibinfo {author} {\bibfnamefont {Philipp}\ \bibnamefont
  {Strack}},\ }\bibfield  {title} {\enquote {\bibinfo {title} {{Keldysh}
  approach for nonequilibrium phase transitions in quantum optics: Beyond the
  {Dicke} model in optical cavities},}\ }\href {\doibase
  10.1103/PhysRevA.87.023831} {\bibfield  {journal} {\bibinfo  {journal}
  {Physical Review A}\ }\textbf {\bibinfo {volume} {87}},\ \bibinfo {pages}
  {023831} (\bibinfo {year} {2013})},\ \Eprint {http://arxiv.org/abs/1210.3623}
  {arXiv:1210.3623 [cond-mat.quant-gas]} \BibitemShut {NoStop}%
\bibitem [{\citenamefont {Maghrebi}\ and\ \citenamefont
  {Gorshkov}(2016)}]{maghrebiGorshkov2016}%
  \BibitemOpen
  \bibfield  {author} {\bibinfo {author} {\bibfnamefont {Mohammad~F.}\
  \bibnamefont {Maghrebi}}\ and\ \bibinfo {author} {\bibfnamefont {Alexey~V.}\
  \bibnamefont {Gorshkov}},\ }\bibfield  {title} {\enquote {\bibinfo {title}
  {Nonequilibrium many-body steady states via {Keldysh} formalism},}\ }\href
  {\doibase 10.1103/PhysRevB.93.014307} {\bibfield  {journal} {\bibinfo
  {journal} {Physical Review B}\ }\textbf {\bibinfo {volume} {93}},\ \bibinfo
  {pages} {014307} (\bibinfo {year} {2016})},\ \Eprint
  {http://arxiv.org/abs/1507.01939} {arXiv:1507.01939 [cond-mat.quant-gas]}
  \BibitemShut {NoStop}%
\bibitem [{\citenamefont {Kelly}\ \emph {et~al.}(2020)\citenamefont {Kelly},
  \citenamefont {Nandkishore},\ and\ \citenamefont
  {Marino}}]{kelly2020exploring}%
  \BibitemOpen
  \bibfield  {author} {\bibinfo {author} {\bibfnamefont {Shane~P}\ \bibnamefont
  {Kelly}}, \bibinfo {author} {\bibfnamefont {Rahul}\ \bibnamefont
  {Nandkishore}}, \ and\ \bibinfo {author} {\bibfnamefont {Jamir}\ \bibnamefont
  {Marino}},\ }\bibfield  {title} {\enquote {\bibinfo {title} {Exploring
  many-body localization in quantum systems coupled to an environment via
  wegner-wilson flows},}\ }\href@noop {} {\bibfield  {journal} {\bibinfo
  {journal} {Nuclear Physics B}\ }\textbf {\bibinfo {volume} {951}},\ \bibinfo
  {pages} {114886} (\bibinfo {year} {2020})}\BibitemShut {NoStop}%
\bibitem [{\citenamefont {Zelle}\ \emph {et~al.}(2024)\citenamefont {Zelle},
  \citenamefont {Daviet}, \citenamefont {Rosch},\ and\ \citenamefont
  {Diehl}}]{zelle2024}%
  \BibitemOpen
  \bibfield  {author} {\bibinfo {author} {\bibfnamefont {Carl~Philipp}\
  \bibnamefont {Zelle}}, \bibinfo {author} {\bibfnamefont {Romain}\
  \bibnamefont {Daviet}}, \bibinfo {author} {\bibfnamefont {Achim}\
  \bibnamefont {Rosch}}, \ and\ \bibinfo {author} {\bibfnamefont {Sebastian}\
  \bibnamefont {Diehl}},\ }\bibfield  {title} {\enquote {\bibinfo {title}
  {Universal phenomenology at critical exceptional points of nonequilibrium
  {O(N)} models},}\ }\href {\doibase 10.1103/PhysRevX.14.021052} {\bibfield
  {journal} {\bibinfo  {journal} {Physical Review X}\ }\textbf {\bibinfo
  {volume} {14}},\ \bibinfo {pages} {021052} (\bibinfo {year} {2024})},\
  \Eprint {http://arxiv.org/abs/2304.09207} {arXiv:2304.09207} \BibitemShut
  {NoStop}%
\bibitem [{\citenamefont {Daviet}\ \emph {et~al.}(2024)\citenamefont {Daviet},
  \citenamefont {Zelle}, \citenamefont {Rosch},\ and\ \citenamefont
  {Diehl}}]{daviet2024}%
  \BibitemOpen
  \bibfield  {author} {\bibinfo {author} {\bibfnamefont {Romain}\ \bibnamefont
  {Daviet}}, \bibinfo {author} {\bibfnamefont {Carl~Philipp}\ \bibnamefont
  {Zelle}}, \bibinfo {author} {\bibfnamefont {Achim}\ \bibnamefont {Rosch}}, \
  and\ \bibinfo {author} {\bibfnamefont {Sebastian}\ \bibnamefont {Diehl}},\
  }\bibfield  {title} {\enquote {\bibinfo {title} {Nonequilibrium criticality
  at the onset of time-crystalline order},}\ }\href {\doibase
  10.1103/PhysRevLett.132.167102} {\bibfield  {journal} {\bibinfo  {journal}
  {Physical Review Letters}\ }\textbf {\bibinfo {volume} {132}},\ \bibinfo
  {pages} {167102} (\bibinfo {year} {2024})},\ \Eprint
  {http://arxiv.org/abs/2312.13372} {arXiv:2312.13372} \BibitemShut {NoStop}%
\bibitem [{\citenamefont {Zelle}\ \emph {et~al.}(2026)\citenamefont {Zelle},
  \citenamefont {John}, \citenamefont {Supple}, \citenamefont {Daviet},\ and\
  \citenamefont {Diehl}}]{zelle2026goldstone}%
  \BibitemOpen
  \bibfield  {author} {\bibinfo {author} {\bibfnamefont {Carl~Philipp}\
  \bibnamefont {Zelle}}, \bibinfo {author} {\bibfnamefont {Gustav}\
  \bibnamefont {John}}, \bibinfo {author} {\bibfnamefont {Orla}\ \bibnamefont
  {Supple}}, \bibinfo {author} {\bibfnamefont {Romain}\ \bibnamefont {Daviet}},
  \ and\ \bibinfo {author} {\bibfnamefont {Sebastian}\ \bibnamefont {Diehl}},\
  }\href@noop {} {\enquote {\bibinfo {title} {Far-from-equilibrium scaling of
  non-abelian {Goldstone} modes},}\ } (\bibinfo {year} {2026}),\ \Eprint
  {http://arxiv.org/abs/2608.13666} {arXiv:2608.13666 [cond-mat.stat-mech]}
  \BibitemShut {NoStop}%
\bibitem [{\citenamefont {Sachdev}(2011)}]{sachdev2011}%
  \BibitemOpen
  \bibfield  {author} {\bibinfo {author} {\bibfnamefont {Subir}\ \bibnamefont
  {Sachdev}},\ }\href {\doibase 10.1017/CBO9780511973765} {\emph {\bibinfo
  {title} {Quantum Phase Transitions}}},\ \bibinfo {edition} {2nd}\ ed.\
  (\bibinfo  {publisher} {Cambridge University Press},\ \bibinfo {address}
  {Cambridge},\ \bibinfo {year} {2011})\BibitemShut {NoStop}%
\bibitem [{\citenamefont {Gring}\ \emph {et~al.}(2012)\citenamefont {Gring},
  \citenamefont {Kuhnert}, \citenamefont {Langen}, \citenamefont {Kitagawa},
  \citenamefont {Rauer}, \citenamefont {Schreitl}, \citenamefont {Mazets},
  \citenamefont {Smith}, \citenamefont {Demler},\ and\ \citenamefont
  {Schmiedmayer}}]{gring2012}%
  \BibitemOpen
  \bibfield  {author} {\bibinfo {author} {\bibfnamefont {Michael}\ \bibnamefont
  {Gring}}, \bibinfo {author} {\bibfnamefont {Maximilian}\ \bibnamefont
  {Kuhnert}}, \bibinfo {author} {\bibfnamefont {Tim}\ \bibnamefont {Langen}},
  \bibinfo {author} {\bibfnamefont {Takuya}\ \bibnamefont {Kitagawa}}, \bibinfo
  {author} {\bibfnamefont {Bernhard}\ \bibnamefont {Rauer}}, \bibinfo {author}
  {\bibfnamefont {Matthias}\ \bibnamefont {Schreitl}}, \bibinfo {author}
  {\bibfnamefont {Igor}\ \bibnamefont {Mazets}}, \bibinfo {author}
  {\bibfnamefont {David~A.}\ \bibnamefont {Smith}}, \bibinfo {author}
  {\bibfnamefont {Eugene}\ \bibnamefont {Demler}}, \ and\ \bibinfo {author}
  {\bibfnamefont {J{\"o}rg}\ \bibnamefont {Schmiedmayer}},\ }\bibfield  {title}
  {\enquote {\bibinfo {title} {Relaxation and pre-thermalization in an isolated
  quantum system},}\ }\href {\doibase 10.1126/science.1224953} {\bibfield
  {journal} {\bibinfo  {journal} {Science}\ }\textbf {\bibinfo {volume}
  {337}},\ \bibinfo {pages} {1318--1322} (\bibinfo {year} {2012})},\ \Eprint
  {http://arxiv.org/abs/1112.0013} {arXiv:1112.0013 [cond-mat.quant-gas]}
  \BibitemShut {NoStop}%
\bibitem [{\citenamefont {Kardar}\ \emph {et~al.}(1986)\citenamefont {Kardar},
  \citenamefont {Parisi},\ and\ \citenamefont {Zhang}}]{kardar1986}%
  \BibitemOpen
  \bibfield  {author} {\bibinfo {author} {\bibfnamefont {Mehran}\ \bibnamefont
  {Kardar}}, \bibinfo {author} {\bibfnamefont {Giorgio}\ \bibnamefont
  {Parisi}}, \ and\ \bibinfo {author} {\bibfnamefont {Yi-Cheng}\ \bibnamefont
  {Zhang}},\ }\bibfield  {title} {\enquote {\bibinfo {title} {Dynamic scaling
  of growing interfaces},}\ }\href {\doibase 10.1103/PhysRevLett.56.889}
  {\bibfield  {journal} {\bibinfo  {journal} {Physical Review Letters}\
  }\textbf {\bibinfo {volume} {56}},\ \bibinfo {pages} {889} (\bibinfo {year}
  {1986})}\BibitemShut {NoStop}%
\bibitem [{\citenamefont {Canet}\ \emph {et~al.}(2011)\citenamefont {Canet},
  \citenamefont {Chat{\'e}}, \citenamefont {Delamotte},\ and\ \citenamefont
  {Wschebor}}]{canet2011}%
  \BibitemOpen
  \bibfield  {author} {\bibinfo {author} {\bibfnamefont {L{\'e}onie}\
  \bibnamefont {Canet}}, \bibinfo {author} {\bibfnamefont {Hugues}\
  \bibnamefont {Chat{\'e}}}, \bibinfo {author} {\bibfnamefont {Bertrand}\
  \bibnamefont {Delamotte}}, \ and\ \bibinfo {author} {\bibfnamefont
  {Nicol{\'a}s}\ \bibnamefont {Wschebor}},\ }\bibfield  {title} {\enquote
  {\bibinfo {title} {Nonperturbative renormalization group for the
  {Kardar--Parisi--Zhang} equation: General framework and first
  applications},}\ }\href {\doibase 10.1103/PhysRevE.84.061128} {\bibfield
  {journal} {\bibinfo  {journal} {Physical Review E}\ }\textbf {\bibinfo
  {volume} {84}},\ \bibinfo {pages} {061128} (\bibinfo {year} {2011})},\
  \bibinfo {note} {erratum: Phys. Rev. E \textbf{86}, 019904 (2012)},\ \Eprint
  {http://arxiv.org/abs/1107.2289} {arXiv:1107.2289 [cond-mat.stat-mech]}
  \BibitemShut {NoStop}%
\bibitem [{\citenamefont {Forster}\ \emph {et~al.}(1977)\citenamefont
  {Forster}, \citenamefont {Nelson},\ and\ \citenamefont
  {Stephen}}]{forster1977}%
  \BibitemOpen
  \bibfield  {author} {\bibinfo {author} {\bibfnamefont {Dieter}\ \bibnamefont
  {Forster}}, \bibinfo {author} {\bibfnamefont {David~R.}\ \bibnamefont
  {Nelson}}, \ and\ \bibinfo {author} {\bibfnamefont {Michael~J.}\ \bibnamefont
  {Stephen}},\ }\bibfield  {title} {\enquote {\bibinfo {title} {Large-distance
  and long-time properties of a randomly stirred fluid},}\ }\href {\doibase
  10.1103/PhysRevA.16.732} {\bibfield  {journal} {\bibinfo  {journal} {Physical
  Review A}\ }\textbf {\bibinfo {volume} {16}},\ \bibinfo {pages} {732--749}
  (\bibinfo {year} {1977})}\BibitemShut {NoStop}%
\bibitem [{\citenamefont {Marino}\ and\ \citenamefont
  {Diehl}(2016{\natexlab{a}})}]{marinoDiehl2016}%
  \BibitemOpen
  \bibfield  {author} {\bibinfo {author} {\bibfnamefont {Jamir}\ \bibnamefont
  {Marino}}\ and\ \bibinfo {author} {\bibfnamefont {Sebastian}\ \bibnamefont
  {Diehl}},\ }\bibfield  {title} {\enquote {\bibinfo {title} {Quantum dynamical
  field theory for nonequilibrium phase transitions in driven open systems},}\
  }\href {\doibase 10.1103/PhysRevB.94.085150} {\bibfield  {journal} {\bibinfo
  {journal} {Physical Review B}\ }\textbf {\bibinfo {volume} {94}},\ \bibinfo
  {pages} {085150} (\bibinfo {year} {2016}{\natexlab{a}})},\ \Eprint
  {http://arxiv.org/abs/1606.00452} {arXiv:1606.00452 [cond-mat.quant-gas]}
  \BibitemShut {NoStop}%
\bibitem [{\citenamefont {Marino}\ and\ \citenamefont
  {Diehl}(2016{\natexlab{b}})}]{marinoDiehlPRL2016}%
  \BibitemOpen
  \bibfield  {author} {\bibinfo {author} {\bibfnamefont {Jamir}\ \bibnamefont
  {Marino}}\ and\ \bibinfo {author} {\bibfnamefont {Sebastian}\ \bibnamefont
  {Diehl}},\ }\bibfield  {title} {\enquote {\bibinfo {title} {Driven
  {Markovian} quantum criticality},}\ }\href {\doibase
  10.1103/PhysRevLett.116.070407} {\bibfield  {journal} {\bibinfo  {journal}
  {Physical Review Letters}\ }\textbf {\bibinfo {volume} {116}},\ \bibinfo
  {pages} {070407} (\bibinfo {year} {2016}{\natexlab{b}})},\ \Eprint
  {http://arxiv.org/abs/1508.02723} {arXiv:1508.02723 [cond-mat.quant-gas]}
  \BibitemShut {NoStop}%
\bibitem [{\citenamefont {Huang}(2008)}]{huang2008statistical}%
  \BibitemOpen
  \bibfield  {author} {\bibinfo {author} {\bibfnamefont {Kerson}\ \bibnamefont
  {Huang}},\ }\href@noop {} {\emph {\bibinfo {title} {Statistical mechanics}}}\
  (\bibinfo  {publisher} {John Wiley \& Sons},\ \bibinfo {year}
  {2008})\BibitemShut {NoStop}%
\bibitem [{\citenamefont {Altland}\ and\ \citenamefont
  {Simons}(2023)}]{altlandSimons2023}%
  \BibitemOpen
  \bibfield  {author} {\bibinfo {author} {\bibfnamefont {Alexander}\
  \bibnamefont {Altland}}\ and\ \bibinfo {author} {\bibfnamefont {Ben}\
  \bibnamefont {Simons}},\ }\href {\doibase 10.1017/9781108781244} {\emph
  {\bibinfo {title} {Condensed Matter Field Theory}}},\ \bibinfo {edition}
  {3rd}\ ed.\ (\bibinfo  {publisher} {Cambridge University Press},\ \bibinfo
  {address} {Cambridge},\ \bibinfo {year} {2023})\BibitemShut {NoStop}%
\bibitem [{\citenamefont {Sieberer}\ \emph {et~al.}(2014)\citenamefont
  {Sieberer}, \citenamefont {Huber}, \citenamefont {Altman},\ and\
  \citenamefont {Diehl}}]{sieberer2014}%
  \BibitemOpen
  \bibfield  {author} {\bibinfo {author} {\bibfnamefont {L.~M.}\ \bibnamefont
  {Sieberer}}, \bibinfo {author} {\bibfnamefont {S.~D.}\ \bibnamefont {Huber}},
  \bibinfo {author} {\bibfnamefont {E.}~\bibnamefont {Altman}}, \ and\ \bibinfo
  {author} {\bibfnamefont {S.}~\bibnamefont {Diehl}},\ }\bibfield  {title}
  {\enquote {\bibinfo {title} {Nonequilibrium functional renormalization for
  driven-dissipative {Bose--Einstein} condensation},}\ }\href {\doibase
  10.1103/PhysRevB.89.134310} {\bibfield  {journal} {\bibinfo  {journal}
  {Physical Review B}\ }\textbf {\bibinfo {volume} {89}},\ \bibinfo {pages}
  {134310} (\bibinfo {year} {2014})},\ \Eprint {http://arxiv.org/abs/1309.7027}
  {arXiv:1309.7027 [cond-mat.quant-gas]} \BibitemShut {NoStop}%
\bibitem [{\citenamefont {Kamenev}(2023)}]{kamenev2023}%
  \BibitemOpen
  \bibfield  {author} {\bibinfo {author} {\bibfnamefont {Alex}\ \bibnamefont
  {Kamenev}},\ }\href {\doibase 10.1017/9781108769266} {\emph {\bibinfo {title}
  {Field Theory of Non-Equilibrium Systems}}},\ \bibinfo {edition} {2nd}\ ed.\
  (\bibinfo  {publisher} {Cambridge University Press},\ \bibinfo {address}
  {Cambridge},\ \bibinfo {year} {2023})\BibitemShut {NoStop}%
\bibitem [{\citenamefont {Mesterh{\'a}zy}\ \emph {et~al.}(2015)\citenamefont
  {Mesterh{\'a}zy}, \citenamefont {Stockemer},\ and\ \citenamefont
  {Tanizaki}}]{mesterhazy2015quantum}%
  \BibitemOpen
  \bibfield  {author} {\bibinfo {author} {\bibfnamefont {D.}~\bibnamefont
  {Mesterh{\'a}zy}}, \bibinfo {author} {\bibfnamefont {J.~H.}\ \bibnamefont
  {Stockemer}}, \ and\ \bibinfo {author} {\bibfnamefont {Y.}~\bibnamefont
  {Tanizaki}},\ }\bibfield  {title} {\enquote {\bibinfo {title} {From quantum
  to classical dynamics: The relativistic $o(n)$ model in the framework of the
  real-time functional renormalization group},}\ }\href {\doibase
  10.1103/PhysRevD.92.076001} {\bibfield  {journal} {\bibinfo  {journal} {Phys.
  Rev. D}\ }\textbf {\bibinfo {volume} {92}},\ \bibinfo {pages} {076001}
  (\bibinfo {year} {2015})}\BibitemShut {NoStop}%
\bibitem [{\citenamefont {Lin}\ \emph {et~al.}(2023)\citenamefont {Lin},
  \citenamefont {Bu},\ and\ \citenamefont {Lei}}]{lin2023non}%
  \BibitemOpen
  \bibfield  {author} {\bibinfo {author} {\bibfnamefont {Shu}\ \bibnamefont
  {Lin}}, \bibinfo {author} {\bibfnamefont {Yanyan}\ \bibnamefont {Bu}}, \ and\
  \bibinfo {author} {\bibfnamefont {Chang}\ \bibnamefont {Lei}},\ }\bibfield
  {title} {\enquote {\bibinfo {title} {Non-gaussianity from schwinger-keldysh
  effective field theory},}\ }\href@noop {} {\bibfield  {journal} {\bibinfo
  {journal} {arXiv preprint arXiv:2301.06703}\ } (\bibinfo {year}
  {2023})}\BibitemShut {NoStop}%
\bibitem [{\citenamefont {T{\"a}uber}(2014)}]{tauber2014}%
  \BibitemOpen
  \bibfield  {author} {\bibinfo {author} {\bibfnamefont {Uwe~C.}\ \bibnamefont
  {T{\"a}uber}},\ }\href {\doibase 10.1017/CBO9781139046213} {\emph {\bibinfo
  {title} {Critical Dynamics: A Field Theory Approach to Equilibrium and
  Non-Equilibrium Scaling Behavior}}}\ (\bibinfo  {publisher} {Cambridge
  University Press},\ \bibinfo {address} {Cambridge},\ \bibinfo {year}
  {2014})\BibitemShut {NoStop}%
\bibitem [{\citenamefont {Frey}\ and\ \citenamefont
  {T{\"a}uber}(1994)}]{freyTauber1994}%
  \BibitemOpen
  \bibfield  {author} {\bibinfo {author} {\bibfnamefont {Erwin}\ \bibnamefont
  {Frey}}\ and\ \bibinfo {author} {\bibfnamefont {Uwe~Claus}\ \bibnamefont
  {T{\"a}uber}},\ }\bibfield  {title} {\enquote {\bibinfo {title} {Two-loop
  renormalization-group analysis of the {Burgers--Kardar--Parisi--Zhang}
  equation},}\ }\href {\doibase 10.1103/PhysRevE.50.1024} {\bibfield  {journal}
  {\bibinfo  {journal} {Physical Review E}\ }\textbf {\bibinfo {volume} {50}},\
  \bibinfo {pages} {1024--1045} (\bibinfo {year} {1994})},\ \Eprint
  {http://arxiv.org/abs/cond-mat/9406068} {arXiv:cond-mat/9406068} \BibitemShut
  {NoStop}%
\bibitem [{\citenamefont {Janssen}\ and\ \citenamefont
  {Schmittmann}(1986)}]{janssen1986field}%
  \BibitemOpen
  \bibfield  {author} {\bibinfo {author} {\bibfnamefont {H.~K.}\ \bibnamefont
  {Janssen}}\ and\ \bibinfo {author} {\bibfnamefont {Beate}\ \bibnamefont
  {Schmittmann}},\ }\bibfield  {title} {\enquote {\bibinfo {title} {Field
  theory of long time behaviour in driven diffusive systems},}\ }\href
  {\doibase 10.1007/BF01726201} {\bibfield  {journal} {\bibinfo  {journal}
  {Zeitschrift f{\"u}r Physik B Condensed Matter}\ }\textbf {\bibinfo {volume}
  {63}},\ \bibinfo {pages} {517--520} (\bibinfo {year} {1986})}\BibitemShut
  {NoStop}%
\bibitem [{\citenamefont {Solon}\ and\ \citenamefont
  {Tailleur}(2013)}]{solonTailleur2013}%
  \BibitemOpen
  \bibfield  {author} {\bibinfo {author} {\bibfnamefont {A.~P.}\ \bibnamefont
  {Solon}}\ and\ \bibinfo {author} {\bibfnamefont {J.}~\bibnamefont
  {Tailleur}},\ }\bibfield  {title} {\enquote {\bibinfo {title} {Revisiting the
  flocking transition using active spins},}\ }\href {\doibase
  10.1103/PhysRevLett.111.078101} {\bibfield  {journal} {\bibinfo  {journal}
  {Physical Review Letters}\ }\textbf {\bibinfo {volume} {111}},\ \bibinfo
  {pages} {078101} (\bibinfo {year} {2013})}\BibitemShut {NoStop}%
\bibitem [{\citenamefont {Solon}\ and\ \citenamefont
  {Tailleur}(2015)}]{solonTailleur2015}%
  \BibitemOpen
  \bibfield  {author} {\bibinfo {author} {\bibfnamefont {A.~P.}\ \bibnamefont
  {Solon}}\ and\ \bibinfo {author} {\bibfnamefont {J.}~\bibnamefont
  {Tailleur}},\ }\bibfield  {title} {\enquote {\bibinfo {title} {Flocking with
  discrete symmetry: The two-dimensional active {Ising} model},}\ }\href
  {\doibase 10.1103/PhysRevE.92.042119} {\bibfield  {journal} {\bibinfo
  {journal} {Physical Review E}\ }\textbf {\bibinfo {volume} {92}},\ \bibinfo
  {pages} {042119} (\bibinfo {year} {2015})}\BibitemShut {NoStop}%
\bibitem [{\citenamefont {Scandolo}\ \emph {et~al.}(2023)\citenamefont
  {Scandolo}, \citenamefont {Pausch},\ and\ \citenamefont
  {Cates}}]{scandolo2023}%
  \BibitemOpen
  \bibfield  {author} {\bibinfo {author} {\bibfnamefont {Mattia}\ \bibnamefont
  {Scandolo}}, \bibinfo {author} {\bibfnamefont {Johannes}\ \bibnamefont
  {Pausch}}, \ and\ \bibinfo {author} {\bibfnamefont {Michael~E.}\ \bibnamefont
  {Cates}},\ }\bibfield  {title} {\enquote {\bibinfo {title} {Active {Ising}
  models of flocking: a field-theoretic approach},}\ }\href {\doibase
  10.1140/epje/s10189-023-00364-w} {\bibfield  {journal} {\bibinfo  {journal}
  {The European Physical Journal E}\ }\textbf {\bibinfo {volume} {46}},\
  \bibinfo {pages} {103} (\bibinfo {year} {2023})},\ \Eprint
  {http://arxiv.org/abs/2306.10791} {arXiv:2306.10791 [cond-mat.stat-mech]}
  \BibitemShut {NoStop}%
\bibitem [{\citenamefont {Kamenev}\ and\ \citenamefont
  {Levchenko}(2009)}]{kamenevLevchenko2009}%
  \BibitemOpen
  \bibfield  {author} {\bibinfo {author} {\bibfnamefont {Alex}\ \bibnamefont
  {Kamenev}}\ and\ \bibinfo {author} {\bibfnamefont {Alex}\ \bibnamefont
  {Levchenko}},\ }\bibfield  {title} {\enquote {\bibinfo {title} {{Keldysh}
  technique and non-linear $\sigma$-model: basic principles and
  applications},}\ }\href {\doibase 10.1080/00018730902850504} {\bibfield
  {journal} {\bibinfo  {journal} {Advances in Physics}\ }\textbf {\bibinfo
  {volume} {58}},\ \bibinfo {pages} {197--319} (\bibinfo {year}
  {2009})}\BibitemShut {NoStop}%
\bibitem [{\citenamefont {Janssen}\ and\ \citenamefont
  {T{\"a}uber}(2005)}]{janssenTauber2005}%
  \BibitemOpen
  \bibfield  {author} {\bibinfo {author} {\bibfnamefont {Hans-Karl}\
  \bibnamefont {Janssen}}\ and\ \bibinfo {author} {\bibfnamefont {Uwe~C.}\
  \bibnamefont {T{\"a}uber}},\ }\bibfield  {title} {\enquote {\bibinfo {title}
  {The field theory approach to percolation processes},}\ }\href {\doibase
  10.1016/j.aop.2004.09.011} {\bibfield  {journal} {\bibinfo  {journal} {Annals
  of Physics}\ }\textbf {\bibinfo {volume} {315}},\ \bibinfo {pages} {147--192}
  (\bibinfo {year} {2005})},\ \Eprint {http://arxiv.org/abs/cond-mat/0409670}
  {arXiv:cond-mat/0409670} \BibitemShut {NoStop}%
\bibitem [{\citenamefont {Canet}\ \emph {et~al.}(2010)\citenamefont {Canet},
  \citenamefont {Chat{\'e}}, \citenamefont {Delamotte},\ and\ \citenamefont
  {Wschebor}}]{canet2010}%
  \BibitemOpen
  \bibfield  {author} {\bibinfo {author} {\bibfnamefont {L{\'e}onie}\
  \bibnamefont {Canet}}, \bibinfo {author} {\bibfnamefont {Hugues}\
  \bibnamefont {Chat{\'e}}}, \bibinfo {author} {\bibfnamefont {Bertrand}\
  \bibnamefont {Delamotte}}, \ and\ \bibinfo {author} {\bibfnamefont
  {Nicol{\'a}s}\ \bibnamefont {Wschebor}},\ }\bibfield  {title} {\enquote
  {\bibinfo {title} {Nonperturbative renormalization group for the
  {Kardar-Parisi-Zhang} equation},}\ }\href {\doibase
  10.1103/PhysRevLett.104.150601} {\bibfield  {journal} {\bibinfo  {journal}
  {Physical Review Letters}\ }\textbf {\bibinfo {volume} {104}},\ \bibinfo
  {pages} {150601} (\bibinfo {year} {2010})},\ \Eprint
  {http://arxiv.org/abs/0905.1025} {arXiv:0905.1025} \BibitemShut {NoStop}%
\bibitem [{\citenamefont {{Dalla Torre}}\ \emph {et~al.}(2012)\citenamefont
  {{Dalla Torre}}, \citenamefont {Demler}, \citenamefont {Giamarchi},\ and\
  \citenamefont {Altman}}]{dallaTorre2012}%
  \BibitemOpen
  \bibfield  {author} {\bibinfo {author} {\bibfnamefont {Emanuele~G.}\
  \bibnamefont {{Dalla Torre}}}, \bibinfo {author} {\bibfnamefont {Eugene}\
  \bibnamefont {Demler}}, \bibinfo {author} {\bibfnamefont {Thierry}\
  \bibnamefont {Giamarchi}}, \ and\ \bibinfo {author} {\bibfnamefont {Ehud}\
  \bibnamefont {Altman}},\ }\bibfield  {title} {\enquote {\bibinfo {title}
  {Dynamics and universality in noise-driven dissipative systems},}\ }\href
  {\doibase 10.1103/PhysRevB.85.184302} {\bibfield  {journal} {\bibinfo
  {journal} {Physical Review B}\ }\textbf {\bibinfo {volume} {85}},\ \bibinfo
  {pages} {184302} (\bibinfo {year} {2012})}\BibitemShut {NoStop}%
\bibitem [{\citenamefont {Fr{\"o}ml}\ \emph {et~al.}(2020)\citenamefont
  {Fr{\"o}ml}, \citenamefont {Muckel}, \citenamefont {Kollath}, \citenamefont
  {Chiocchetta},\ and\ \citenamefont {Diehl}}]{froml2020}%
  \BibitemOpen
  \bibfield  {author} {\bibinfo {author} {\bibfnamefont {Heinrich}\
  \bibnamefont {Fr{\"o}ml}}, \bibinfo {author} {\bibfnamefont {Christopher}\
  \bibnamefont {Muckel}}, \bibinfo {author} {\bibfnamefont {Corinna}\
  \bibnamefont {Kollath}}, \bibinfo {author} {\bibfnamefont {Alessio}\
  \bibnamefont {Chiocchetta}}, \ and\ \bibinfo {author} {\bibfnamefont
  {Sebastian}\ \bibnamefont {Diehl}},\ }\bibfield  {title} {\enquote {\bibinfo
  {title} {Ultracold quantum wires with localized losses: Many-body quantum
  {Zeno} effect},}\ }\href {\doibase 10.1103/PhysRevB.101.144301} {\bibfield
  {journal} {\bibinfo  {journal} {Physical Review B}\ }\textbf {\bibinfo
  {volume} {101}},\ \bibinfo {pages} {144301} (\bibinfo {year}
  {2020})}\BibitemShut {NoStop}%
\bibitem [{\citenamefont {Marino}(2022)}]{marino2022}%
  \BibitemOpen
  \bibfield  {author} {\bibinfo {author} {\bibfnamefont {Jamir}\ \bibnamefont
  {Marino}},\ }\bibfield  {title} {\enquote {\bibinfo {title} {Universality
  class of {Ising} critical states with long-range losses},}\ }\href {\doibase
  10.1103/PhysRevLett.129.050603} {\bibfield  {journal} {\bibinfo  {journal}
  {Physical Review Letters}\ }\textbf {\bibinfo {volume} {129}},\ \bibinfo
  {pages} {050603} (\bibinfo {year} {2022})}\BibitemShut {NoStop}%
\bibitem [{\citenamefont {Coleman}\ and\ \citenamefont
  {Weinberg}(1973)}]{colemanWeinberg1973}%
  \BibitemOpen
  \bibfield  {author} {\bibinfo {author} {\bibfnamefont {Sidney}\ \bibnamefont
  {Coleman}}\ and\ \bibinfo {author} {\bibfnamefont {Erick}\ \bibnamefont
  {Weinberg}},\ }\bibfield  {title} {\enquote {\bibinfo {title} {Radiative
  corrections as the origin of spontaneous symmetry breaking},}\ }\href
  {\doibase 10.1103/PhysRevD.7.1888} {\bibfield  {journal} {\bibinfo  {journal}
  {Physical Review D}\ }\textbf {\bibinfo {volume} {7}},\ \bibinfo {pages}
  {1888--1910} (\bibinfo {year} {1973})}\BibitemShut {NoStop}%
\bibitem [{\citenamefont {Halperin}\ \emph {et~al.}(1974)\citenamefont
  {Halperin}, \citenamefont {Lubensky},\ and\ \citenamefont
  {Ma}}]{halperin1974}%
  \BibitemOpen
  \bibfield  {author} {\bibinfo {author} {\bibfnamefont {B.~I.}\ \bibnamefont
  {Halperin}}, \bibinfo {author} {\bibfnamefont {T.~C.}\ \bibnamefont
  {Lubensky}}, \ and\ \bibinfo {author} {\bibfnamefont {Shang-keng}\
  \bibnamefont {Ma}},\ }\bibfield  {title} {\enquote {\bibinfo {title}
  {First-order phase transitions in superconductors and smectic-{A} liquid
  crystals},}\ }\href {\doibase 10.1103/PhysRevLett.32.292} {\bibfield
  {journal} {\bibinfo  {journal} {Physical Review Letters}\ }\textbf {\bibinfo
  {volume} {32}},\ \bibinfo {pages} {292--295} (\bibinfo {year}
  {1974})}\BibitemShut {NoStop}%
\bibitem [{\citenamefont {Bak}\ \emph {et~al.}(1976)\citenamefont {Bak},
  \citenamefont {Krinsky},\ and\ \citenamefont {Mukamel}}]{bak1976}%
  \BibitemOpen
  \bibfield  {author} {\bibinfo {author} {\bibfnamefont {Per}\ \bibnamefont
  {Bak}}, \bibinfo {author} {\bibfnamefont {S.}~\bibnamefont {Krinsky}}, \ and\
  \bibinfo {author} {\bibfnamefont {D.}~\bibnamefont {Mukamel}},\ }\bibfield
  {title} {\enquote {\bibinfo {title} {First-order transitions, symmetry, and
  the $\epsilon$ expansion},}\ }\href {\doibase 10.1103/PhysRevLett.36.52}
  {\bibfield  {journal} {\bibinfo  {journal} {Physical Review Letters}\
  }\textbf {\bibinfo {volume} {36}},\ \bibinfo {pages} {52--55} (\bibinfo
  {year} {1976})}\BibitemShut {NoStop}%
\bibitem [{\citenamefont {Grassberger}(2006)}]{grassberger2006}%
  \BibitemOpen
  \bibfield  {author} {\bibinfo {author} {\bibfnamefont {Peter}\ \bibnamefont
  {Grassberger}},\ }\bibfield  {title} {\enquote {\bibinfo {title} {Tricritical
  directed percolation in $2+1$ dimensions},}\ }\href {\doibase
  10.1088/1742-5468/2006/01/P01004} {\bibfield  {journal} {\bibinfo  {journal}
  {Journal of Statistical Mechanics: Theory and Experiment}\ }\textbf {\bibinfo
  {volume} {2006}},\ \bibinfo {pages} {P01004} (\bibinfo {year} {2006})},\
  \Eprint {http://arxiv.org/abs/cond-mat/0510428} {arXiv:cond-mat/0510428}
  \BibitemShut {NoStop}%
\bibitem [{\citenamefont {Marcuzzi}\ \emph {et~al.}(2014)\citenamefont
  {Marcuzzi}, \citenamefont {Levi}, \citenamefont {Diehl}, \citenamefont
  {Garrahan},\ and\ \citenamefont {Lesanovsky}}]{marcuzzi2014}%
  \BibitemOpen
  \bibfield  {author} {\bibinfo {author} {\bibfnamefont {Matteo}\ \bibnamefont
  {Marcuzzi}}, \bibinfo {author} {\bibfnamefont {Emanuele}\ \bibnamefont
  {Levi}}, \bibinfo {author} {\bibfnamefont {Sebastian}\ \bibnamefont {Diehl}},
  \bibinfo {author} {\bibfnamefont {Juan~P.}\ \bibnamefont {Garrahan}}, \ and\
  \bibinfo {author} {\bibfnamefont {Igor}\ \bibnamefont {Lesanovsky}},\
  }\bibfield  {title} {\enquote {\bibinfo {title} {Universal nonequilibrium
  properties of dissipative {Rydberg} gases},}\ }\href {\doibase
  10.1103/PhysRevLett.113.210401} {\bibfield  {journal} {\bibinfo  {journal}
  {Physical Review Letters}\ }\textbf {\bibinfo {volume} {113}},\ \bibinfo
  {pages} {210401} (\bibinfo {year} {2014})}\BibitemShut {NoStop}%
\bibitem [{\citenamefont {Overbeck}\ \emph {et~al.}(2017)\citenamefont
  {Overbeck}, \citenamefont {Maghrebi}, \citenamefont {Gorshkov},\ and\
  \citenamefont {Weimer}}]{overbeck2017}%
  \BibitemOpen
  \bibfield  {author} {\bibinfo {author} {\bibfnamefont {Vincent~R.}\
  \bibnamefont {Overbeck}}, \bibinfo {author} {\bibfnamefont {Mohammad~F.}\
  \bibnamefont {Maghrebi}}, \bibinfo {author} {\bibfnamefont {Alexey~V.}\
  \bibnamefont {Gorshkov}}, \ and\ \bibinfo {author} {\bibfnamefont {Hendrik}\
  \bibnamefont {Weimer}},\ }\bibfield  {title} {\enquote {\bibinfo {title}
  {Multicritical behavior in dissipative {Ising} models},}\ }\href {\doibase
  10.1103/PhysRevA.95.042133} {\bibfield  {journal} {\bibinfo  {journal}
  {Physical Review A}\ }\textbf {\bibinfo {volume} {95}},\ \bibinfo {pages}
  {042133} (\bibinfo {year} {2017})}\BibitemShut {NoStop}%
\bibitem [{\citenamefont {Gr{\'e}goire}\ and\ \citenamefont
  {Chat{\'e}}(2004)}]{gregoireChate2004}%
  \BibitemOpen
  \bibfield  {author} {\bibinfo {author} {\bibfnamefont {Guillaume}\
  \bibnamefont {Gr{\'e}goire}}\ and\ \bibinfo {author} {\bibfnamefont {Hugues}\
  \bibnamefont {Chat{\'e}}},\ }\bibfield  {title} {\enquote {\bibinfo {title}
  {Onset of collective and cohesive motion},}\ }\href {\doibase
  10.1103/PhysRevLett.92.025702} {\bibfield  {journal} {\bibinfo  {journal}
  {Physical Review Letters}\ }\textbf {\bibinfo {volume} {92}},\ \bibinfo
  {pages} {025702} (\bibinfo {year} {2004})}\BibitemShut {NoStop}%
\bibitem [{\citenamefont {Solon}\ \emph {et~al.}(2015)\citenamefont {Solon},
  \citenamefont {Chat{\'e}},\ and\ \citenamefont {Tailleur}}]{solon2015}%
  \BibitemOpen
  \bibfield  {author} {\bibinfo {author} {\bibfnamefont {Alexandre~P.}\
  \bibnamefont {Solon}}, \bibinfo {author} {\bibfnamefont {Hugues}\
  \bibnamefont {Chat{\'e}}}, \ and\ \bibinfo {author} {\bibfnamefont {Julien}\
  \bibnamefont {Tailleur}},\ }\bibfield  {title} {\enquote {\bibinfo {title}
  {From phase to microphase separation in flocking models: The essential role
  of nonequilibrium fluctuations},}\ }\href {\doibase
  10.1103/PhysRevLett.114.068101} {\bibfield  {journal} {\bibinfo  {journal}
  {Physical Review Letters}\ }\textbf {\bibinfo {volume} {114}},\ \bibinfo
  {pages} {068101} (\bibinfo {year} {2015})},\ \Eprint
  {http://arxiv.org/abs/1406.6088} {arXiv:1406.6088 [cond-mat.stat-mech]}
  \BibitemShut {NoStop}%
\bibitem [{\citenamefont {Larkin}\ and\ \citenamefont
  {Pikin}(1969)}]{larkinPikin1969}%
  \BibitemOpen
  \bibfield  {author} {\bibinfo {author} {\bibfnamefont {A.~I.}\ \bibnamefont
  {Larkin}}\ and\ \bibinfo {author} {\bibfnamefont {S.~A.}\ \bibnamefont
  {Pikin}},\ }\bibfield  {title} {\enquote {\bibinfo {title} {Phase transitions
  of the first order but nearly of the second},}\ }\href@noop {} {\bibfield
  {journal} {\bibinfo  {journal} {Soviet Physics JETP}\ }\textbf {\bibinfo
  {volume} {29}},\ \bibinfo {pages} {891} (\bibinfo {year} {1969})},\ \bibinfo
  {note} {[Zh. Eksp. Teor. Fiz. \textbf{56}, 1664 (1969)]}\BibitemShut
  {NoStop}%
\bibitem [{\citenamefont {Berges}\ \emph {et~al.}(2002)\citenamefont {Berges},
  \citenamefont {Tetradis},\ and\ \citenamefont {Wetterich}}]{berges2002frg}%
  \BibitemOpen
  \bibfield  {author} {\bibinfo {author} {\bibfnamefont {J.}~\bibnamefont
  {Berges}}, \bibinfo {author} {\bibfnamefont {N.}~\bibnamefont {Tetradis}}, \
  and\ \bibinfo {author} {\bibfnamefont {C.}~\bibnamefont {Wetterich}},\
  }\bibfield  {title} {\enquote {\bibinfo {title} {Non-perturbative
  renormalization flow in quantum field theory and statistical physics},}\
  }\href {\doibase 10.1016/S0370-1573(01)00098-9} {\bibfield  {journal}
  {\bibinfo  {journal} {Physics Reports}\ }\textbf {\bibinfo {volume} {363}},\
  \bibinfo {pages} {223--386} (\bibinfo {year} {2002})}\BibitemShut {NoStop}%
\bibitem [{\citenamefont {Dupuis}\ \emph {et~al.}(2021)\citenamefont {Dupuis},
  \citenamefont {Canet}, \citenamefont {Eichhorn}, \citenamefont {Metzner},
  \citenamefont {Pawlowski}, \citenamefont {Tissier},\ and\ \citenamefont
  {Wschebor}}]{dupuis2021}%
  \BibitemOpen
  \bibfield  {author} {\bibinfo {author} {\bibfnamefont {N.}~\bibnamefont
  {Dupuis}}, \bibinfo {author} {\bibfnamefont {L.}~\bibnamefont {Canet}},
  \bibinfo {author} {\bibfnamefont {A.}~\bibnamefont {Eichhorn}}, \bibinfo
  {author} {\bibfnamefont {W.}~\bibnamefont {Metzner}}, \bibinfo {author}
  {\bibfnamefont {J.~M.}\ \bibnamefont {Pawlowski}}, \bibinfo {author}
  {\bibfnamefont {M.}~\bibnamefont {Tissier}}, \ and\ \bibinfo {author}
  {\bibfnamefont {N.}~\bibnamefont {Wschebor}},\ }\bibfield  {title} {\enquote
  {\bibinfo {title} {The nonperturbative functional renormalization group and
  its applications},}\ }\href {\doibase 10.1016/j.physrep.2021.01.001}
  {\bibfield  {journal} {\bibinfo  {journal} {Physics Reports}\ }\textbf
  {\bibinfo {volume} {910}},\ \bibinfo {pages} {1--114} (\bibinfo {year}
  {2021})}\BibitemShut {NoStop}%
\bibitem [{\citenamefont {Adzhemyan}\ \emph {et~al.}(2022)\citenamefont
  {Adzhemyan}, \citenamefont {Evdokimov}, \citenamefont {Hnati{\v{c}}},
  \citenamefont {Ivanova}, \citenamefont {Kompaniets}, \citenamefont {Kudlis},\
  and\ \citenamefont {Zakharov}}]{adzhemyan2022}%
  \BibitemOpen
  \bibfield  {author} {\bibinfo {author} {\bibfnamefont {L.~Ts.}\ \bibnamefont
  {Adzhemyan}}, \bibinfo {author} {\bibfnamefont {D.~A.}\ \bibnamefont
  {Evdokimov}}, \bibinfo {author} {\bibfnamefont {M.}~\bibnamefont
  {Hnati{\v{c}}}}, \bibinfo {author} {\bibfnamefont {E.~V.}\ \bibnamefont
  {Ivanova}}, \bibinfo {author} {\bibfnamefont {M.~V.}\ \bibnamefont
  {Kompaniets}}, \bibinfo {author} {\bibfnamefont {A.}~\bibnamefont {Kudlis}},
  \ and\ \bibinfo {author} {\bibfnamefont {D.~V.}\ \bibnamefont {Zakharov}},\
  }\bibfield  {title} {\enquote {\bibinfo {title} {The dynamic critical
  exponent $z$ for 2d and 3d {Ising} models from five-loop $\varepsilon$
  expansion},}\ }\href {\doibase 10.1016/j.physleta.2021.127870} {\bibfield
  {journal} {\bibinfo  {journal} {Physics Letters A}\ }\textbf {\bibinfo
  {volume} {425}},\ \bibinfo {pages} {127870} (\bibinfo {year} {2022})},\
  \Eprint {http://arxiv.org/abs/2111.04719} {arXiv:2111.04719} \BibitemShut
  {NoStop}%
\bibitem [{\citenamefont {Wiese}(1998)}]{wiese1998}%
  \BibitemOpen
  \bibfield  {author} {\bibinfo {author} {\bibfnamefont {Kay~J{\"o}rg}\
  \bibnamefont {Wiese}},\ }\bibfield  {title} {\enquote {\bibinfo {title} {On
  the perturbation expansion of the {KPZ} equation},}\ }\href {\doibase
  10.1023/B:JOSS.0000026730.76868.c4} {\bibfield  {journal} {\bibinfo
  {journal} {Journal of Statistical Physics}\ }\textbf {\bibinfo {volume}
  {93}},\ \bibinfo {pages} {143} (\bibinfo {year} {1998})},\ \Eprint
  {http://arxiv.org/abs/cond-mat/9802068} {arXiv:cond-mat/9802068} \BibitemShut
  {NoStop}%
\bibitem [{\citenamefont {L{\"a}ssig}(1995)}]{lassig1995}%
  \BibitemOpen
  \bibfield  {author} {\bibinfo {author} {\bibfnamefont {Michael}\ \bibnamefont
  {L{\"a}ssig}},\ }\bibfield  {title} {\enquote {\bibinfo {title} {On the
  renormalization of the {Kardar--Parisi--Zhang} equation},}\ }\href {\doibase
  10.1016/0550-3213(95)00268-W} {\bibfield  {journal} {\bibinfo  {journal}
  {Nuclear Physics B}\ }\textbf {\bibinfo {volume} {448}},\ \bibinfo {pages}
  {559--574} (\bibinfo {year} {1995})},\ \Eprint
  {http://arxiv.org/abs/cond-mat/9501094} {arXiv:cond-mat/9501094} \BibitemShut
  {NoStop}%
\bibitem [{\citenamefont {Wetterich}(1993)}]{wetterich1993}%
  \BibitemOpen
  \bibfield  {author} {\bibinfo {author} {\bibfnamefont {Christof}\
  \bibnamefont {Wetterich}},\ }\bibfield  {title} {\enquote {\bibinfo {title}
  {Exact evolution equation for the effective potential},}\ }\href {\doibase
  10.1016/0370-2693(93)90726-X} {\bibfield  {journal} {\bibinfo  {journal}
  {Physics Letters B}\ }\textbf {\bibinfo {volume} {301}},\ \bibinfo {pages}
  {90--94} (\bibinfo {year} {1993})}\BibitemShut {NoStop}%
\bibitem [{\citenamefont {Jentsch}\ and\ \citenamefont
  {Lee}(2023)}]{jentschLee2023}%
  \BibitemOpen
  \bibfield  {author} {\bibinfo {author} {\bibfnamefont {Patrick}\ \bibnamefont
  {Jentsch}}\ and\ \bibinfo {author} {\bibfnamefont {Chiu~Fan}\ \bibnamefont
  {Lee}},\ }\bibfield  {title} {\enquote {\bibinfo {title} {Critical phenomena
  in compressible polar active fluids: Dynamical and functional renormalization
  group studies},}\ }\href {\doibase 10.1103/PhysRevResearch.5.023061}
  {\bibfield  {journal} {\bibinfo  {journal} {Physical Review Research}\
  }\textbf {\bibinfo {volume} {5}},\ \bibinfo {pages} {023061} (\bibinfo {year}
  {2023})}\BibitemShut {NoStop}%
\bibitem [{\citenamefont {Blaizot}\ \emph {et~al.}(2006)\citenamefont
  {Blaizot}, \citenamefont {M{\'e}ndez-Galain},\ and\ \citenamefont
  {Wschebor}}]{blaizot2006}%
  \BibitemOpen
  \bibfield  {author} {\bibinfo {author} {\bibfnamefont {Jean-Paul}\
  \bibnamefont {Blaizot}}, \bibinfo {author} {\bibfnamefont {Ram{\'o}n}\
  \bibnamefont {M{\'e}ndez-Galain}}, \ and\ \bibinfo {author} {\bibfnamefont
  {Nicol{\'a}s}\ \bibnamefont {Wschebor}},\ }\bibfield  {title} {\enquote
  {\bibinfo {title} {A new method to solve the non-perturbative renormalization
  group equations},}\ }\href {\doibase 10.1016/j.physletb.2005.10.086}
  {\bibfield  {journal} {\bibinfo  {journal} {Physics Letters B}\ }\textbf
  {\bibinfo {volume} {632}},\ \bibinfo {pages} {571--578} (\bibinfo {year}
  {2006})},\ \Eprint {http://arxiv.org/abs/hep-th/0503103}
  {arXiv:hep-th/0503103} \BibitemShut {NoStop}%
\bibitem [{\citenamefont {Benitez}\ \emph {et~al.}(2012)\citenamefont
  {Benitez}, \citenamefont {Blaizot}, \citenamefont {Chat{\'e}}, \citenamefont
  {Delamotte}, \citenamefont {M{\'e}ndez-Galain},\ and\ \citenamefont
  {Wschebor}}]{benitez2012}%
  \BibitemOpen
  \bibfield  {author} {\bibinfo {author} {\bibfnamefont {F.}~\bibnamefont
  {Benitez}}, \bibinfo {author} {\bibfnamefont {J.-P.}\ \bibnamefont
  {Blaizot}}, \bibinfo {author} {\bibfnamefont {H.}~\bibnamefont {Chat{\'e}}},
  \bibinfo {author} {\bibfnamefont {B.}~\bibnamefont {Delamotte}}, \bibinfo
  {author} {\bibfnamefont {R.}~\bibnamefont {M{\'e}ndez-Galain}}, \ and\
  \bibinfo {author} {\bibfnamefont {N.}~\bibnamefont {Wschebor}},\ }\bibfield
  {title} {\enquote {\bibinfo {title} {Nonperturbative renormalization group
  preserving full-momentum dependence: Implementation and quantitative
  evaluation},}\ }\href {\doibase 10.1103/PhysRevE.85.026707} {\bibfield
  {journal} {\bibinfo  {journal} {Physical Review E}\ }\textbf {\bibinfo
  {volume} {85}},\ \bibinfo {pages} {026707} (\bibinfo {year} {2012})},\
  \Eprint {http://arxiv.org/abs/1110.2665} {arXiv:1110.2665} \BibitemShut
  {NoStop}%
\bibitem [{\citenamefont {Kloss}\ \emph {et~al.}(2012)\citenamefont {Kloss},
  \citenamefont {Canet},\ and\ \citenamefont {Wschebor}}]{kloss2012}%
  \BibitemOpen
  \bibfield  {author} {\bibinfo {author} {\bibfnamefont {Thomas}\ \bibnamefont
  {Kloss}}, \bibinfo {author} {\bibfnamefont {L{\'e}onie}\ \bibnamefont
  {Canet}}, \ and\ \bibinfo {author} {\bibfnamefont {Nicol{\'a}s}\ \bibnamefont
  {Wschebor}},\ }\bibfield  {title} {\enquote {\bibinfo {title}
  {Nonperturbative renormalization group for the stationary
  {Kardar--Parisi--Zhang} equation: Scaling functions and amplitude ratios in
  $1+1$, $2+1$, and $3+1$ dimensions},}\ }\href {\doibase
  10.1103/PhysRevE.86.051124} {\bibfield  {journal} {\bibinfo  {journal}
  {Physical Review E}\ }\textbf {\bibinfo {volume} {86}},\ \bibinfo {pages}
  {051124} (\bibinfo {year} {2012})}\BibitemShut {NoStop}%
\bibitem [{\citenamefont {Canet}(2025)}]{canet2025}%
  \BibitemOpen
  \bibfield  {author} {\bibinfo {author} {\bibfnamefont {L{\'e}onie}\
  \bibnamefont {Canet}},\ }\bibfield  {title} {\enquote {\bibinfo {title} {The
  non-perturbative sides of the {Kardar--Parisi--Zhang} equation},}\ }\href
  {\doibase 10.1088/1742-5468/ae1e75} {\bibfield  {journal} {\bibinfo
  {journal} {Journal of Statistical Mechanics: Theory and Experiment}\ }\textbf
  {\bibinfo {volume} {2025}},\ \bibinfo {pages} {124003} (\bibinfo {year}
  {2025})},\ \Eprint {http://arxiv.org/abs/2509.01472} {arXiv:2509.01472}
  \BibitemShut {NoStop}%
\bibitem [{\citenamefont {Canet}\ \emph {et~al.}(2016)\citenamefont {Canet},
  \citenamefont {Delamotte},\ and\ \citenamefont {Wschebor}}]{canet2016}%
  \BibitemOpen
  \bibfield  {author} {\bibinfo {author} {\bibfnamefont {L{\'e}onie}\
  \bibnamefont {Canet}}, \bibinfo {author} {\bibfnamefont {Bertrand}\
  \bibnamefont {Delamotte}}, \ and\ \bibinfo {author} {\bibfnamefont
  {Nicol{\'a}s}\ \bibnamefont {Wschebor}},\ }\bibfield  {title} {\enquote
  {\bibinfo {title} {Fully developed isotropic turbulence: Nonperturbative
  renormalization group formalism and fixed-point solution},}\ }\href {\doibase
  10.1103/PhysRevE.93.063101} {\bibfield  {journal} {\bibinfo  {journal}
  {Physical Review E}\ }\textbf {\bibinfo {volume} {93}},\ \bibinfo {pages}
  {063101} (\bibinfo {year} {2016})},\ \Eprint {http://arxiv.org/abs/1411.7780}
  {arXiv:1411.7780} \BibitemShut {NoStop}%
\bibitem [{\citenamefont {Polkovnikov}(2010)}]{polkovnikov2010}%
  \BibitemOpen
  \bibfield  {author} {\bibinfo {author} {\bibfnamefont {Anatoli}\ \bibnamefont
  {Polkovnikov}},\ }\bibfield  {title} {\enquote {\bibinfo {title} {Phase space
  representation of quantum dynamics},}\ }\href {\doibase
  10.1016/j.aop.2010.02.006} {\bibfield  {journal} {\bibinfo  {journal} {Annals
  of Physics}\ }\textbf {\bibinfo {volume} {325}},\ \bibinfo {pages}
  {1790--1852} (\bibinfo {year} {2010})}\BibitemShut {NoStop}%
\bibitem [{\citenamefont {Hosseinabadi}\ \emph {et~al.}(2025)\citenamefont
  {Hosseinabadi}, \citenamefont {Chelpanova},\ and\ \citenamefont
  {Marino}}]{hosseinabadi2025}%
  \BibitemOpen
  \bibfield  {author} {\bibinfo {author} {\bibfnamefont {Hossein}\ \bibnamefont
  {Hosseinabadi}}, \bibinfo {author} {\bibfnamefont {Oksana}\ \bibnamefont
  {Chelpanova}}, \ and\ \bibinfo {author} {\bibfnamefont {Jamir}\ \bibnamefont
  {Marino}},\ }\bibfield  {title} {\enquote {\bibinfo {title} {User-friendly
  truncated {Wigner} approximation for dissipative spin dynamics},}\ }\href
  {\doibase 10.1103/1wwv-k7hg} {\bibfield  {journal} {\bibinfo  {journal} {PRX
  Quantum}\ }\textbf {\bibinfo {volume} {6}},\ \bibinfo {pages} {030344}
  (\bibinfo {year} {2025})},\ \Eprint {http://arxiv.org/abs/2503.17443}
  {arXiv:2503.17443} \BibitemShut {NoStop}%
\bibitem [{\citenamefont {Hartmann}\ \emph {et~al.}(2026)\citenamefont
  {Hartmann}, \citenamefont {Schlegel}, \citenamefont {Noel}, \citenamefont
  {Mink},\ and\ \citenamefont {Fleischhauer}}]{hartmann2026}%
  \BibitemOpen
  \bibfield  {author} {\bibinfo {author} {\bibfnamefont {Jens}\ \bibnamefont
  {Hartmann}}, \bibinfo {author} {\bibfnamefont {Tom}\ \bibnamefont
  {Schlegel}}, \bibinfo {author} {\bibfnamefont {Viktoria}\ \bibnamefont
  {Noel}}, \bibinfo {author} {\bibfnamefont {Christopher~D.}\ \bibnamefont
  {Mink}}, \ and\ \bibinfo {author} {\bibfnamefont {Michael}\ \bibnamefont
  {Fleischhauer}},\ }\href {\doibase 10.48550/arXiv.2607.17295} {\enquote
  {\bibinfo {title} {Truncated {Wigner} approximation for spins in continuous
  phase space},}\ } (\bibinfo {year} {2026}),\ \Eprint
  {http://arxiv.org/abs/2607.17295} {arXiv:2607.17295 [quant-ph]} \BibitemShut
  {NoStop}%
\bibitem [{\citenamefont {Plenio}\ and\ \citenamefont
  {Knight}(1998)}]{plenioKnight1998}%
  \BibitemOpen
  \bibfield  {author} {\bibinfo {author} {\bibfnamefont {M.~B.}\ \bibnamefont
  {Plenio}}\ and\ \bibinfo {author} {\bibfnamefont {P.~L.}\ \bibnamefont
  {Knight}},\ }\bibfield  {title} {\enquote {\bibinfo {title} {The quantum-jump
  approach to dissipative dynamics in quantum optics},}\ }\href {\doibase
  10.1103/RevModPhys.70.101} {\bibfield  {journal} {\bibinfo  {journal}
  {Reviews of Modern Physics}\ }\textbf {\bibinfo {volume} {70}},\ \bibinfo
  {pages} {101--144} (\bibinfo {year} {1998})}\BibitemShut {NoStop}%
\bibitem [{\citenamefont {Daley}(2014)}]{daley2014}%
  \BibitemOpen
  \bibfield  {author} {\bibinfo {author} {\bibfnamefont {Andrew~J.}\
  \bibnamefont {Daley}},\ }\bibfield  {title} {\enquote {\bibinfo {title}
  {Quantum trajectories and open many-body quantum systems},}\ }\href {\doibase
  10.1080/00018732.2014.933502} {\bibfield  {journal} {\bibinfo  {journal}
  {Advances in Physics}\ }\textbf {\bibinfo {volume} {63}},\ \bibinfo {pages}
  {77--149} (\bibinfo {year} {2014})}\BibitemShut {NoStop}%
\bibitem [{\citenamefont {Schollw{\"o}ck}(2011)}]{schollwock2011}%
  \BibitemOpen
  \bibfield  {author} {\bibinfo {author} {\bibfnamefont {Ulrich}\ \bibnamefont
  {Schollw{\"o}ck}},\ }\bibfield  {title} {\enquote {\bibinfo {title} {The
  density-matrix renormalization group in the age of matrix product states},}\
  }\href {\doibase 10.1016/j.aop.2010.09.012} {\bibfield  {journal} {\bibinfo
  {journal} {Annals of Physics}\ }\textbf {\bibinfo {volume} {326}},\ \bibinfo
  {pages} {96--192} (\bibinfo {year} {2011})}\BibitemShut {NoStop}%
\bibitem [{\citenamefont {Weimer}\ \emph {et~al.}(2021)\citenamefont {Weimer},
  \citenamefont {Kshetrimayum},\ and\ \citenamefont {Or{\'u}s}}]{weimer2021}%
  \BibitemOpen
  \bibfield  {author} {\bibinfo {author} {\bibfnamefont {Hendrik}\ \bibnamefont
  {Weimer}}, \bibinfo {author} {\bibfnamefont {Augustine}\ \bibnamefont
  {Kshetrimayum}}, \ and\ \bibinfo {author} {\bibfnamefont {Rom{\'a}n}\
  \bibnamefont {Or{\'u}s}},\ }\bibfield  {title} {\enquote {\bibinfo {title}
  {Simulation methods for open quantum many-body systems},}\ }\href {\doibase
  10.1103/RevModPhys.93.015008} {\bibfield  {journal} {\bibinfo  {journal}
  {Reviews of Modern Physics}\ }\textbf {\bibinfo {volume} {93}},\ \bibinfo
  {pages} {015008} (\bibinfo {year} {2021})}\BibitemShut {NoStop}%
\bibitem [{\citenamefont {dos Prazeres}\ \emph {et~al.}(2026)\citenamefont {dos
  Prazeres}, \citenamefont {Hosseinabadi},\ and\ \citenamefont
  {Marino}}]{dos2026kinetically}%
  \BibitemOpen
  \bibfield  {author} {\bibinfo {author} {\bibfnamefont {Luis~Fernando}\
  \bibnamefont {dos Prazeres}}, \bibinfo {author} {\bibfnamefont {Hossein}\
  \bibnamefont {Hosseinabadi}}, \ and\ \bibinfo {author} {\bibfnamefont
  {Jamir}\ \bibnamefont {Marino}},\ }\bibfield  {title} {\enquote {\bibinfo
  {title} {Kinetically constrained superradiance},}\ }\href@noop {} {\bibfield
  {journal} {\bibinfo  {journal} {arXiv e-prints}\ ,\ \bibinfo {pages}
  {arXiv--2605}} (\bibinfo {year} {2026})}\BibitemShut {NoStop}%
\end{thebibliography}%
\end{document}